[ english version / português abaixo ]

# **Disinformation about autism in Latin America and the Caribbean:** Mapping 150 false causes and 150 false cures of ASD in conspiracy theory communities on Telegram

*Ergon Cugler de Moraes Silva* [1]

Getulio Vargas Foundation
São Paulo, São Paulo, Brazil

contato@ergoncugler.com
www.ergoncugler.com

*Arthur Ataide Ferreira Garcia* [2]

Metropolitan Univ. of Santos
Santos, São Paulo, Brazil

arthur.afg.2003@gmail.com
www.autistas.org.br

*Guilherme de Almeida* [3]

State University of Campinas
Campinas, São Paulo, Brazil

g229669@dac.unicamp.br
www.autistas.org.br

*Julie Ricard* [4]

Getulio Vargas Foundation
São Paulo, São Paulo, Brazil

juliec.ricard@gmail.com
www.eureka.club

## Abstract

How do conspiracy theory communities in Latin America and the Caribbean structure, articulate, and sustain the dissemination of disinformation about autism? To answer this question, this research investigates the structuring, articulation, and promotion of autism-related disinformation in conspiracy theory communities in Latin America and the Caribbean. By analyzing publications from 1,659 Telegram communities over ten years (2015 – 2025) and examining more than 58 million pieces of shared content from approximately 5.3 million users, this study explores how false narratives about autism are promoted, including unfounded claims about its causes and promises of miraculous cures. The adopted methodology combines network analysis, time series analysis, thematic clustering, and content analysis, enabling the identification of dissemination patterns, key influencers, and interconnections with other conspiracy theories. Among the key findings, Brazilian communities stand out as the leading producers and distributors of these narratives in the region, accounting for 46% of the analyzed content. Additionally, there has been an exponential 15,000% (x151) increase in the volume of autism-related disinformation since the COVID-19 pandemic in Latin America and the Caribbean, highlighting the correlation between health crises and the rise of conspiracy beliefs. The research also reveals that false cures, such as chlorine dioxide (CDS), ozone therapy, and extreme diets, are widely promoted within these communities and commercially exploited, often preying on desperate families in exchange for money. By addressing the research question, this study aims to contribute to the understanding of the disinformation ecosystem and proposes critical reflections on how to confront these harmful narratives.

[ english version / português abaixo ]

**Key findings:**

➔ **Over 100 million views and 4 million users reached:** Conspiracy theories involving autism have reached at least 4,186,031 users in Latin America and the Caribbean between 2015 and 2025, totaling 99,318,993 views, 107,880 reactions in 47,261 mapped and categorized posts across the continent.

➔ **Brazil ranked #1 in Latin-America and the Caribbean:** Brazilian conspiracy theory communities account for 46% of autism-related content in the region, totaling 22,007 posts, reaching up to 1,726,364 users and 13,944,477 views. Following Brazil, Argentina, Mexico, Venezuela, and Colombia also stand out among the countries that produce the most conspiratorial content about autism.

➔ **150 false causes of autism, from parasites to Doritos:** Among the mapped versions, explanations for the "cause" of autism range from serotonin deficiency and aluminum exposure to claims involving Doritos consumption, Earth's magnetic field reversal, and chemtrail influence. Other theories resort to moral panic and scientific denialism, attributing autism to 5G, Wi-Fi, microwaves, and even vaccines.

➔ **150 false cures for autism, from chlorine dioxide to Tesla electroshock:** The false promise of a cure for autism has become a lucrative business, driven by disinformation and opportunism. Among the 150 identified "cures", dangerous practices stand out, such as the consumption of chlorine dioxide (CDS), known as "MMS", a toxic substance promoted as a miraculous solution. Additionally, absurd methods like ozone therapy, Tesla electroshock therapy, and even the ingestion of colloidal silver and methylene blue are marketed as supposedly effective treatments. Many of these products and practices are openly sold by groups that exploit the desperation of families, profiting from the monetization of lies and putting lives at risk.

➔ **15,000% growth in autism-related conspiracy theories post-pandemic:** The COVID-19 pandemic was the gateway to the explosion of autism-related disinformation on the continent. Between 2019 and 2024 (five years), the volume of misleading content grew by more than 15,000% (x150), with a significant 635% (x7.35) increase during the pandemic period (2020 – 2021) alone. This accelerated growth demonstrates how the health crisis created space for conspiracy narratives that continued to expand in the following years.

## 1. Introduction

Over the past years, conspiracy theory communities on Telegram have become a significant vector for the dissemination of disinformation across Latin America and the Caribbean. Building upon previous research focused exclusively on Brazil, this study expands the scope of analysis to a broader regional perspective, covering multiple countries and linguistic variations. By systematically navigating through **1,659 conspiracy theory communities** and analyzing **58,521,152 pieces of content** shared by **5,310,728 users**, this research seeks to map, characterize, and understand the structure and dynamics of disinformation networks concerning autism.

Unlike previous studies that concentrated on vaccine hesitancy and off-label treatments, this investigation specifically examines the spread of falsehoods related to autism. Across multiple digital environments, conspiracy narratives surrounding autism range from baseless claims about its causes — such as links to vaccines, heavy metals, or dietary factors

— to misleading or outright fraudulent "cures" or "treatments" that promote harmful treatments like chelation therapy, hyperbaric oxygen chambers, and chlorine dioxide consumption (commonly known as "MMS" or "CDS"). By analyzing the evolution, interconnections, and ideological overlaps of these narratives, this study aims to provide a comprehensive view of how autism-related disinformation spreads, who its main promoters are, and what strategies they employ to gain credibility among different audiences.

To accomplish this, we apply a systematic methodology consistent across a series of studies, ensuring comparability and reproducibility of findings. This approach involves network analysis, time series evaluations, thematic clustering, and content classification to determine how conspiracy communities engage with and reinforce autism-related disinformation. Moreover, the study highlights regional specificities, considering the sociopolitical and cultural nuances that shape the adoption of conspiracy theories in different Latin American and Caribbean contexts.

The dataset comprises groups and users from a diverse set of countries, including **Brazil, Uruguay, Colombia, Peru, Mexico, Venezuela, Ecuador, Paraguay, Argentina, Chile, Bolivia, Costa Rica, Honduras, Panama, the Dominican Republic, Puerto Rico, Cuba, Guatemala, and Jamaica**. While some groups operate within national boundaries, others function as transnational hubs, facilitating the rapid dissemination of disinformation across linguistic and geopolitical borders. This transnational dimension underscores the need for comparative analysis and a deeper understanding of how digital disinformation infrastructures transcend individual nations.

A critical aspect of this study is its ethical commitment to data protection and user privacy, adhering to regional legislation such as Brazil's General Data Protection Law (Law No. 13,709/2018) and equivalent frameworks across other Latin American and Caribbean countries. The study does not seek to expose individuals but rather to analyze patterns, networks, and strategies used to propagate disinformation.

Given the widespread consequences of autism-related disinformation — including its impact on public health, parental decision-making, and neurodivergent individuals — this study raises a central question: **How do conspiracy theory communities across Latin America and the Caribbean structure, articulate, and promote the spread of disinformation about autism?** By addressing this, the research contributes to the broader fight against disinformation and helps inform policies and digital literacy efforts aimed at mitigating the harm caused by these narratives.

## 2. Literature review and rationale

Conspiracy thinking has become a pervasive force in contemporary public discourse, shaping attitudes not only toward politics and institutions but also toward science and health. Scholars have shown that such thinking is often fueled by low institutional trust, perceived loss of control, and a polarized media environment. In particular, Enders (2019) and Albertson & Guiler (2020) have identified that belief in conspiracy theories correlates strongly with partisanship, ideological extremism, and distrust of democratic processes. The COVID-19 pandemic accelerated this trend, thriving on and fueling conspiratorial thinking and distrust about scientific processes and institutions.

Studies in conspiracy theory literature emphasize how conspiracy beliefs can foster authoritarian attitudes, electoral distrust, and rejection of pluralism (Stoeckel, 2023; Jiang, 2021). For instance, Pickel et al. found that COVID-related conspiracies were associated with rising support for authoritarian governance, while Czech (2022) examined how Polish Catholic nationalism integrates religious identity with conspiracy logics that oppose secular democratic norms. Similarly, Yendell & Herbert (2022) demonstrate that religious fundamentalism is often predictive of conspiratorial thinking, especially when framed in terms of spiritual warfare or cosmic struggle between good and evil. These religious and political dimensions are especially relevant in Latin America and among Latino diasporas, where Cortina & Rottinghaus (2020) found high levels of conspiratorial thinking linked to institutional marginalization, low civic trust, and ethnicized political discourse.

Crucially, conspiracy thinking often provides fertile ground for the adoption of pseudoscientific health beliefs. These two epistemic systems are deeply interconnected: both challenge mainstream authority, rely on anecdotal or intuitive reasoning, and thrive in digital environments. This relationship is evident in the context of health-related conspiracy theories, which often thrive on distrust in scientific processes and institutions. Such thinking can lead to the endorsement of pseudoscientific claims, as seen during the COVID-19 pandemic, where misinformation and conspiracy theories about the virus and vaccines proliferated.

Pseudoscientific claims are frequently presented as narrative messaging, offering emotionally satisfying stories of miraculous cures, hidden truths, or elite cover-ups. Underlying these beliefs is often a perception of cause – effect relationships that feel intuitively true but lack empirical grounding. Chow et al. (2021) found that individuals frequently endorse health myths based on perceived contingency — the belief that one action causes a particular health outcome, even when evidence is absent or contradictory. These perceptions are not purely cognitive but are often tied to paranormal or magical thinking, revealing a deeper desire for control, meaning, or hope in situations of uncertainty, such as caring for an autistic child.

In the case of autism-related disinformation, these systems converge into powerful narratives that offer false causes and cures, at the expense of scientific evidence and public health – as evidenced ahead. Such beliefs may dangerously affect health-related behavior and public health more broadly. As seen in the case of vaccines, conspiracy beliefs have been shown to shape health behaviors during the pandemic (Kowalska-Duplaga & Duplaga, 2023), and reduce adherence to preventive measures – such as vaccine uptake – (Prooijen, 2022).

## 3. Methodology

The methodology of this study is organized into three subsections: **2.1. Data extraction**, which describes the process and tools used to collect information from Telegram communities; **2.2. Data processing**, which discusses the criteria and methods applied to classify and anonymize the collected data; and **2.3. Approaches to data analysis**, which details the techniques used to investigate the connections, temporal series, content, and thematic overlaps within conspiracy theory communities.

### 3.1. Data extraction

This project began in February 2023 with the publication of the first version of TelegramScrap (Silva, 2023), a proprietary, free, and open-source tool that utilizes Telegram's Application Programming Interface (API) by Telethon library and organizes data extraction cycles from groups and open channels on Telegram. Over the months, the database was expanded and refined using four approaches to identifying conspiracy theory communities:

**(i) Use of keywords:** At the project's outset, a list of keywords was compiled to facilitate the direct identification of conspiracy theory groups and channels on Telegram across Latin America and the Caribbean. These keywords included *apocalypse*, *survivalism*, *climate change*, *flat earth*, *conspiracy theory*, *globalism*, *new world order*, *occultism*, *esotericism*, *alternative cures*, *QAnon*, *reptilians*, *revisionism*, *aliens*, among others. The search was conducted in both Portuguese and Spanish to ensure broad regional coverage. Initially, this approach enabled the identification of communities whose titles and/or descriptions explicitly referenced conspiracy theories. However, over time, it became evident that many groups deliberately used variations in spelling, special characters, or coded language to avoid detection. To address this challenge, the keyword list was refined iteratively based on observed linguistic patterns in different countries, expanding the dataset beyond Brazilian groups to include communities from Uruguay, Colombia, Peru, Mexico, Venezuela, Ecuador, Paraguay, Argentina, Chile, Bolivia, Costa Rica, Honduras, Panama, the Dominican Republic, Puerto Rico, Cuba, Guatemala, and Jamaica.

**(ii) Telegram channel recommendation mechanism:** A key discovery during the investigation was the use of Telegram's recommendation system for channels (but not groups). Telegram automatically suggests up to ten "similar channels" when viewing a specific channel, based on its content. This mechanism was instrumental in identifying additional Latin American and Caribbean conspiracy theory communities beyond those detected through keyword searches. By leveraging these recommendations, it was possible to uncover a wider network of interconnected communities, many of which had no explicit references to conspiracy theories in their descriptions but were thematically aligned.

**(iii) Snowball approach for invitation identification:** After identifying an initial set of communities for extraction, a proprietary algorithm was developed to detect and analyze URLs containing the prefix "t.me/", which is used for invitations to Telegram groups and channels. This process allowed for the accumulation of hundreds of thousands of links, which were then ranked based on frequency and cross-referenced to identify new groups. The snowball approach was crucial in revealing additional conspiracy theory communities across Latin America and the Caribbean, as many invitations circulated internally within already identified groups. The iterative repetition of this method ensured continuous discovery and expansion of the dataset.

**(iv) Expansion to tweets published on X mentioning invitations:** To further diversify the sources of conspiracy theory communities, a proprietary search query was developed to identify Telegram invitations shared on X (formerly Twitter). The search focused on conspiracy theory-related keywords combined with the "t.me/" prefix, using the syntax: https://x.com/search?q=lang%3Aes%20%22t.me%2F%22%20SEARCH-TERM and https://x.com/search?q=lang%3Apt%20%22t.me%2F%22%20SEARCH-TERM. This method proved particularly effective for capturing links to transnational conspiracy theory groups, which often use X as a dissemination platform to attract new members.

By implementing these multiple identification strategies — developed over months of methodological refinement — it was possible to construct a comprehensive database of **1,649 conspiracy theory communities** across Latin America and the Caribbean. Collectively, these communities have **published 58,637,137 pieces of content** between **December 2015 and January 2025**, with a combined total of **5,345,332 users**. It is important to note that this figure includes two considerations: first, the user count is variable, as members frequently join and leave communities, meaning the total represents a snapshot from the data extraction period; second, the same user may be a member of multiple groups, leading to possible duplication in the overall count. While the actual number of unique individuals engaging with conspiracy theory content may be lower due to overlapping memberships, the scale of participation remains significant across the Latin American and Caribbean digital landscape.

### 3.2. Data processing

With all the conspiracy theory groups and channels on Telegram extracted from across Latin America and the Caribbean, a manual classification process was conducted based on the title and description of each community. If there was an explicit mention in the title or description referring to a specific theme, the community was categorized into one of the following classifications: (i) "Anti-Science"; (ii) "Anti-Woke and Gender"; (iii) "Antivax"; (iv) "Apocalypse and Survivalism"; (v) "Climate Changes"; (vi) "Flat Earth"; (vii) "Globalism"; (viii) "New World Order"; (ix) "Occultism and Esotericism"; (x) "Off Label and Quackery"; (xi) "QAnon"; (xii) "Reptilians and Creatures"; (xiii) "Revisionism and Hate Speech"; (xiv) "UFO and Universe". If no explicit reference to these themes was found in the title or description, the community was classified under (xv) General Conspiracy. Given the expanded scope of this study, the classification process accounted for variations in language and terminology across different Latin American and Caribbean countries, particularly in Spanish- and Portuguese-speaking communities. Additionally, certain themes exhibited regional nuances, such as the presence of specific political conspiracy narratives more prevalent in particular nations. In the following tables, we present the metrics related to the classification of conspiracy theory communities across Latin America and the Caribbean.

**Table 01.** Conspiracy theory community countries (metrics up to January 2025).

[ english version / português abaixo ]

| Country | Groups | Users | Contents | Comments | Total |
|---|---|---|---|---|---|
| Argentina | 62 | 545,594 | 1,459,065 | 4,796,166 | 6,255,231 |
| Bolivia | 09 | 4,622 | 96,010 | 2,347 | 98,357 |
| Brasil | 1,000 | 2,537,760 | 15,779,699 | 16,110,578 | 31,890,277 |
| Chile | 43 | 76,375 | 469,916 | 431,154 | 901,070 |
| Colombia | 71 | 152,946 | 1,121,331 | 2,075,572 | 3,196,903 |
| Costa Rica | 13 | 7,728 | 128,306 | 11,095 | 139,401 |
| Cuba | 01 | 71 | 439 | 00 | 439 |
| Ecuador | 29 | 15,559 | 111,280 | 662,661 | 773,941 |
| Guatemala | 03 | 97 | 159 | 22 | 181 |
| Honduras | 02 | 299 | 1,551 | 76 | 1,627 |
| Jamaica | 01 | 25 | 22 | 00 | 22 |
| México | 59 | 318,740 | 692,847 | 2,171,015 | 2,863,862 |
| Panamá | 07 | 4,343 | 15,704 | 3,695 | 19,399 |
| Paraguay | 12 | 9,430 | 49,317 | 16,210 | 65,527 |
| Perú | 35 | 47,341 | 1,149,962 | 305,918 | 1,455,880 |
| Puerto Rico | 03 | 2,026 | 3,080 | 626 | 3,706 |
| República Dominicana | 01 | 12 | 29 | 10 | 39 |
| Transnacional | 291 | 1,555,733 | 3,648,008 | 6,289,729 | 9,937,737 |
| Uruguay | 08 | 14,987 | 115,012 | 223,401 | 338,413 |
| Venezuela | 09 | 17,040 | 73,699 | 505,441 | 579,140 |
| **Total** | **1,659** | **5,310,728** | **24,915,436** | **33,605,716** | **58,521,152** |

Source: Own elaboration (2025).

**Table 02.** Conspiracy theory community categories (metrics up to January 2025).

| Category | Groups | Users | Contents | Comments | Total |
|---|---|---|---|---|---|
| Anti-Science | 31 | 102,163 | 323,012 | 982,414 | 1,305,426 |
| Anti-Woke and Gender | 57 | 181,257 | 636,734 | 1,981,732 | 2,618,466 |
| Antivax | 280 | 985,438 | 3,562,816 | 3,775,987 | 7,338,803 |
| Apocalypse and Survivalism | 44 | 169,568 | 1,503,161 | 583,290 | 2,086,451 |
| Climate Changes | 43 | 46,154 | 504,505 | 171,285 | 675,790 |
| Flat Earth | 52 | 48,725 | 556,063 | 1,337,079 | 1,893,142 |
| General Conspiracy | 219 | 1,016,301 | 5,102,585 | 7,595,481 | 12,698,066 |
| Globalism | 62 | 504,759 | 1,246,500 | 1,097,291 | 2,343,791 |
| NWO | 195 | 607,731 | 3,985,405 | 5,888,434 | 9,873,839 |
| Occultism and Esotericism | 83 | 208,393 | 1,801,951 | 2,285,810 | 4,087,761 |
| Off Label and Quackery | 343 | 966,010 | 3,013,077 | 6,839,001 | 9,852,078 |
| QAnon | 42 | 126,342 | 893,353 | 273,350 | 1,166,703 |
| Reptilians and Creatures | 28 | 128,543 | 180,786 | 70,342 | 251,128 |
| Revisionism and Hate Speech | 108 | 111,628 | 495,319 | 265,194 | 760,513 |
| UFO and Universe | 72 | 107,716 | 1,110,169 | 459,026 | 1,569,195 |
| **Total** | **1,659** | **5,310,728** | **24,915,436** | **33,605,716** | **58,521,152** |

Source: Own elaboration (2025).

Additionally, it is important to emphasize that only open communities were extracted for this study. These are groups and channels that are not only publicly identifiable but also allow unrestricted access to their content, meaning that any Telegram user can join and view the discussions without requiring approval or an invitation. Furthermore, in compliance with regional data protection regulations, particularly Brazil's Lei Geral de Proteção de Dados Pessoais (LGPD – Law No. 13,709/2018), all extracted data were fully anonymized to ensure privacy and ethical research standards. This anonymization process extends beyond user-level data to include community identification, meaning that no specific group, channel, or participant can be traced through this study. Thus, while the data analyzed originate from publicly available content, additional layers of privacy protection were implemented to prevent direct attribution, reinforcing the ethical commitment to user confidentiality.

### 3.3. Approaches to data analysis

**(i) Descriptive Analysis of Autism-Related Claims:** This study systematically examines **150 alleged causes** and **150 alleged cures** for autism, as propagated within conspiracy theory communities on Telegram. Through a structured classification framework, each claim is categorized based on its thematic origin (e.g., environmental factors, vaccines, dietary influences, genetic theories) and the nature of the purported cure (e.g., alternative medicine, detox protocols, pharmaceutical claims, pseudoscientific interventions). By mapping these narratives, the analysis aims to identify patterns, ideological motivations, and cross-referencing mechanisms that sustain and amplify autism-related disinformation. This descriptive approach provides a foundational understanding of how disinformation is structured within these communities and how it evolves over time.

**(ii) Network Analysis:** This methodological framework allows us to assess whether these communities serve as self-referential sources of legitimacy or whether they systematically cross-promote other conspiracy theories, broadening their users' engagement with disinformation networks. Additionally, this study builds upon the network analysis approach of Rocha *et al.* (2024), who applied a similar technique to Telegram communities by identifying content similarities based on unique message IDs.

**(iii) Time Series Analysis:** To structure and analyze temporal trends, the *Pandas* library (McKinney, 2010) was employed to organize data frames, facilitating the observation of: (a) the volume of publications over time and (b) user engagement dynamics, including views, reactions, and comments extracted from the platform's metadata. Beyond raw volumetric analysis, the *Plotly* library (Plotly Technologies Inc., 2015) was utilized to generate visual representations, making it possible to track temporal fluctuations in content production and interaction levels.

**(iv) Content Analysis:** In addition to general word frequency analysis, time series were applied to assess the evolution of the most frequently used terms over time. The dataset spans from **July 2017** (earliest publications about autism) to **January 2025** (study period). Using the *Pandas* (McKinney, 2010) and *WordCloud* (Mueller, 2020) libraries, the study presents both volumetric and graphical representations of term usage patterns, offering insights into thematic shifts and the persistence of specific narratives.

In summary, the methodology applied encompassed the entire research process, from data extraction — conducted using the proprietary TelegramScrap tool (Silva, 2023) — to the processing and analysis of collected data. A diverse set of analytical approaches was employed to systematically identify, classify, and examine conspiracy theory communities on Telegram across Latin America and the Caribbean. Each stage of the study was designed to uphold data integrity and ensure strict adherence to regional data protection regulations, particularly Brazil's Lei Geral de Proteção de Dados (LGPD – Law No. 13,709/2018). The following sections present the study's findings, providing an in-depth exploration of the structural dynamics, thematic trends, and interconnections within the analyzed communities.

## 4. Results

The geographical distribution of mapped publications on autism reveals a significant concentration of groups and content in Brazil, followed by a substantial number of interactions in transnational communities. Countries such as Colombia, Argentina, and Mexico also show a considerable volume of publications and discussions on the topic, highlighting the reach and relevance of autism-related narratives in different national contexts. The high number of users and content recorded in Brazilian groups indicates intense participation and engagement in the dissemination of information — or disinformation — about autism, making it essential to conduct a qualitative analysis of these interactions to understand their implications.

**Table 03.** Mapped publications on autism by country (metrics up to January 2025).

| Country | Groups | Users | Contents | Comments | Total |
|---|---|---|---|---|---|
| Brasil | 487 | 1,726,364 | 10,591 | 11,416 | 22,007 |
| Transnacional | 194 | 1,385,420 | 5,039 | 3,126 | 8,165 |
| Colombia | 57 | 112,453 | 1,567 | 1,179 | 2,746 |
| Argentina | 45 | 528,190 | 1,517 | 4,299 | 5,816 |
| México | 42 | 261,407 | 1,178 | 1,770 | 2,948 |
| Perú | 21 | 43,543 | 675 | 331 | 1,006 |
| Chile | 26 | 62,086 | 582 | 558 | 1,140 |
| Venezuela | 06 | 16,869 | 262 | 2,276 | 2,538 |
| Bolivia | 04 | 4,295 | 184 | 06 | 190 |
| Costa Rica | 06 | 6,085 | 160 | 06 | 166 |
| Ecuador | 17 | 13,339 | 126 | 51 | 177 |
| Paraguay | 05 | 4,780 | 73 | 19 | 92 |
| Uruguay | 04 | 14,857 | 56 | 174 | 230 |
| Puerto Rico | 01 | 1,869 | 23 | 04 | 27 |
| Panamá | 06 | 4,339 | 09 | 01 | 10 |
| Honduras | 01 | 135 | 02 | 01 | 03 |
| **Total** | **922** | **4,186,031** | **22,044** | **25,217** | **47,261** |

Source: Own elaboration (2025).

In addition to the country-based analysis, the classification by thematic categories provides a more detailed view of the main approaches within discussions on autism. As shown in Table 04, there is a significant intersection between this topic and issues related to the anti-vaccine movement, "globalism", and broader conspiracy theories, such as the "New World Order" (NWO). The high volume of content and interactions within these categories suggests that autism frequently appears in debates marked by disinformation and controversial narratives, reinforcing the need for monitoring and effective strategies to counter the spread of discourses that may negatively impact society.

**Table 04.** Mapped publications on autism by category (metrics up to January 2025).

| Category | Groups | Users | Contents | Comments | Total |
|---|---|---|---|---|---|
| Off Label and Quackery | 200 | 803,150 | 7,321 | 7,805 | 15,126 |
| Antivax | 183 | 877,305 | 5,140 | 3,915 | 9,055 |
| General Conspiracy | 129 | 849,959 | 2,827 | 2,150 | 4,977 |
| NWO | 119 | 456,193 | 2,414 | 4,293 | 6,707 |
| Occultism and Esotericism | 41 | 138,320 | 929 | 2,984 | 3,913 |
| Apocalypse and Survivalism | 24 | 138,720 | 683 | 230 | 913 |
| Globalism | 43 | 363,933 | 549 | 609 | 1,158 |
| Anti-Woke and Gender | 30 | 105,511 | 425 | 960 | 1,385 |
| Climate Changes | 28 | 30,957 | 358 | 23 | 381 |
| UFO and Universe | 19 | 67,227 | 349 | 148 | 497 |
| QAnon | 28 | 115,696 | 301 | 133 | 434 |
| Flat Earth | 17 | 25,827 | 238 | 632 | 870 |
| Revisionism and Hate Speech | 36 | 61,551 | 231 | 205 | 436 |
| Anti-Science | 17 | 53,846 | 213 | 1,117 | 1,330 |
| Reptilians and Creatures | 08 | 97,836 | 66 | 13 | 79 |
| **Total** | **922** | **4,186,031** | **22,044** | **25,217** | **47,261** |

Source: Own elaboration (2025).

The results are detailed below in the order outlined in the methodology.

## 4.1. Descriptive

### 4.1.1. False causes of autism

One of the most widespread narratives associates autism with electromagnetic radiation, the Earth's magnetic field reversal, and pesticide use. These claims suggest that technological modernization and intensive chemical product usage are directly linked to the increase in ASD diagnoses. The electromagnetic radiation emitted by cell phones, Wi-Fi networks, and 5G antennas is non-ionizing, meaning it does not have the capacity to modify human DNA or cause brain alterations. Scientific studies have demonstrated that this form of radiation does not pose a risk to neurological development.

The idea that the Earth's magnetic field reversal causes autism lacks any evidence. This phenomenon occurs in geological cycles, and there is no historical record of neurological impacts on humans. Pesticides, in turn, are often pointed to as one of the environmental causes of autism. Although prolonged exposure to pesticides can have health impacts, there are no studies proving a causal relationship between pesticide use and ASD development.

Furthermore, a concerning branch of this disinformation suggests that parasites are responsible for autism, promoting so-called deworming protocols. This type of practice is dangerous, as it leads parents to administer toxic substances, such as chlorine dioxide (MMS), along with other harmful deworming protocols, giving false hope to caregivers of autistic children for a miraculous cure. Autism is not caused by parasites, and there is no scientific basis to justify antiparasitic treatments for this condition.

[ english version / português abaixo ]

**Figure 01.** Examples of conspiracy theories about autism causes.

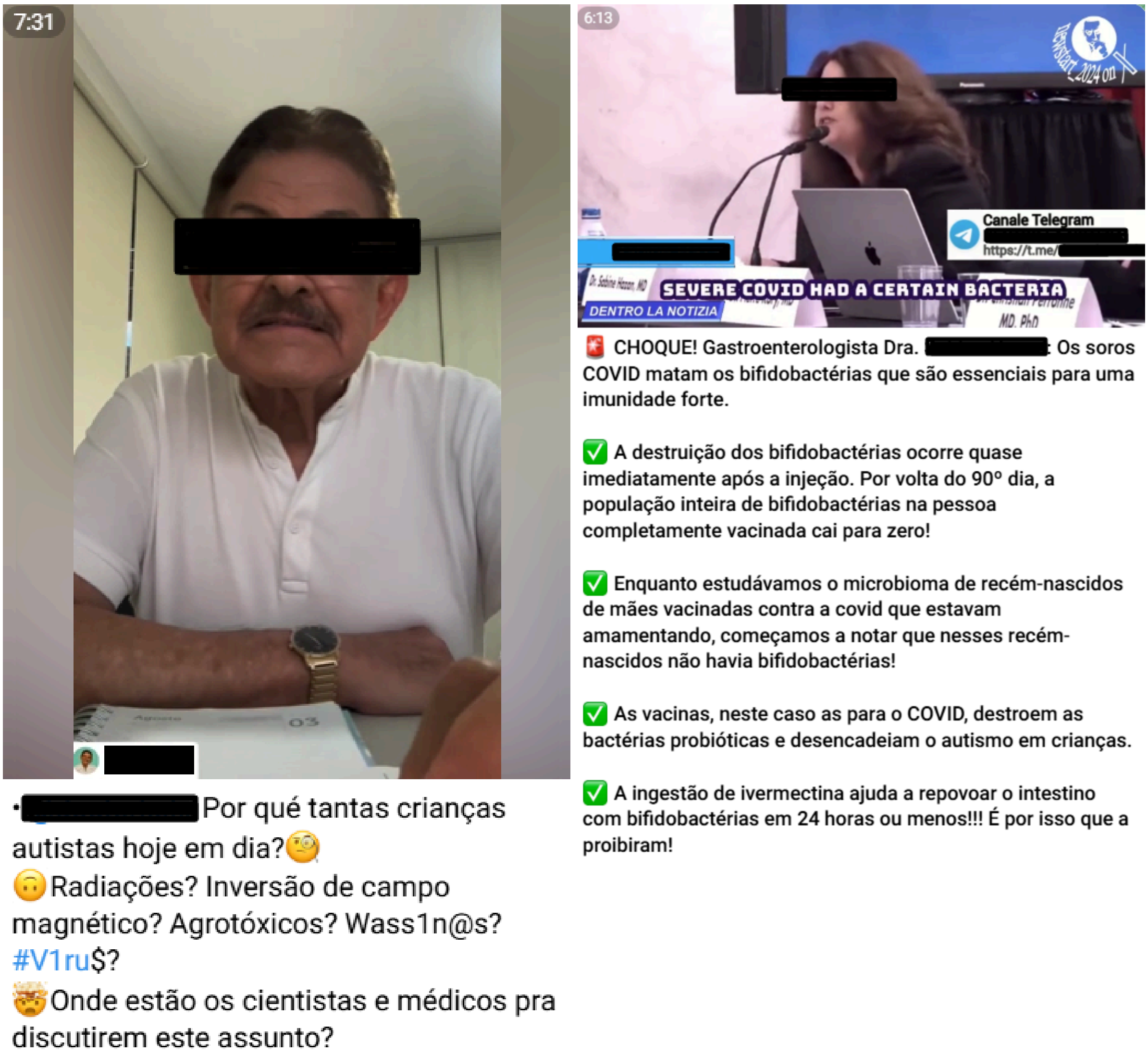

Source: Screenshot (2025).

Another widely circulated narrative in conspiracy communities suggests that vegetable oils and ultra-processed foods, such as Doritos, could cause autism. The argument is based on the claim that omega-6 fatty acids, found in oils like soybean, canola, and sunflower, act as a "metabolic poison", promoting brain inflammation. Although a balance between omega-3 and omega-6 is important for overall health, there is no study directly linking the consumption of these oils to autism. Similarly, the presence of genetically modified ingredients in processed foods, such as transgenic corn in snacks, has no scientific correlation with ASD development. This narrative fits into a broader trend of food-related disinformation, which demonizes certain food groups without scientific basis, diverting attention from real discussions about nutrition and health.

**Figure 02.** Examples of conspiracy theories about autism causes.

**Este óleo vegetal está vinculado a doenças crônicas**

**Óleo de soja pode causar mudanças irreversíveis em seu cérebro**

**Resumo da matéria**:
• Óleos vegetais e óleos de sementes como soja, canola, girassol, semente de uva, milho, cártamo, amendoim e óleo de farelo de arroz são carregados com ácido linoleico ômega-6 (LA), que atua como um veneno metabólico quando consumido em excesso. Qualquer coisa acima de 10 gramas por dia causará problemas a longo prazo;
• Os óleos de sementes são incrivelmente pró-inflamatórios e promovem a oxidação em seu corpo. Essa oxidação, por sua vez, desencadeia a disfunção mitocondrial que então impulsiona o processo da doença;
• O óleo de soja demonstrou causar alterações genéticas irreversíveis no cérebro de camundongos. Isso foi verdade tanto para o óleo de soja não modificado quanto para o óleo de soja modificado para ser baixo em AL.; • Ambos produziram efeitos pronunciados no hipotálamo, que regula o metabolismo e as respostas ao estresse;
• Vários genes nos camundongos que foram alimentados com óleo de soja não estavam funcionando corretamente, incluindo um gene que produz ocitocina, o "hormônio do amor". Cerca de 100 outros genes também foram afetados. Essas mudanças podem ter ramificações para o metabolismo energético, função cerebral adequada e doenças como autismo e doença de Parkinson;

Por que você não deveria comer Doritos de qualquer maneira

1. Doritos contém ingredientes OGM, como o primeiro ingrediente, milho

Os OGM usados para fazer Doritos são considerados cancerígenos que têm sido associados ao câncer de mama, autismo, alergias ao glúten, diabetes, inflamação e distúrbios que afetam os sistemas digestivo e reprodutivo.

2. É feito com uma grande quantidade de óleos hidrogenados processados comercialmente

Isso pode levar a um aumento de radicais livres no corpo. Eles também são geneticamente modificados e carregados com gorduras trans, que podem causar inflamação, imunidade comprometida, aumento da circulação de estrogênio ruim e falta de nutrientes.

3. É feito com corantes, incluindo corante amarelo 5, corante amarelo 6 e corante vermelho 40.

Os efeitos colaterais de longo prazo da ingestão de corantes incluem distúrbios imunológicos, A.D.D. e TDAH, especialmente em crianças.

Source: Screenshot (2025).

As it relates to vaccines, the supposed release of chemical substances into the atmosphere by airplanes (chemtrails), and the HAARP project, a U.S. atmospheric research program. Aluminum is one of the most abundant elements in nature and is naturally present in food, water, and cosmetic products. The small amount used as an adjuvant in vaccines poses no risk to human health and is rapidly eliminated by the body.

Chemtrails are another conspiracy theory that does not provide any scientific proof. The trails left by airplanes are merely water vapor condensation and have no relation to human health. HAARP (High-Frequency Active Auroral Research Program), frequently mentioned in conspiracy theories, is a research program on the Earth's ionosphere. There is no evidence that it has the capacity to manipulate minds or influence neurological conditions such as autism.

**Figure 03.** Examples of conspiracy theories about autism causes.

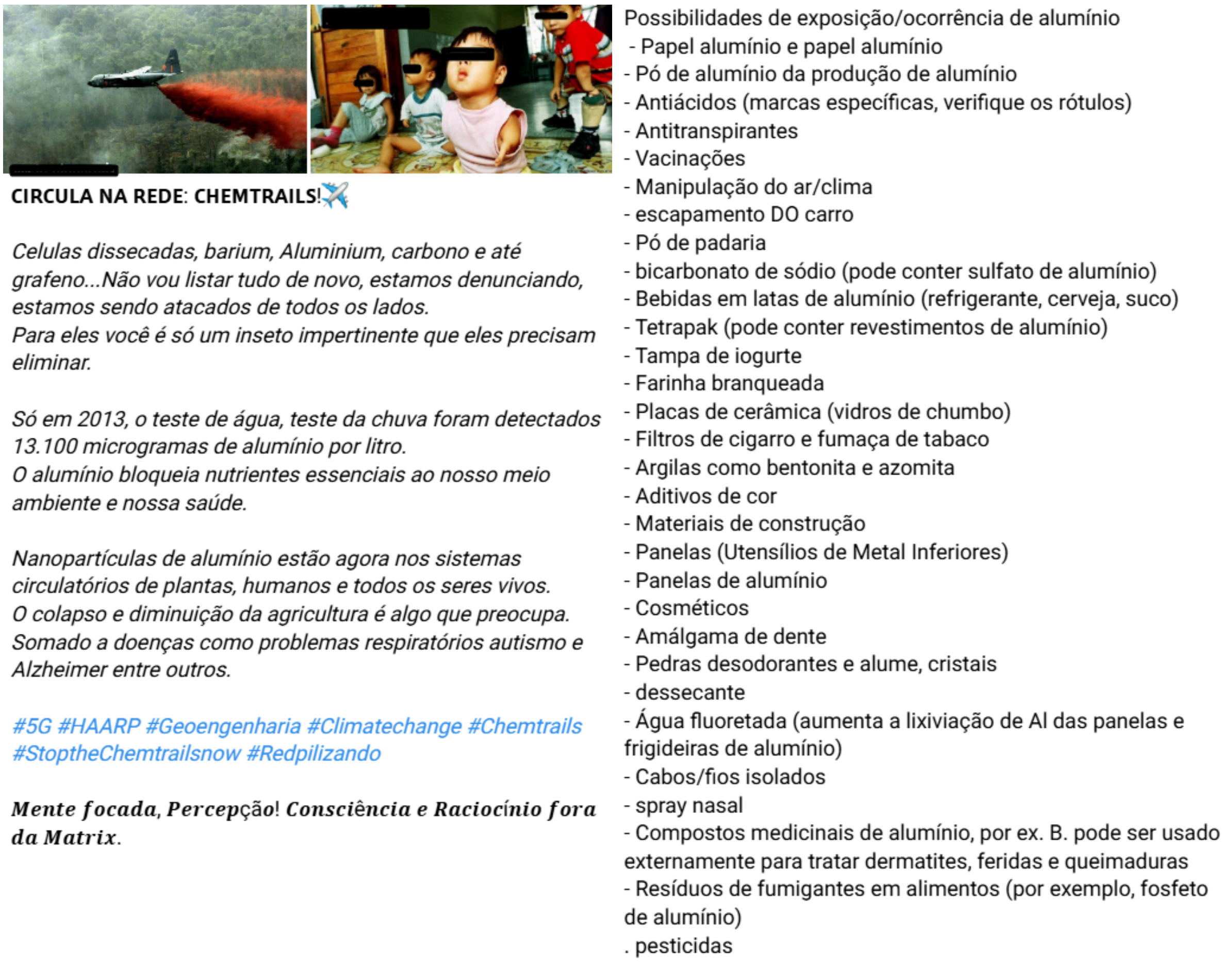

Source: Screenshot (2025).

In addition to the elements already mentioned, some conspiracy communities circulate a list of chemical substances that supposedly cause autism, including:

- Arsenic, N-Hexane, Organochlorines, Organophosphates, Toluene, and Trichloroethylene. Although these chemical compounds can be toxic in high doses, there is no proof that they cause autism.
- Alcohol, Baking Soda, Borax, and Caffeine. Some of these substances can affect the nervous system, but there is no causal relationship with ASD.
- Chemtrails, Sulfur, Petroleum, Wireless, and WPO. Frequently cited as agents that alter the human brain, but there is no scientific basis to support this hypothesis.

The dissemination of these claims creates a scenario of paranoia and disinformation, distancing families and caregivers from reliable medical information.

**Figure 04.** Examples of conspiracy theories about autism causes.

❤️❤️**EPIGENÉTICA Y ENFERMEDADES CEREBRALES: PAPEL DE LA EXPOSICIÓN TEMPRANA A CONTAMINANTES.**❤️❤️

La zeolita es un anti-veneno, un quelador de metales pesados y pesticidas. Se sospecha que son los causantes de numerosas enfermedades. Podemos mencionar las enfermedades neurológicas porque ciertos metales como el aluminio y el mercurio "aman" el cerebro. Los científicos han descubierto que la exposición temprana a los contaminantes, es decir, en el feto y en los niños pequeños, favorece la aparición de lesiones de Alzheimer y Parkinson. Aquí es donde entra la noción de epigenética, una nueva ciencia que se define como el silenciamiento de un gen a lo largo de la vida o de varias generaciones. Esto se debe a las modificaciones genéticas adquiridas a través de los venenos. Así que hay "interruptores" epigenéticos, ¿cómo evitarlos? ¿Es útil un quelante de arcilla o zeolita para las mujeres embarazadas?

❤️**La epigenética en la aparición de enfermedades neurológicas**

Hay pruebas de la exposición temprana a los contaminantes y la aparición de enfermedades neurológicas. Estos factores ambientales alteran la genética del individuo; estamos entrando en el campo de la epigenética, una ciencia que ha cobrado impulso. Todo lo que hacemos, comemos, absorbemos como alimento o veneno puede influir en el ADN. Dependiendo de las alteraciones y la predisposición de los individuos; tendremos cánceres, enfermedades neurológicas y muchas otras dolencias.

- Transmissão deliberada de parasitas através do ar, vacinação e alimentos - Morgellons, autismo (verme), vermes, tênias, vermes, fungos como candida e muitos mais.
- Radiação, celular, wireless - altas frequências discordantes, criam desequilíbrio
- Digitalização humana - chips RFID, nanobots, parasitas sintéticos, realidade virtual
- Microondas – especialmente o popular micro-ondas na cozinha, destrói completamente a estrutura dos alimentos, deixando apenas o enchimento tóxico.
- Chemtrails - Morgellons, escurecimento global, manipulação do clima, pulverização de toxinas em geral.
- Medicamentos que contêm aditivos nocivos e/ou foram modificados quimicamente - Muitas substâncias são benéficas mesmo em sua forma pura ou por que mais foram colocadas sob o BTMG e o álcool está disponível gratuitamente?
- Limpadores de corpo antigos e comprovados, como petróleo, WPO, bórax, bicarbonato de sódio, enxofre, terebintina, são fornecidos com avisos de perigo e caveiras - porque sabem o quão forte é o efeito de limpeza
- Bactérias (microrganismos) são nossos "inimigos" - quando na verdade são nossos "amigos"
- Drogas pesadas como açúcar, cafeína e álcool estão disponíveis gratuitamente - uma verdadeira planta medicinal (cânhamo), por outro lado, foi transformada em uma droga "ruim" e criminalizada, que foi massivamente geneticamente modificada

Source: Screenshot (2025).

Another recurring theory suggests that autism is caused by a deficiency in the gut microbiome or insufficient vitamin B12. This narrative has led many people to believe that using probiotics, restrictive diets, and intense vitamin B12 supplementation could "reverse" ASD. While studies show that the microbiome may play a role in neurological health, there is no evidence that an intestinal imbalance causes autism. Some autistic children may have differences in their microbiota, but this may be an effect of the condition rather than a cause. Vitamin B12 deficiency can indeed cause neurological problems, but there is no scientific relationship between B12 deficiency and ASD. Excessive supplementation can lead to adverse effects, including metabolic imbalances.

**Figure 05.** Examples of conspiracy theories about autism causes.

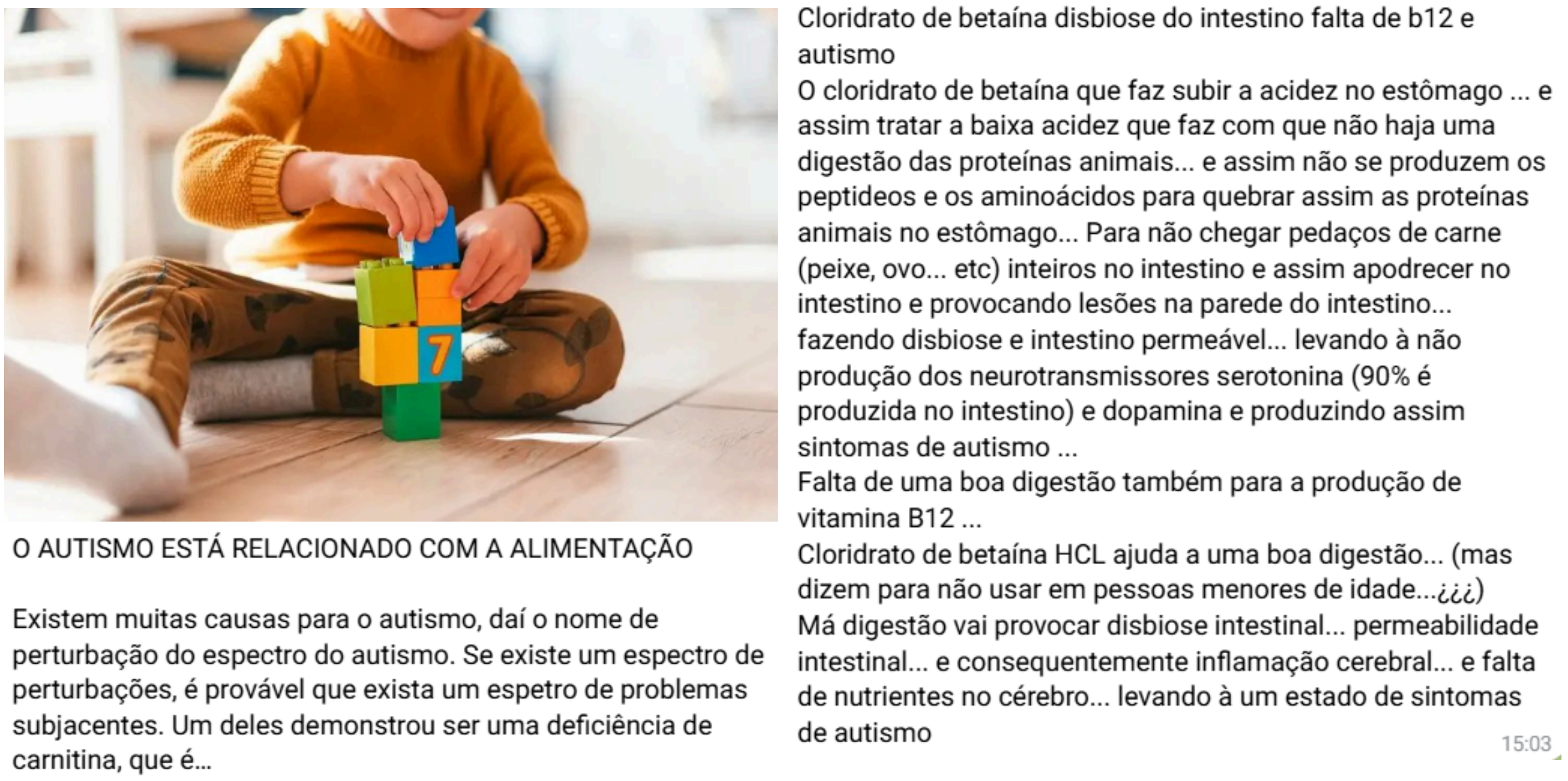

Source: Screenshot (2025).

With the advancement of COVID-19 vaccines, some conspiracy communities began associating the spike protein generated by vaccines with autism development. This claim, besides being unfounded, fuels an anti-vaccine discourse that has already been widely debunked by science. The spike protein in mRNA vaccines is temporary and eliminated by the body within days, with no evidence that it causes neurological damage or genetic alterations. Furthermore, the theory that vaccines cause autism originated from a fraudulent study published in 1998, which was later debunked. Multiple large-scale epidemiological studies have confirmed that there is no relationship between vaccination and ASD. Unfortunately, these narratives encourage the anti-vaccine movement, putting not only autistic children at risk but also the entire population by increasing the spread of preventable diseases.

[ english version / português abaixo ]

**Figure 06.** Examples of conspiracy theories about autism causes.

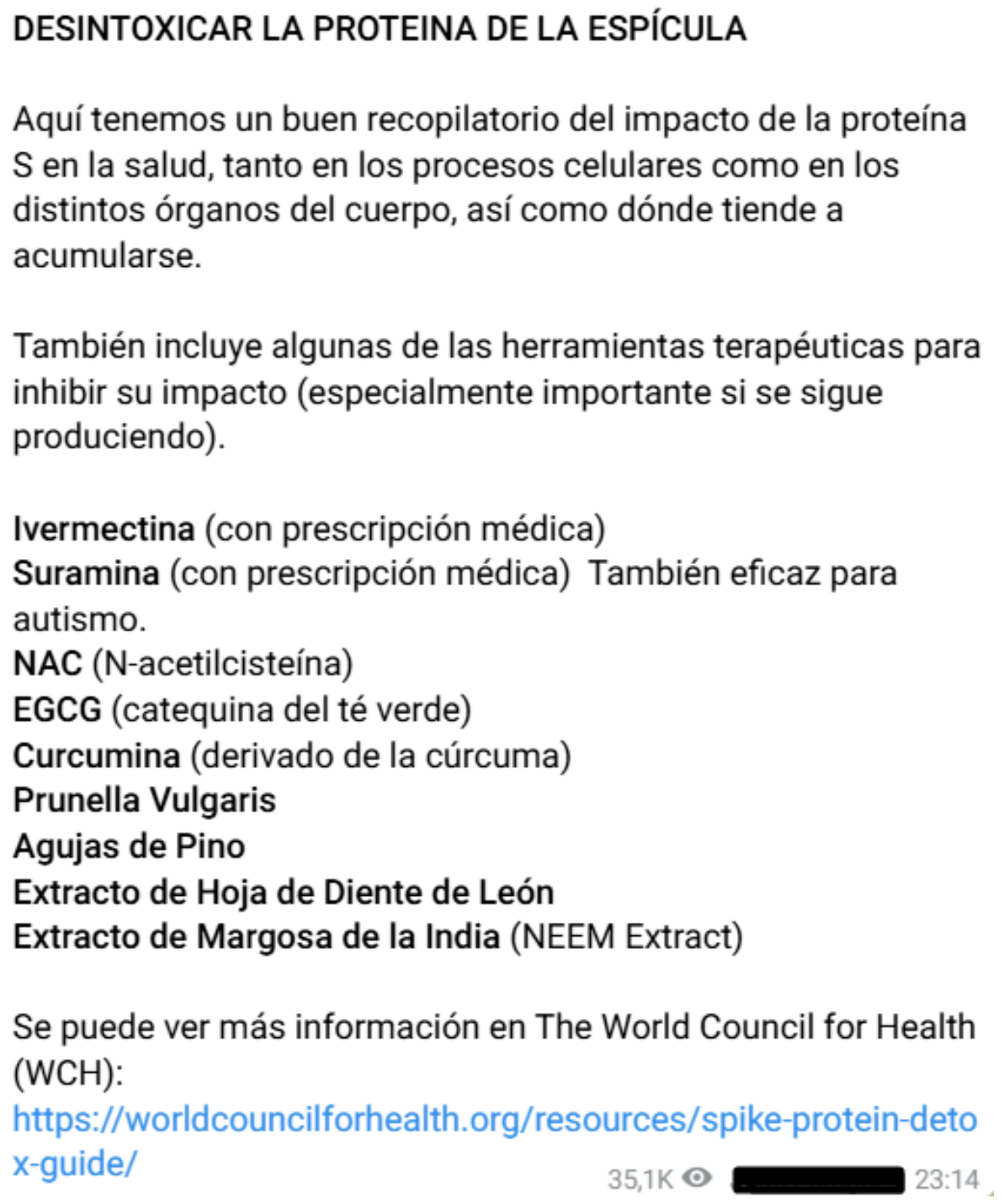
DESINTOXICAR LA PROTEINA DE LA ESPÍCULA

Aquí tenemos un buen recopilatorio del impacto de la proteína S en la salud, tanto en los procesos celulares como en los distintos órganos del cuerpo, así como dónde tiende a acumularse.

También incluye algunas de las herramientas terapéuticas para inhibir su impacto (especialmente importante si se sigue produciendo).

**Ivermectina** (con prescripción médica)
**Suramina** (con prescripción médica)  También eficaz para autismo.
**NAC** (N-acetilcisteína)
**EGCG** (catequina del té verde)
**Curcumina** (derivado de la cúrcuma)
**Prunella Vulgaris**
**Agujas de Pino**
**Extracto de Hoja de Diente de León**
**Extracto de Margosa de la India** (NEEM Extract)

Se puede ver más información en The World Council for Health (WCH):
https://worldcouncilforhealth.org/resources/spike-protein-detox-guide/
35,1K [redacted] 23:14

La FDA ha publicado pruebas concluyentes en su sitio web de que la vacuna DTap puede causar autismo. De acuerdo con el documento en línea de Vacunas de Sangre Biológicas de la FDA, un fabricante de vacunas admite en su paquete que su vacunación puede causar autismo como una de las muchas reacciones adversas.

- Estos eventos adversos informados durante el uso posterior a la aprobación de la vacuna Tripedia incluyen púrpura trombocitopénica idiopática, SIDS, reacción anafiláctica, celulitis, autismo, convulsión / convulsión de gran mal, encefalopatía, hipotonía, neuropatía, somnolencia y apnea. Los eventos se incluyeron en esta lista debido a la gravedad o frecuencia de los informes. Debido a que estos eventos se informan voluntariamente de una población de tamaño incierto, no siempre es posible estimar de manera confiable sus frecuencias o establecer una relación causal con los componentes de la vacuna Tripedia.

Source: Screenshot (2025).

Although these communities claim to have discovered the true causes of autism, the broad and often contradictory variety of explanations reveals the incoherence of these narratives. If autism is supposedly caused by everything — from electromagnetic radiation, processed foods, and vaccines to vitamin deficiencies and heavy metal exposure — this demonstrates not only the lack of scientific basis but also the internal inconsistency of these discourses. Below, a table presents the full list of the 150 supposed causes of autism circulating in these conspiracy communities, highlighting how disinformation adapts and expands to reinforce different fears and pseudoscientific beliefs.

**Table 05.** List of false causes of autism mapped in the communities (A – Z).

| | Supposed cause | Example |
|---|---|---|
| **1** | 5G | Some conspiracy theories claim that radiation from 5G networks interferes with neurological development and causes autism. However, there is no scientific evidence to support this claim. The radiation emitted by 5G is non-ionizing, meaning it does not have enough energy to cause alterations in DNA or the human brain. The spread of this theory reflects a broader pattern of disinformation about technology and health, without any scientific basis. |
| **2** | Acetaminophen | Some communities claim that the use of paracetamol causes autism by "intoxicating" the infant brain. However, there is no scientific evidence linking the medication to the disorder, and it is safe when used correctly. |
| **3** | Alcohol | Some narratives associate alcohol consumption during pregnancy with autism, but this is a confusion with Fetal Alcohol Syndrome, which is a different disorder unrelated to ASD. |

[ english version / português abaixo ]

| | | |
|---|---|---|
| **4** | Aluminum | Anti-vaccine communities claim that aluminum present in vaccine adjuvants causes autism. However, there is no evidence that aluminum in safe amounts causes any neurological alterations. |
| **5** | Aluminum sulfate | There are claims that this chemical compound, used in drinking water treatment, causes autism by affecting the brain. However, there is no scientific evidence supporting this claim, and aluminum levels in water are strictly controlled to ensure safety. |
| **6** | Amino acid deficiency | Some communities claim that a lack of essential amino acids interferes with brain formation and leads to autism. While inadequate nutrition can impact overall development, no studies associate amino acid deficiency with ASD. |
| **7** | Ammonium sulfate | Some communities claim that ammonium sulfate, used as a fertilizer, may be present in food and trigger autism. However, there is no scientific proof linking this compound to ASD. |
| **8** | Amoxicillin | Some theories claim that antibiotics alter the gut microbiota and trigger autism. While antibiotics can impact gut flora, there is no evidence that they cause ASD. |
| **9** | Antacids | It is suggested that the use of antacids during pregnancy interferes with the absorption of essential nutrients and leads to autism. However, there is no scientific basis for this claim. |
| **10** | Antiperspirants | It is claimed that aluminum in deodorants clogs lymphatic glands and affects neurological development. However, dermal absorption of this aluminum is insignificant. |
| **11** | Arsenic | Some studies associate excessive arsenic exposure with neurological disorders, but there is no proof that it is a direct cause of autism. |
| **12** | Ascorbic Acid | Vitamin C is mistakenly pointed to as a factor in autism development for supposedly altering metabolic processes. However, vitamin C deficiency can cause scurvy, but there is no link to ASD. |
| **13** | Aspartame | Artificial sweeteners are frequently targeted by disinformation and are wrongly linked to ASD for supposedly altering neurotransmitters. However, studies show that moderate consumption is safe. |
| **14** | Autoimmune disease | There are claims that maternal autoimmune reactions during pregnancy attack the fetal brain and cause ASD. While autoimmune diseases can influence fetal development, there is no scientific evidence that they are the cause of autism. |
| **15** | Azomite | Some pseudoscientific communities claim that exposure to rare minerals from azomite may influence autism development. However, there is no scientific basis for this claim. |
| **16** | B12 deficiency | Some theories claim that a lack of vitamin B12 during pregnancy or childhood leads to autism. While B12 is essential for neurological development, its deficiency can cause other problems, such as anemia, but it is not linked to ASD. |
| **17** | Baking powder | Some pseudoscientific communities claim that ingredients used in baking, such as chemical leavening agents, affect the brain and cause autism. However, there is no scientific basis for this claim. |
| **18** | Baking soda | Proponents of "natural cures" claim that baking soda reverses autism by alkalizing the body, but this theory has no scientific basis. |

[ english version / português abaixo ]

| | | |
|---|---|---|
| **19** | Beef extract | Some diets claim that compounds derived from beef interfere with brain function and lead to autism. However, there is no scientific basis for this claim. |
| **20** | Bentonite | Clays like bentonite are promoted as "detoxifiers" to prevent autism, but there is no scientific evidence that toxin elimination prevents or reverses the disorder. |
| **21** | Biotin (in Ethanol) | It is claimed that biotin in certain chemical forms causes ASD by interfering with metabolism. However, biotin is essential for various bodily functions, and there is no connection to autism. |
| **22** | Bisphenol | Bisphenol (BPA), found in plastics, has been studied for possible endocrine effects, but there is no evidence that it causes autism. |
| **23** | Bleached flour | Proponents of natural diets claim that bleaching agents used in wheat flour affect the brain and cause autism. However, there are no studies proving this relationship. |
| **24** | Borax | Some pseudoscientific communities promote borax as a "detoxifier", suggesting that its absence leads to ASD. However, borax can be toxic and has no relation to autism. |
| **25** | Brain inflammation | Some theories suggest that brain inflammation is a direct cause of autism. While inflammatory processes may occur in some neurological conditions, there is no evidence that they are the origin of ASD, which has a genetic basis. |
| **26** | Caffeine | Some theories claim that caffeine consumption during pregnancy alters neurological development and causes ASD, but existing studies do not indicate such a relationship. |
| **27** | Calcium chloride | Some pseudoscientific claims suggest that mineral imbalances cause autism, but this is not supported by evidence. |
| **28** | Calcium deficiency | Pseudoscientific narratives suggest that calcium deficiency impairs neuronal communication and causes autism. However, while calcium is essential for bone and muscle health, there is no evidence that its absence is related to ASD. |
| **29** | Calcium pantothenate (in ethanol) | Some claims suggest that the presence of ethanol in vitamin B5 supplements causes autism. However, there is no scientific basis for this claim, as calcium pantothenate is an essential nutrient for cellular metabolism. |
| **30** | Calf heart infusion | Some pseudoscientific narratives claim that this type of infusion can alter neurological functions and lead to autism. However, there is no scientific basis for this claim. |
| **31** | Candida | It is claimed that intestinal Candida infections cause autism by releasing toxins into the brain. However, there is no evidence to support this hypothesis. |
| **32** | Canola oil | Some theories claim that consuming canola oil, due to its genetically modified origin and industrial processing, leads to autism. However, there is no scientific evidence that vegetable oils influence ASD development. |
| **33** | Car exhaust | Some theories suggest that exposure to pollutants from vehicle exhaust gases affects neurodevelopment and leads to autism. While air pollution has negative health impacts, there is no evidence that it is a cause of ASD. |
| **34** | Carnitine deficiency | Some communities claim that low levels of carnitine, a compound involved in energy metabolism, cause autism. However, studies do not demonstrate any causal relationship between carnitine and ASD. |
| **35** | Casein | Some restrictive diets claim that casein, a milk protein, worsens autism symptoms. While intolerances may exist, there is no causal relationship with ASD. |
| **36** | Ceramic tiles | Some narratives suggest that inhaling ceramic particles or their chemical components can affect the brain and lead to autism. However, there is no study linking ceramic exposure to ASD. |
| **37** | Chemtrails | The conspiracy theory claims that chemical trails from airplanes alter neurological development and cause autism, but there is no scientific basis for this claim. |
| **38** | Cholesterol sulfate deficiency | Some theories suggest that a lack of this compound in the brain interferes with neuron myelination and leads to autism. However, there is no scientific proof that cholesterol sulfate deficiency is a causal factor of ASD. |
| **39** | Cigarette smoke | Some narratives suggest that cigarette exposure during pregnancy causes autism. While smoking during pregnancy is associated with various health problems, there is no evidence that it is a causal factor for ASD. |
| **40** | Color additives | Some communities claim that artificial food dyes, such as tartrazine, affect neurodevelopment and lead to ASD. While some dyes may cause hyperactivity in sensitive children, there is no scientific evidence linking them to autism. |
| **41** | Construction materials | Some claims suggest that substances in paints, cement, and treated wood cause autism. While certain chemicals can be toxic, there is no proof that they are causal factors of ASD. |
| **42** | Corn | Some narratives suggest that genetically modified corn contains toxins that affect the brain and cause ASD. However, no study has proven this relationship, making it one of many disinformation claims about genetically modified foods. |
| **43** | Cosmetics | There are theories that chemical substances in cosmetic products, such as parabens and phthalates, interfere with neurological development. While these compounds are regulated, no study has proven their link to autism. |
| **44** | COVID | Some narratives suggest that infection with the SARS-CoV-2 virus can cause autism, particularly in pregnant women. However, ASD is predominantly genetic in origin, and there is no evidence that viral infections are associated with its development. |
| **45** | Cysteine hydrochloride | There are claims that this amino acid, used in supplements and foods, affects brain function and causes autism. However, cysteine is necessary for protein synthesis, and there is no evidence that it causes ASD. |
| **46** | Dental Amalgam | There is a myth that mercury in dental amalgams leads to autism. However, studies show that mercury exposure from fillings does not alter neurological development. |
| **47** | Deodorant stones | Some narratives suggest that natural deodorants based on potassium alum release aluminum into the body and cause ASD. However, there is no scientific evidence supporting this hypothesis. |
| **48** | Desiccant | Some claims suggest that chemical substances used to remove moisture from food and products cause autism. However, there is no scientific proof that desiccants have any impact on neurodevelopment. |
| **49** | Disodium phosphate | This chemical compound, used in food and medicine, is claimed to cause autism by supposedly affecting brain chemistry. However, there is no scientific evidence linking disodium phosphate to ASD. |
| **50** | Doritos | Some theories claim that genetically modified corn and preservatives in snacks like Doritos lead to autism. However, there is no scientific proof that processed or genetically modified foods cause ASD. |

[ english version / português abaixo ]

| | | |
|---|---|---|
| **51** | Eggs | Some communities claim that egg consumption, due to its cholesterol content or antibiotic residues, can lead to autism. However, there is no scientific evidence linking egg consumption to ASD. |
| **52** | Epigene P450 | Some pseudoscientific communities claim that alterations in the P450 enzyme system, which metabolizes substances in the liver, are linked to autism. However, there is no scientific evidence to support this hypothesis. |
| **53** | Feldspar | Some communities claim that exposure to feldspar, a common mineral in the Earth's crust, causes autism. However, there is no scientific basis for this claim. |
| **54** | Ferrous sulfate | There are claims that ferrous sulfate, used to treat anemia, may interfere with neurodevelopment and cause ASD. However, it is essential for health, and there is no proven relationship with autism. |
| **55** | Ferrous sulfate heptahydrate | Some pseudoscientific narratives suggest that this chemical compound, used in iron supplementation, could cause ASD. However, it is safe and essential for preventing anemia, with no connection to autism. |
| **56** | Fluoride | Conspiracy theories claim that fluoride in water and dental products causes autism by affecting brain function. However, scientific studies show that fluoride is safe and essential for oral health. |
| **57** | Formaldehyde | Some theories claim that exposure to formaldehyde in furniture, cosmetics, and vaccines causes autism. While large amounts of formaldehyde are toxic, the doses present in the environment and vaccines are extremely low and pose no risk of causing ASD. |
| **58** | Fungi | Some theories claim that fungal infections in the gut or other parts of the body affect the brain and lead to autism. However, there is no scientific proof linking fungal infections to ASD. |
| **59** | Gelatin | Some claims suggest that gelatin, due to its animal collagen content and additives, can cause autism. However, there is no scientific evidence linking gelatin to ASD. |
| **60** | Genetically modified organisms (GMOs) | Conspiracy theories claim that genetically modified foods (GMOs) affect the brain and cause autism. However, scientific studies show that GMOs have no connection to ASD. |
| **61** | Glutathione | Some communities claim that low levels of glutathione, a natural antioxidant, cause autism. While glutathione plays important roles in the body, there is no evidence that its deficiency is a causal factor of ASD. |
| **62** | Gluten | Many alternative diets claim that gluten consumption can cause or worsen autism, but there is no scientific evidence to support this relationship. Gluten is only problematic for individuals with celiac disease or gluten sensitivity. |
| **63** | Glyphosate deficiency | This false claim originates from conspiracy groups that believe glyphosate, a herbicide, is essential for metabolic processes and that its absence leads to autism. There is no scientific basis for this claim. |
| **64** | Grape seed | Some theories suggest that antioxidant substances in grape seeds affect the brain and may cause ASD. However, there is no scientific evidence supporting this claim. |
| **65** | Graphene | Conspiracy theories claim that graphene particles in vaccines or the environment cause autism. However, there is no scientific proof that graphene is harmful to the brain or present in doses significant enough to cause damage. |

[ english version / português abaixo ]

| | | |
|---|---|---|
| **66** | HAARP | The High-Frequency Active Auroral Research Program (HAARP) is the target of conspiracy theories claiming that it alters the human mind and causes diseases, including autism. There is no scientific basis for this theory, making it a classic example of disinformation. |
| **67** | Hydrated magnesium sulfate | Some claims suggest that this compound, used in supplements and medications, may interfere with brain development. However, it is safe, and there is no evidence that it causes ASD. |
| **68** | IgG Food Allergy | Proponents of gluten- and lactose-free diets claim that food allergies increase the risk of autism. However, IgG reactions have no causal relationship with ASD, and many of these intolerances are not scientifically validated. |
| **69** | Intestinal Bacteria | There are claims that imbalances in gut microbiota cause autism. While gut flora can influence overall health, there is no evidence that it is a causal factor of ASD. |
| **70** | Intestinal inflammation | Some communities claim that gut disorders are the root cause of autism. While gastrointestinal issues are common in some people with ASD, they are associated conditions, not causes of the disorder. |
| **71** | Iodine | Some claims suggest that iodine deficiency or excess during pregnancy causes autism. While iodine is essential for thyroid development, there is no proof that it is involved in ASD. |
| **72** | Lactose | There are claims that lactose, present in milk, can cause or worsen autism. However, there is no proven link between lactose consumption and ASD. |
| **73** | Lanthanides | Some conspiracy theorists suggest that these rare metals, found in electronics, interfere with brain function and cause autism. However, there is no scientific evidence supporting this hypothesis. |
| **74** | Lead | Lead exposure can have neurotoxic effects, but there is no proof that it is a direct factor in ASD development. There is no solid evidence, much less consensus in the scientific community, that heavy metals such as mercury or lead are responsible for autism. |
| **75** | Lidocaine | Some theories claim that lidocaine, a common anesthetic, causes autism when used in medical procedures. However, no study has linked lidocaine to ASD. |
| **76** | Linoleic Acid Omega-6 (LA) | Some narratives suggest that excessive consumption of this fatty acid promotes brain inflammation and leads to autism. However, the balance between omega-6 and omega-3 is important for health, with no evidence of a relationship with ASD. |
| **77** | Magnesium chloride | While some claim that magnesium chloride causes autism, others promote magnesium supplements as an "autism cure" by supposedly correcting deficiencies, but there is no scientific basis for this claim. |
| **78** | Magnetic field reversal | Some theories claim that changes in the Earth's magnetic field influence neurodevelopment and cause ASD. However, there is no evidence that Earth's magnetism has any effect on autism. |
| **79** | Mercury | The theory that mercury causes autism stems from disinformation about vaccines and dental amalgams. Scientific studies show that the mercury present in vaccines (thimerosal) has never been associated with ASD. |
| **80** | Metal-induced inflammation | Conspiracy theories suggest that "heavy metal toxicity" causes brain inflammation and leads to autism. While heavy metals can be toxic in high doses, there is no evidence that they are the cause of ASD. |

[ english version / português abaixo ]

| | | |
|---|---|---|
| **81** | Mica | Some communities suggest that exposure to mica, a mineral used in cosmetics and paints, causes autism. However, there is no evidence linking mica to ASD. |
| **82** | Microwaves | There are theories claiming that radiation from microwave ovens alters cell structures and causes autism. However, microwave radiation is non-ionizing and has no effects on DNA or the human brain. |
| **83** | Mineral deficiency | Some alternative diets claim that a lack of minerals such as zinc, iron, and magnesium is related to autism. While these minerals are important for various bodily functions, ASD is linked to genetic factors rather than mineral deficiency. |
| **84** | Mineral salts | Some communities claim that imbalances in minerals such as sodium, potassium, and magnesium lead to autism. While these minerals are important for health, there is no proof linking them to ASD. |
| **85** | Mitochondrial dysfunction | Some theories suggest that failures in cellular energy production lead to autism. While certain mitochondrial dysfunctions can be found in individuals with ASD, they are consequences of the disorder in some cases, not its cause. |
| **86** | Monopotassium phosphate | Claims suggest that this food additive harms the nervous system and leads to autism. However, there is no scientific basis to support this claim. |
| **87** | Monosodium glutamate (MSG) | MSG, a flavor enhancer, is frequently targeted by disinformation, including claims that it harms the brain and causes autism. However, scientific studies show that its consumption in normal amounts is safe and unrelated to ASD. |
| **88** | N-Hexane | Some theories claim that this solvent, found in adhesives and paints, can cause autism by affecting the nervous system. While N-hexane can be neurotoxic in prolonged and high-dose exposures, there is no scientific evidence linking it to ASD. |
| **89** | Nasal spray | Some theories suggest that nasal decongestants and other sprays may contain substances that affect the brain and lead to autism. However, there are no studies demonstrating any relationship between nasal spray use and ASD. |
| **90** | Niacin | Some narratives claim that a deficiency or excess of niacin (vitamin B3) can trigger autism. While niacin is essential for various metabolic functions, there is no causal link between its levels and ASD. |
| **91** | Opioids | Some narratives claim that opioid use during pregnancy can cause autism. While opioid use during pregnancy may be linked to adverse effects on fetal development, there are no conclusive studies directly linking them to ASD. |
| **92** | Organochlorines | Some claims suggest that organochlorine pesticides, such as DDT, are responsible for the rise in autism cases. While these compounds have toxic effects on the nervous system in high exposures, there is no scientific evidence that they are a direct cause of ASD. |
| **93** | Organophosphates | There are theories claiming that organophosphate pesticides, used in agriculture, cause autism by affecting the nervous system. While excessive exposure may have neurological impacts, there is no evidence that it causes ASD. |
| **94** | Palmitic Acid | It is claimed that this fatty acid, present in breast milk and vegetable oils, is linked to autism by altering brain function. However, it is an essential component of infant nutrition. |
| **95** | Paracetamol | Some theories claim that using paracetamol during pregnancy or childhood may trigger ASD. However, scientific studies do not demonstrate any causal relationship between the use of this medication and autism. |

[ english version / português abaixo ]

| | | |
|---|---|---|
| **96** | Parasite | One of the most dangerous and widely spread conspiracy theories claims that autism is caused by parasitic infections in the body and that it can therefore be "cured" through deworming protocols. This narrative is based on the false idea that parasites, supposedly residing in the intestines or brain, release toxins that alter a child's neurological development. These claims have led to extremely harmful practices, such as the use of toxic substances for so-called "parasite cleansing". Among the most dangerous methods are the use of chlorine dioxide (MMS/CDS), inappropriate doses of ivermectin, castor oil, highly toxic herbal therapies, excessive enema use, and even protocols based on turpentine (an industrial solvent). These methods not only lack any scientific basis but can also lead to irreversible health damage, including severe poisoning, gastrointestinal injuries, liver failure, and even death. |
| **97** | Peanut seed | Some claims suggest that substances found in peanuts, such as aflatoxins, cause autism. While aflatoxins can be toxic in large quantities, there is no evidence that they are responsible for ASD. |
| **98** | Persistent screen use | Some theories claim that screen use in childhood can cause autism. Although heavy screen use can harm childhood development, there is no significant evidence that it can cause autism. |
| **99** | Petroleum | Some theories claim that exposure to petroleum-derived products in plastics and fuels causes autism. While certain chemical compounds may have health effects, there is no proof that petroleum or its derivatives cause ASD. |
| **100** | Plastic / Microplastic | There are claims that ingesting or being exposed to microplastics interferes with neurological development and causes autism. While microplastics are an environmental concern, there are no studies proving they cause ASD. |
| **101** | Polyethylene glycol (PEG) | Some claims state that PEG, used as an excipient in medications and vaccines, affects the blood-brain barrier and causes autism. However, there is no scientific evidence supporting this relationship. |
| **102** | Polysorbate | Polysorbate, an emulsifier found in food and vaccines, is frequently cited by conspiracy theorists as a substance that causes autism. However, studies show that it is safe and has no connection to ASD. |
| **103** | Potassium chloride | Some narratives allege that electrolyte imbalances, such as excess or deficiency of potassium, cause autism. However, potassium chloride is an essential nutrient for cellular functions and has no causal relationship with ASD. |
| **104** | Potassium sulfate | Some conspiracy theories suggest that fertilizers containing potassium sulfate alter brain function and cause autism. However, there is no scientific proof supporting this claim. |
| **105** | Preservatives | There is a popular belief that artificial preservatives, especially those used in processed foods, alter brain function and cause autism. However, no reliable study has established this relationship. |
| **106** | Proline | Some theories claim that proline, a non-essential amino acid, affects neurotransmitters and leads to autism. However, there is no scientific evidence supporting this claim. |
| **107** | Pyridoxine hydrochloride (in ethanol) | Some theories claim that vitamin B6 in the form of pyridoxine, especially when diluted in ethanol, can cause autism. However, vitamin B6 is essential for neurological functions, and its deficiency — not its consumption — can cause neurological problems. |

[ english version / português abaixo ]

| | | |
|---|---|---|
| **108** | Riboflavin (in ethanol) | Some theories claim that vitamin B2 (riboflavin), when diluted in ethanol, affects neurodevelopment and causes autism. However, riboflavin is essential for cellular metabolism, and there is no evidence that it causes ASD. |
| **109** | Rice bran oil | Some claims suggest that this oil, due to certain antioxidants it contains, can interfere with neurological development. However, there is no scientific evidence linking rice bran oil to autism. |
| **110** | Safflower seed | Some narratives claim that safflower seed oil or extract interferes with neurotransmitters and causes autism. However, there is no scientific basis for this claim. |
| **111** | School meals | Some narratives claim that food served in schools contains additives that cause autism. They support a conspiracy theory that the government is making children autistic. However, there is no scientific basis for this claim. |
| **112** | Serotonin deficiency | There are claims that low serotonin levels in the brain cause autism. Although differences in serotonin levels are observed in some individuals with ASD, this is an effect of the condition, not its cause. |
| **113** | Silica nanoparticles | Some communities claim that silica nanoparticles, used in supplements and food, can interfere with neurological development and cause ASD. However, there is no scientific proof for this claim. |
| **114** | Sodium chloride | There are claims that table salt alters neurotransmitters and leads to autism. However, sodium is essential for nervous system function, and there is no scientific evidence to support this hypothesis. |
| **115** | Sodium hydroxide | Also known as caustic soda, it is used in the food and pharmaceutical industries in minimal and safe quantities. There is no evidence that it causes ASD. |
| **116** | Soft drinks | Some claims suggest that consuming soft drinks, especially those containing artificial sweeteners, causes autism. However, there is no scientific evidence supporting this hypothesis. |
| **117** | Soy | Some narratives claim that phytoestrogens in soy affect neurodevelopment and cause autism. However, there is no scientific evidence linking soy consumption to ASD. Phytoestrogens are natural compounds that do not negatively impact neurological development. |
| **118** | Soybean oil | Some conspiracy theories claim that soybean oil contains phytoestrogens that affect the developing brain and cause ASD. While soy contains bioactive compounds, there is no evidence that its consumption causes autism. |
| **119** | Spike protein | Anti-vaccine narratives claim that the spike protein, produced by the body after COVID-19 vaccination, causes autism. However, there is no evidence linking the COVID-19 vaccine to ASD. |
| **120** | Sugars | Some communities claim that excessive sugar consumption causes brain inflammation and leads to autism. However, there is no scientific evidence that sugar influences ASD development. |
| **121** | Sulfur | There are claims that sulfur present in certain foods and supplements interferes with neurological development and causes autism. Sulfur is an essential mineral for various bodily functions, with no relation to ASD. |
| **122** | Sunflower oil | There are claims that consuming this oil, rich in omega-6, can increase inflammation and trigger autism. However, there are no scientific studies proving this hypothesis. |

[ english version / português abaixo ]

| | | |
|---|---|---|
| **123** | Tap Water | It is claimed that substances such as fluoride or heavy metals in drinking water cause autism. However, water quality regulations prevent harmful levels of these compounds. |
| **124** | Testosterone | Some theories claim that altered testosterone levels during pregnancy increase the risk of autism. While hormones influence fetal development, there is no conclusive evidence that testosterone is a causal factor for ASD. |
| **125** | Tetrapak | There are claims that products packaged in Tetrapak contain chemical compounds that cause autism. However, there is no scientific evidence proving this relationship. |
| **126** | Thiamine (in ethanol) | Some narratives claim that thiamine (vitamin B1) in ethanol solution affects the brain and causes ASD. However, thiamine is essential for energy metabolism, and there is no scientific evidence linking its chemical form to autism. |
| **127** | Thimerosal (Mercury derivative, ethylmercury) | This is one of the most widely spread claims among anti-vaccine movements, suggesting that thimerosal, a preservative used in some past vaccines, causes autism. However, extensive scientific studies have shown that there is no relationship between thimerosal and ASD, and it has been removed from nearly all childhood vaccines as a precaution. |
| **128** | Titanium dioxide nanoparticles | Conspiratorial claims suggest that these nanoparticles, used in cosmetics and food, cross the blood-brain barrier and affect the brain. However, no studies associate titanium dioxide exposure with autism. |
| **129** | Toluene | Some claims suggest that exposure to toluene, a chemical solvent, causes autism. While prolonged exposure to solvents can be harmful to health, there is no scientific evidence linking toluene to ASD. |
| **130** | Toothpaste | There are claims that substances found in toothpaste, such as fluoride and triclosan, cause autism. However, fluoride is safe in recommended amounts and essential for oral health, with no relation to ASD. |
| **131** | Trichloroethylene | Some claims suggest that exposure to this industrial solvent may trigger ASD. While trichloroethylene can be toxic at high exposure levels, there is no evidence that it is a causal factor for autism. |
| **132** | Tris base | Some pseudoscientific communities suggest that this chemical compound, used in molecular biology, affects neurological development. However, there is no study proving this hypothesis. |
| **133** | Tris hydrochloride | Some theories suggest that this chemical buffer, used in laboratories, may be present in consumer products and cause autism. However, there is no scientific evidence for this claim. |
| **134** | Tylenol | Like paracetamol, there are claims that Tylenol causes autism when used during pregnancy. However, scientific studies do not show any causal relationship between this medication and ASD. |
| **135** | Tyrosine hydrochloride | Some claims suggest that this essential amino acid affects neurotransmitters and causes ASD. However, tyrosine is crucial for the production of dopamine and other neurotransmitters, with no connection to autism. |
| **136** | Uracil hydrochloride | Some pseudoscientific theories claim that chemical compounds unknown to the public can cause autism. However, uracil is a nitrogenous base essential for RNA, and no studies suggest any connection to ASD. |
| **137** | Vaccines | The claim that vaccines cause autism is one of the most dangerous and widely debunked by science, yet it continues to be promoted by anti-vaccine groups and disinformation spreaders. This theory originated in 1998 when former doctor Andrew Wakefield published a fraudulent study in The Lancet claiming a link between the MMR vaccine (measles, mumps, and rubella) and autism. This study was completely discredited, Wakefield lost his medical license for fraud, and the publication was retracted. However, the disinformation generated by this hoax persists and has led to a global vaccine hesitancy crisis, directly impacting public health. Numerous large-scale scientific studies, involving millions of children worldwide, have confirmed that there is no relationship between vaccines and autism. Some of the most robust research was conducted in Denmark, the United States, and the United Kingdom, following vaccinated and unvaccinated children over the years and finding no difference in ASD incidence between the groups. Despite this evidence, conspiracy groups continue spreading myths about the relationship between vaccines and autism, often associating the alleged risk with thimerosal, a mercury-based preservative once used in some vaccines. However, the type of mercury in thimerosal (ethylmercury) is rapidly eliminated by the body and does not accumulate in the brain, making it completely safe. Furthermore, thimerosal was removed from most childhood vaccines decades ago, with no impact on autism prevalence, reinforcing that there is no connection between the two. |
| **138** | Viruses | Some communities claim that viral infections, especially during pregnancy, cause autism. While certain infections may have effects on pregnancy, there is no evidence that viruses are a direct cause of ASD. |
| **139** | Vitamin A deficiency | Narratives claim that low vitamin A levels during pregnancy cause autism. While this vitamin is essential for fetal development, its deficiency can cause night blindness and other problems, but it is not linked to ASD. |
| **140** | Vitamin B1 deficiency | Some theories claim that a lack of vitamin B1 (thiamine) impairs neurological development and causes autism. However, B1 deficiency can cause beriberi, but there is no evidence that it leads to ASD. |
| **141** | Vitamin B2 deficiency | There are claims that B2 (riboflavin) deficiency is associated with autism. However, this vitamin plays a role in cellular energy production, and its absence can cause fatigue and dermatological problems, but it is not linked to ASD. |
| **142** | Vitamin B6 deficiency | Some communities claim that low B6 levels affect neurotransmitters and lead to autism. While B6 is important for brain function, its deficiency can cause irritability and neuropathy, but it is not related to ASD. |
| **143** | Vitamin B8 deficiency | Biotin (B8) is essential for cellular metabolism, but there is no evidence that its deficiency causes autism. |
| **144** | Vitamin B9 deficiency | Some theories claim that B9 (folic acid) deficiency leads to ASD. While folic acid supplementation during pregnancy is important for preventing neural tube defects, there is no evidence that its absence causes autism. |
| **145** | Vitamin C deficiency | Some groups claim that a lack of vitamin C affects the brain and leads to autism. However, its deficiency causes scurvy, but it is not associated with ASD. |
| **146** | Vitamin D deficiency | Narratives claim that low vitamin D levels during pregnancy or childhood cause autism. While this vitamin is important for bone and immune health, there is no scientific evidence to support this claim. |
| **147** | Vitamin E deficiency | Some theories suggest that a lack of vitamin E affects the nervous system and leads to autism. While this vitamin has antioxidant properties, its deficiency can cause neurological damage, but it is not linked to ASD. |
| **148** | Wireless | Conspiracy theories claim that Wi-Fi signals and electromagnetic radiation from mobile devices alter brain function and cause autism. However, Wi-Fi waves are low-frequency and non-ionizing, having no effect on neurodevelopment. |
| **149** | WPO | Some claims suggest that industrial cleaning products labeled WPO contain substances that trigger autism. However, there is no scientific evidence supporting this relationship. |
| **150** | X-ray | Some theories claim that radiation exposure from X-rays during pregnancy causes autism. While high doses of radiation are harmful, medical exams use safe levels, and there is no proof they cause ASD. |

Source: Own elaboration (2025).

#### 4.1.2. False cures and treatments for autism

Many messages exploit families' faith, suggesting that autism could be overcome through religious devotion and lifestyle changes. These narratives spread the idea that autistic children can be "cured" simply by believing more in God, promoting a discourse of guilt toward parents and caregivers. Additionally, these same messages often encourage distancing from conventional medicine, advocating extreme dietary changes, food restrictions without professional guidance, and the abandonment of household appliances such as microwaves. The promotion of "detox" to cleanse the body of supposed toxins linked to autism is particularly dangerous, as many of these practices involve harmful substances, especially for children. Homeopathy also appears as a miraculous solution, promoting "isotherapeutics" — treatments that claim to reverse the effects of vaccines and chemical substances. The recurring denial of vaccines and encouragement of unproven therapies put lives at risk.

**Figure 07.** Examples of conspiracy theories about autism cures.

Vale ressaltar que esses tratamentos são alguns das várias possibilidades para tentar fechar o quebra cabeça que é o autismo.

Para o iniciantes:

1 -Ter fé em Deus que nossas crianças serão curadas.

2- Ler muito sobre o assunto das dietas sem gluten, lactose (caseina),milho,soja, conservantes e começar o mais rapido possivel independente de medicos. Sugestão livro: Autismo esperança pela nutrição.

3- Substituir as panelas de aluminios por panelas inox.

4 - Não utilizar mais o forno Microondas.

5- Tentar marcar uma consulta com um medico para solicitar os exames de metais pesados, alergia, fungos e bacterias do GP ou do Brasil e começar a eliminação destes com alguns medicos apropriados.

06- : Cease ( Detox). Limpar o organismo de metais ou substâncias indejadas atraves de homeopatia.

- Quelação Andrew Cutler (AC): reponsavel pela eliminação dos metais pesados ( aluminios, mercurio, chumbos) acusado no exame de metais.

A **homeopatia** tem uma linha de ISOTERÁPICOS que ajudam o corpo a se desintoxicar dos efeitos negativos de substâncias e medicamentos, como a amoxicilina, no seu caso.

Nem todos os homeopatas trabalham com os isoterápicos, mas estes são a base para muitos tratamentos importantes. Ex.:

🔷 reverter o **autismo** (como fazem os homeopatas dos Paises Baixos) procedendo ao detox do excesso de vacinação da criança e de medicamentos tomados durante a gravidez, além de tratar a permeabilidade intestinal (disbiose);
🔷 para detox do corpo após **anestesia** (ex, lidocaína do dentista);
🔷 reduzir efeitos tóxicos de **vacinas** como febre amarela, tríplice, xingling, etc.;
🔷 reduzir efeitos tóxicos após exames com **contraste** (ex. iodo e raio-x, após a tomografia);
🔷 detox de substâncias que causam **alergia**;
🔷 detox de pool de **antibióticos** (amoxicilina e outros);
🔷 ajudar o **desmame** de cigarros ou de medicamentos como os **benzodiazepínicos** (alprazolam ou "frontal", clonazepam ou "rivotril", etc.).

Source: Screenshot (2025).

Another axis of false cures involves the idea that autism is caused by the accumulation of heavy metals and toxins, which could supposedly be eliminated through "detoxification". Treatments with Ethylenediaminetetraacetic Acid (EDTA) and Zeolite are promoted as effective solutions for removing these substances from the body, despite the lack of scientific evidence linking them to autism. Additionally, these groups encourage a series of protocols, including "dairy detox", excessive consumption of amino acids, digestive enzymes, intravenous infusions of NAD+, the use of lithium orotate, and hyperbaric oxygen sessions, claiming that these procedures would reverse autism. In many cases, these products are sold directly by the same people spreading the disinformation, profiting from families' desperation. One particularly dangerous substance promoted in these circles is Suramin, an antiparasitic that, besides lacking any scientific validation, can cause severe side effects.

[ english version / português abaixo ]

**Figure 08.** Examples of conspiracy theories about autism cures.

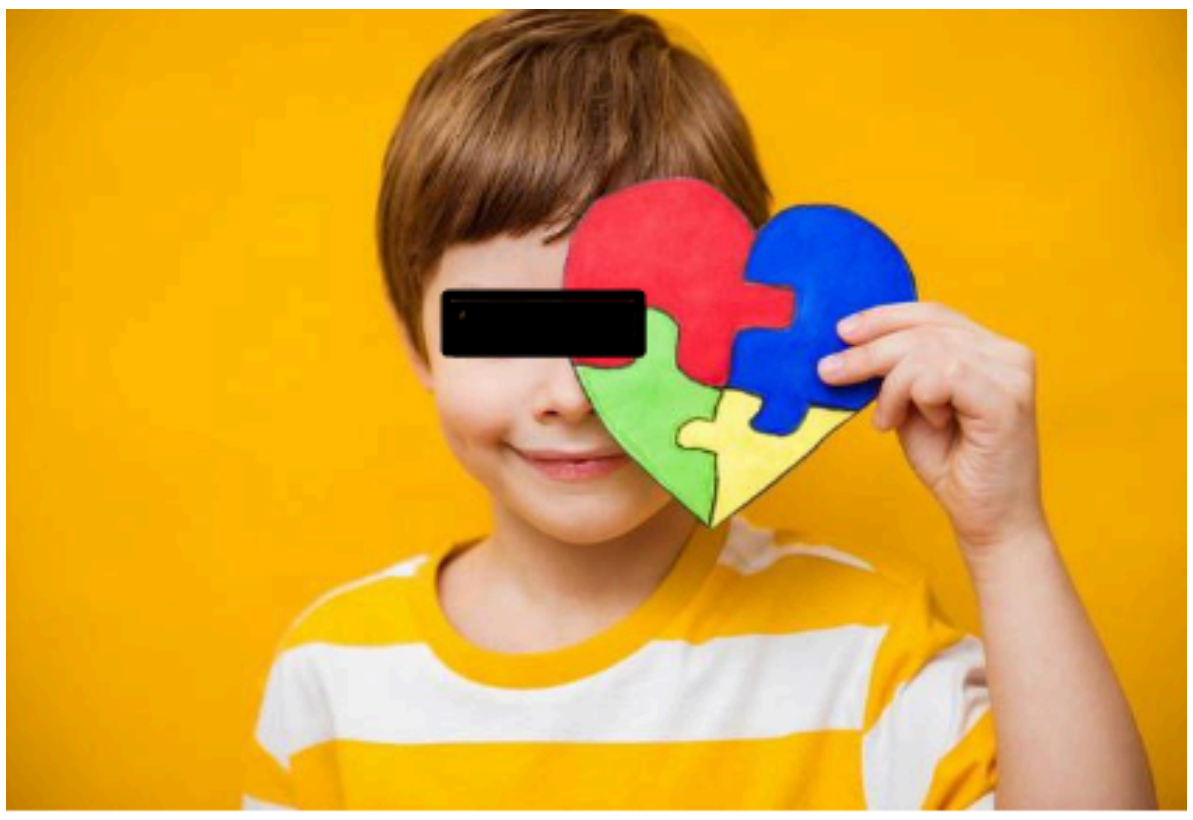

**Protocolo do doutor [redacted] para combater o autismo:**

- Desintoxicação (zeólita)

- Desintoxicação (EDTA)

- Tratamento com Oxigênio Hiperbárico

- NAD+ IV 500 mg

- Orotato de Lítio 20 mg

- Remover laticínios e glúten

- Fones de ouvido

- Mag Phos 1 mil

- Enzimas Digestivas

- Probióticos

- Aminoácidos

- CESSAR terapia

- NÃO VACINAR MAIS

SURAMINA

1 LITRO = 130,00+FRETE
#VENDAS NO MEU PRIVADO OU WATSAP [redacted]

**Pesquisadores testam droga que pode reverter sintomas do autismo**

**Os cientistas descobriram que medicamento corrige 17 tipos de anormalidades ligadas ao autismo, incluindo problemas de comportamento social**

**O que é esse medicamento?**

**A suramina é um fármaco antiparasitário desenvolvido nos laboratórios da Bayer em 1917 e utilizado para tratar infecções por tripanossoma no homem e em algumas espécies animais**

A Suramina, Ácido Shikimico, é um composto encontrado nas Agulhas de Pinheiro, o Antidoto para detox do corpo dos metais pesados

O ANTIDOTO DOS 100 ANOS -

O pinheiro é antioxidante, antidepressivo, antibacteriano, antiviral, antitumoral, anti-inflamatório, estimulador do sistema imunológico, protetor cardiovascular, triglicerídeo reduzindo e muito mais.
Suas substâncias também mata os parasitas e ajuda o autismo.

Source: Screenshot (2025).

The promise of curing autism through chemical products such as Chlorine Dioxide ($ClO_2$) — also known as MMS (Miracle Mineral Solution) — is one of the most severe cases of disinformation. This compound, promoted as a panacea for various diseases, is actually a highly toxic agent, often compared to ingesting bleach. In conspiracy groups, it is sold as a solution to "cleanse" the body, being applied in enemas or administered orally to autistic children, often causing severe intoxications. Furthermore, other substances, such as crystals and quartz, are marketed as aids in autism treatment, reinforcing the commercial nature of these communities. The so-called "deworming protocols" are also concerning, as they are based on the false belief that autism is caused by hidden parasitic infections. This leads parents to subject their children to aggressive antiparasitic treatments without necessity, potentially resulting in serious health damage.

**Figure 09.** Examples of conspiracy theories about autism cures.

**27 NIÑOS COLOMBIANOS CON AUTISMO, CURADOS CON CDS DIÓXIDO DE CLORO** 🙏🏽

Durante décadas hemos escuchado q el autismo es una enfermedad hereditaria e incurable, pero esto está demasiado alejado de la verdad

Desde q se inició con la vacunación en niños, esta enfermedad ha aumentado de manera exponencial

Hay muchos estudios q demuestran q estas "vacunas" son las q producen el autismo. La mal llamada ciencia es solo un negocio ☠️

El CDS llega a cualquier lugar del cuerpo por el torrente sanguíneo y produce la curación casi q de cualquier enfermedad

✅ **Tenemos Disponible:**

✔DIÓXIDO DE CLORO A 3000PPM

✔ZEOLITA MICRONIZADA CLINOPTILOLITA GRADO FARMACÉUTICO

✔TIERRA DE DIATOMEAS GRADO ALIMENTICIO

✔DMSO GRADO FARMACÉUTICO AL 99%

✔AGUA DE MAR HIPERTONICA

✔ ORMUS ORO MONOATÓMICO

✔TREMENTINA RESINA DE PINO

✔ GHEE ORGÁNICO

✔ORGONITAS CONTRA ONDAS ELECTROMAGNÉTICAS, 5G

✔CRISTALES CUARZOS

Fig. 31: Parásitos dentro de biofilm, también llamado de magma parasitario.
En la práctica en todos los niños afectados de autismo y en la mayoría de
las enfermedades crónicas, se ha podido ver una cantidad grande de mucosas, a
veces difícil de identificar ya que se asemeja a un Áscaris muerto o según dicen
algunos, a una mucosidad intestinal. Se encontraron mucosidades intestinales
por encima de 1 metro y por lo tanto es poco probable que sean mucosidades
del propio paciente. La Universidad de Bolonia, en Italia, afirma que es una
mucosidad propia del cuerpo. Sin embargo el Dr. Volinsky de la Universidad de
Florida ha podido hacer un análisis del ADN de la mucosidad, y opina que es
ajeno al cuerpo humano. Por lo tanto de momento, opino que es una forma de
'magma parasitario' no clasificado y por ello, tampoco aparece en los análisis de
los laboratorios. Las evidencias están dadas por los resultados.
Se han podido recuperar a más de 350 niños de autismo, basándose en este
protocolo, y todos expulsaron grandes cantidades de este plasma parasitario
(biofilm) y también otros parásitos. Después de cada expulsión, mejoraron
considerablemente. Lo mismo ocurre en muchas enfermedades crónicas,

Source: Screenshot (2025).

Autism is also included in the list of conditions supposedly curable by the so-called "Med Beds" — fictitious devices that claim to regenerate tissues, cure diseases, and even reverse people's biological age. The idea of these medical beds is widely spread in conspiracy groups, promising to eliminate diseases such as cancer, Alzheimer's, and autism through unproven technologies. These messages often take on an almost messianic tone, suggesting that the "ultimate cure" for autism already exists but is being hidden by global elites. This type of disinformation not only gives false hope to vulnerable families but also creates a lucrative market for selling courses, devices, and access to nonexistent treatments. Additionally, some variations of these narratives include the use of enemas with toxic substances such as Chlorine Dioxide, reinforcing dangerous practices already widely condemned by health experts.

**Figure 10.** Examples of conspiracy theories about autism cures.

16. Autismo: Los niños con autismo también recibirán ayuda con el tratamiento de las camas.

17. Instrumentos ortopédicos: También se abordarán cuestiones de ortopedia como colocar huesos y editar los huesos existentes en el cuerpo.

18. Depresión: La depresión irá sanando poco a poco. En última instancia, un sujeto tendrá que afrontar el trauma de forma positiva. Habrá muchos consejeros capacitados para ayudar a cualquier persona con depresión.

19. Mejoras en general: Med Beds pueden hacer que alguien sea más empático, más inteligente, etc. También se puede aprender o descargar idiomas adicionales. Sin embargo, es importante tener una razón para utilizar las mejoras que solicita descargar. Por ejemplo, no hay razón para descargar todos los idiomas del planeta si no planean utilizarlos a todos. Y una parte importante de la experiencia en este planeta, es el proceso de adquirir los conocimientos que necesitas.

20. Salud perfecta: Las camas devolverán a tu cuerpo una salud óptima y esto incluye a eliminar todo lo negativo que tenga que ver con cualquier vacuna que se haya aplicado.

21. Sana la mente: Cuando sanas la mente, sanas tu cuerpo.

22. Vitalidad: Las Med Beds devuelve a las personas a su estado óptimo de salud. Por ejemplo, si tienes 80 años, tendrás la mejor salud para una persona de 80 años. También si desea regresar la edad y lucir con apariencia más joven lo puede hacer y eso incluye la decisión de tener hijos, si desea.

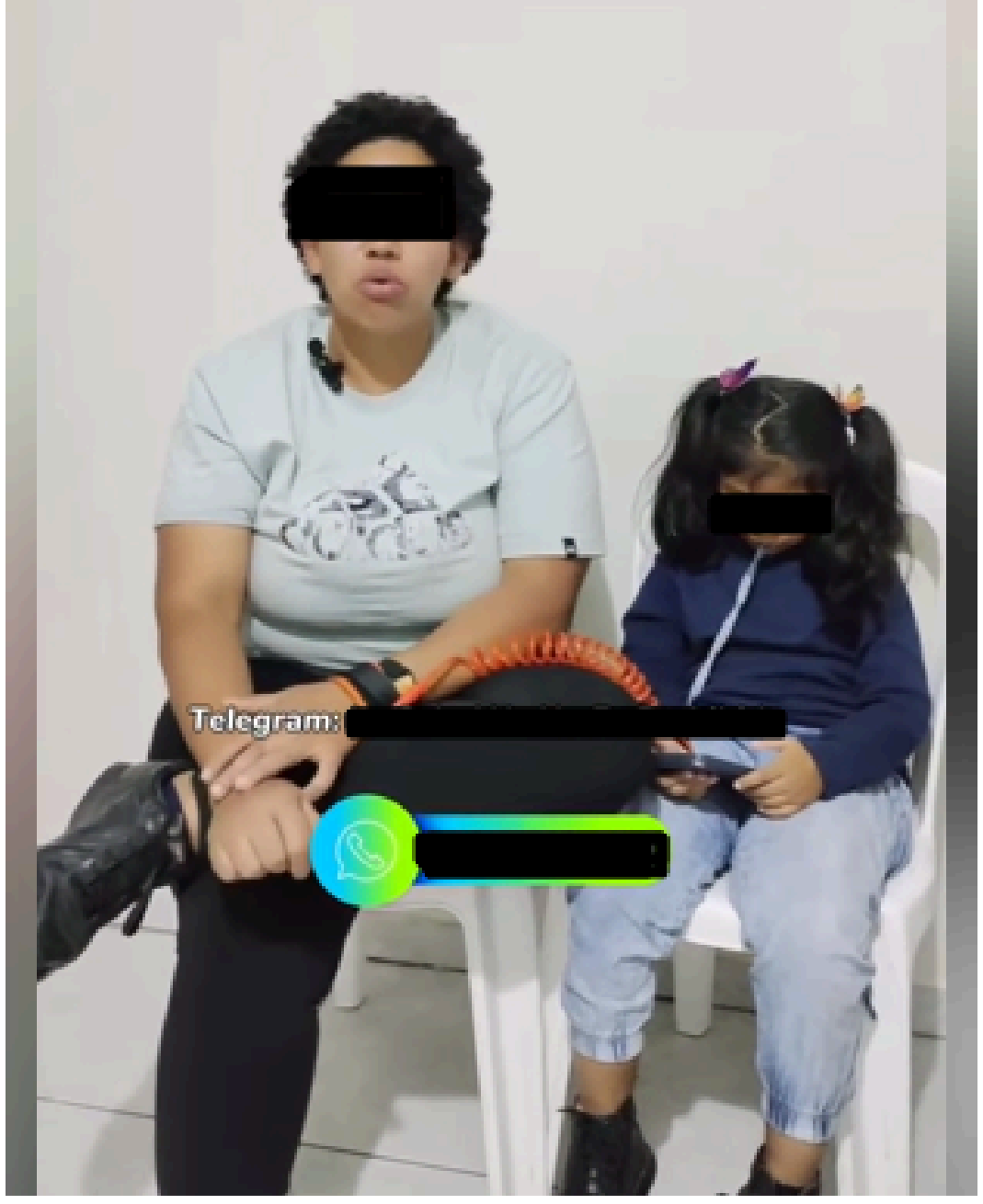

LOS ENEMAS CON CDS ACELERAN LA CURACIÓN DEL AUTISMO Y DE CUALQUIER ENFERMEDAD🤩❤️

El CDS ha sido probado en miles de niños con autismo con excelentes resultados y esta madre nos cuenta cómo los enemas aceleran mucho más la curación

Queda demostrado q los fármacos no están hechos para curar a nadie, solo es cuestión de abrir tu mente y probar lo natural ❤️

✅ TENEMOS DISPONIBLE:

✔️MELATONINA 30mg

✔️DIÓXIDO DE CLORO A 3000PPM

Source: Screenshot (2025).

In some groups more focused on "naturalism", autism is treated as an energetic imbalance that could be corrected through exposure to chemicals and infrared light. These messages promote the idea that the sun has specific healing properties for ASD and that a combination of certain supplements — sold by the same advertisers — with "electromagnetic frequencies" could restore brain function. An even more extreme approach includes the promotion of devices such as "orgonites" and "electromagnetic biofilters", claiming that 5G waves and artificial magnetic fields are responsible for autism. Additionally, some conspiracy communities spread lists of antiparasitic treatments as solutions to alleviate symptoms, reinforcing the disinformation already observed in deworming protocols. This pseudoscientific approach not only diverts attention from evidence-based treatments but also creates a market for ineffective products and therapies.

**Figure 11.** Examples of conspiracy theories about autism cures.

O vermelho no arco iris curva a luz por ter mais eletrons mais massa do q o azul.

Libera na fricção a "massa"= eletrons = eletricidade porque somos seres elétricos, seres que podem se alimentar da luz do Sol. A Vita D a melanina e a melatonina não são as únicas expressões deste FATO.

Sal +água +luz e gordura animal = eletricidade =movimento.

Ainda estou testando ESSA COMBINAÇÃO Azul de Metileno e a Luz Infravermelha.

Mas já sei a resposta.

Entenda eu sou cientista.

Eu testo.

Para mim o clue é saber q ajuda no movimento acima de 4.000 metros de altura.

Carne e sol então

**Infravermelho é o Sol da manhã** e da tarde. Sem ele vc tem doenças. O Sol elimina o mofo q traz os íons negativos e impede a luz do sol vir a ser intramolecular.

Sem Sol não tem vida! Só mofo.

**Antiparasitários:**

1. Terebintina
2. CDS. Enemas de dióxido de cloro.
3. H2O2 10 gotas - 10 obtidas - 10 obtidas
4. Terapia com ozônio
5. Ártemis Anua
6. Infusão de Artemisa absentis (absinto)
7. Tintura de Nozes
8. Terra diatomácea de qualidade alimentar (não adequada para vermes ou oxiúros)
9. Comprimidos ou cápsulas de cravo
10. Limpeza do fígado
11. Juba de leão e Polyporus.. para proteger a microbiota do tratamento e melhorar a disbiose.
12. Clorofila líquida em parasitas hepáticos (dicrocoellium e fasciola)
13. Alho cru com o estômago vazio.
14. Extrato de cominho preto
15. Sementes de abóbora
16. Óleo de orégano
17. Casca de psyllium
18. Maria Treben. Saúde da farmácia, senhor.
19. Medicina Antroposófica. Carmelo Bizkarra
20. Óleos essenciais: canela, tea tree, louro, orégano grego.
21. As argilas Pancho, se ingeridas, também são antiparasitárias.
22. Disfania Ambrosioides (Epazote/Paico). Tome 2 folhas em infusão com o estômago vazio.

Source: Screenshot (2025).

In some groups, autism is associated with a supposed "energetic misalignment" that could be corrected with Tesla frequencies, electroshocks, and other practices without any scientific foundation. Some messages even go as far as to claim that there is a "spiritual heart muscle" located behind the brain that can be "reactivated" to cure autism. This narrative mixes esoteric elements with unrealistic technological promises, often linked to the sale of equipment or sessions of "bioenergetic recalibration". These theories are especially concerning because, by steering families away from evidence-based medical approaches, they put children at risk by promoting harmful practices. The combination of miraculous promises with the commercialization of hope turns these communities into real markets of emotional exploitation, where families' tragedies are leveraged for the financial benefit of those promoting these false cures.

[ english version / português abaixo ]

**Figure 12.** Examples of conspiracy theories about autism cures.

➟CAMA MÉDICA - Camas Médicas

Trump diz: Em cerca de um ano, quase todos os procedimentos hospitalares estarão obsoletos...

Cada cidade terá muitos leitos médicos e câmaras de Tesla capazes de curar e reparar o DNA e curar todas as doenças....

Idade menor (até 30 anos)
O câncer se foi...
cura do autismo...

👉 Nunca mais fibromialgia
👉 Não há necessidade de vacinas
👉 Chega de Alzheimer
👉 Sem dor
👉 Não há mais tribunais

Haverá 5 tipos de camas médicas que desempenharão diferentes funções...

Pods médicos holográficos: podem ser portáteis e a versão menor de um laboratório sobre rodas.

Coloque os feridos na cápsula e coloque-os para dormir para que parem de degenerar e enfraquecer (ficar acordado é perigoso para o nosso bem-estar.

Todos os órgãos e partes do corpo ausentes serão reconstruídos usando sua sequência genética e funções corporais.

Olhos, pernas, braços, rins, etc. ; tudo será gerado à medida que o corpo preserva a memória muscular de todas as partes do seu ser humano.

Os lasers utilizados terão uma unidade métrica tridimensional e potência de plasma para reparar ou substituir.

Cama de Rejuvenescimento: Esta cama usa seus códigos genéticos para rejuvenescer seu corpo e combater o envelhecimento.

Você pode voltar sua idade com um gel biométrico e remover memórias indesejadas. Você pode eliminar áreas de sua vida que estavam erradas ou coisas que o incomodavam.

Leitos Antipatogênicos – Esses leitos removem metais pesados, detritos, parasitas e todos os outros invasores que causam problemas de saúde.

Aqui está o leito espiritual que irá reconectar você com seu eu superior, consertando e consertando o fio de prata.

A parte de trás do cérebro, onde está localizado o músculo cardíaco que conecta você com seu eu superior; pode e irá restaurar sua conexão espiritual com uma versão maior de sua consciência...

Source: Screenshot (2025).

**Table 06.** List of false autism cures and treatments mapped in the communities (A-Z).

| | **Supposed cure and/or treatment** | **Example** |
|---|---|---|
| **1** | Activated Charcoal | Activated charcoal is promoted as a "detoxifier" that removes heavy metals and toxins from the body. However, activated charcoal can interfere with nutrient and medication absorption, as well as cause intestinal obstruction and severe dehydration. Improper use can be dangerous. |
| **2** | Aerochell | This term appears to be related to conspiracy theories about "chemtrails" and airborne chemicals. Some theories claim that exposure to aerochell causes autism and that "purifications" with natural substances can reverse the condition. This is completely false and has no scientific basis. |
| **3** | Alpha-Lipoic Acid | This antioxidant is promoted as a "detoxifier" that supposedly removes heavy metals from the body and "cures" autism. However, there is no scientific evidence that alpha-lipoic acid has any impact on ASD symptoms. Excessive use can cause side effects such as drastic drops in blood sugar and neurological problems. |
| **4** | Amino Acids | Although amino acids are essential for the body, there is no evidence that supplementation can reverse or treat autism. Unnecessary supplements can cause metabolic imbalances. |

[ english version / português abaixo ]

| | | |
|---|---|---|
| **5** | Anthroposophic Medicine | Anthroposophic medicine is an alternative approach based on spiritual and holistic concepts. While some therapies may be useful as supportive care, there is no scientific evidence that anthroposophic medicine treats or cures autism. |
| **6** | Apple | Apples are healthy foods, but they do not cure autism. Some claims state that their pectin content helps "detoxify heavy metals", but there is no scientific evidence supporting this theory. |
| **7** | Arginine | Arginine is an amino acid essential for various bodily functions, but there is no scientific evidence that its supplementation has any impact on autism. |
| **8** | Artemisia Absinthium Infusion (Wormwood) | Wormwood is a plant used in herbal medicine, but its use in high doses can be toxic. Some claims suggest that it "cleanses the body" and improves autism, but there is no scientific proof for this claim. Additionally, wormwood contains thujone, a substance that can cause seizures and neurological damage. |
| **9** | Artemisia Annua | The Artemisia annua plant is used to treat malaria, but pseudoscientific groups promote it as a "natural cure" for autism. There is no scientific evidence supporting this claim, and improper use can be toxic. |
| **10** | Ashwagandha | Withania somnifera, known as ashwagandha, is an herb used in Ayurvedic medicine to reduce stress and improve cognition. Some groups claim it can "regulate neurotransmitters" and treat autism. However, there is no scientific evidence that ashwagandha has any effect on ASD. Additionally, excessive use can cause gastrointestinal problems, intense drowsiness, and negative interactions with psychiatric medications. |
| **11** | Astaxanthin | This antioxidant found in algae and seafood is promoted as a "brain protector" that could help with autism. Although antioxidants are important for overall health, there is no scientific evidence that astaxanthin has any impact on ASD. |
| **12** | Bay Leaf Oil | Bay leaf oil is promoted by some groups as a body purifier, but there is no scientific evidence supporting this claim. |
| **13** | Bed Meds | This term may refer to various medications promoted for use before bedtime, including sedatives and anxiolytics. Although some autistic individuals may experience sleep difficulties, there is no such thing as an "autism medication", and any prescription should be made by a healthcare professional. Improper use of sedatives can lead to dependence, excessive drowsiness, and overdose risk. |
| **14** | Bitter Melon Tincture (from São Caetano) | Bitter melon is promoted as a "body cleanser", but there is no proof that it has any effect on ASD. Excessive consumption can be toxic to the liver and cause gastrointestinal issues. |
| **15** | Black Cumin | Black cumin (Nigella sativa) is used in traditional medicine for various purposes, but there is no evidence that it can treat or cure autism. |
| **16** | Black Cumin Extract | Black cumin is promoted as a "natural anti-inflammatory" that supposedly reduces ASD symptoms. However, there is no scientific evidence supporting this idea. |
| **17** | Black Currant | Black currant is promoted as an antioxidant, but there is no scientific evidence that its consumption has any impact on ASD. |

[ english version / português abaixo ]

| | | |
|---|---|---|
| **18** | Bleach | This is one of the most dangerous and criminal "cures" promoted by conspiracy groups. Chlorine dioxide, also known as MMS (Miracle Mineral Solution) or CDS (Chlorine Dioxide Solution), is a highly toxic industrial bleach that these groups claim "cleanses the body" of substances supposedly causing autism. Its use can cause irreversible gastrointestinal damage, vomiting, severe diarrhea, liver failure, respiratory problems, and even death. Health agencies, including the FDA (USA), Anvisa (Brazil), and WHO, have issued warnings against MMS/CDS, as there is no proof that it works for autism, but there is evidence that it is a poison. |
| **19** | Bone Broth | Promoted as a "natural cure" for autism, bone broth is rich in collagen and minerals, but there is no evidence that it impacts ASD. Additionally, there are concerns about the presence of heavy metals, such as lead, in broths made from animal bones. |
| **20** | Boron | Boron supplements are promoted as a way to "improve brain function" in autism. However, there is no scientific evidence supporting this claim, and excessive boron intake can cause toxicity, digestive problems, and kidney damage. |
| **21** | Broccoli | Broccoli is a healthy food rich in antioxidants, but it is not a cure for autism. The claim that it improves ASD comes from a small and preliminary study on sulforaphane, a compound found in broccoli, but there is no robust evidence that broccoli consumption impacts ASD. |
| **22** | Calcium | Calcium is essential for bone health, but there is no relationship between calcium deficiency and autism. Excess calcium can lead to kidney problems and tissue calcification. |
| **23** | Camel Milk | Camel milk is promoted as an autism treatment by some alternative groups due to its protein and antibody composition. However, there is no scientific evidence that consuming this milk impacts ASD. Additionally, unpasteurized camel milk consumption can pose health risks, such as bacterial infections. |
| **24** | Carnitine | Carnitine is a compound involved in energy metabolism, and its deficiency can cause muscle weakness. However, there is no evidence that its supplementation has any effect on autism. |
| **25** | Casein Detox | This practice involves eliminating dairy products from the diet, based on the claim that casein interferes with brain function and causes autism. However, there is no scientific proof that casein has any relationship with ASD. |
| **26** | Chelaton (Chelation Therapy) | Chelation therapy involves using substances such as EDTA to remove heavy metals from the body. Pseudoscientific groups claim that this "cures" autism, but there is no scientific evidence supporting this claim. Additionally, chelation can be extremely dangerous, causing kidney damage, seizures, cardiac alterations, and even death. The FDA has already issued warnings against using chelation for autism. |
| **27** | Chlorella (Chlorella Vulgaris) | Chlorella is an algae used as a supplement and promoted as a "detoxifier". However, there are no studies proving that it treats autism. Excessive use can cause digestive issues and allergic reactions. |
| **28** | Chlorine Dioxide | Chlorine dioxide, also known as MMS (Miracle Mineral Solution) or CDS, is a highly toxic industrial bleach that conspiracy theorists promote as an "autism cure". Its use can cause irreversible gastrointestinal damage, liver failure, respiratory problems, seizures, and even death. Organizations such as WHO, FDA, and Anvisa have already warned against this dangerous practice. |
| **29** | Cholesterol Sulfate | Some claims suggest that cholesterol sulfate regulates metabolic processes and that its deficiency is linked to ASD. However, there is no scientific evidence that supplementing this compound has any impact on autism. |
| **30** | Cilantro | Cilantro is promoted as a "natural chelator" of heavy metals, supposedly capable of reversing autism. However, there is no scientific evidence that it affects ASD. Additionally, high doses of cilantro can cause gastrointestinal discomfort. |
| **31** | Cinnamon Oil | Cinnamon oil is promoted as a "metabolic regulator" that could treat autism. However, there is no scientific evidence to support this claim. Additionally, excessive consumption may cause gastric irritation and allergic reactions. |
| **32** | Citric Acid | There are claims that the citric acid found in citrus fruits aids in detoxification and "cures" autism. However, citric acid is a common dietary compound and has no effect on ASD. Excessive consumption can cause gastric irritation. |
| **33** | Clove | Like other herbs promoted as "purifiers", clove has no proven impact on ASD. Additionally, clove essential oil can be toxic in high doses. |
| **34** | Coconut Oil | Although it is a healthy fat, coconut oil has no properties that treat or cure autism. Excessive use can lead to weight gain and increased cholesterol levels. |
| **35** | Colloidal Gold | Colloidal gold is promoted as a "cognitive enhancer" for autism, but there is no scientific evidence proving this claim. Excessive consumption can cause metal accumulation in the body and toxicity. |
| **36** | Colloidal Silver | Colloidal silver is promoted as a "natural antiseptic" capable of eliminating agents that supposedly cause autism. However, there is no scientific evidence supporting this claim, and prolonged consumption can cause severe toxicity and silver accumulation in tissues (argyria), permanently turning the skin blue. |
| **37** | Copper | Some narratives claim that copper supplementation balances neurotransmitters in autism. However, excess copper in the body can be toxic, causing liver, neurological, and metabolic damage. |
| **38** | Creatine | Creatine is a supplement used to improve muscle and energy performance, but there is no evidence that it treats or improves autism symptoms. |
| **39** | Crystals / Quartz | The belief that crystals and quartz can balance energies and cure autism is based on pseudoscience. There is no evidence that crystals have any therapeutic effect on ASD or any medical condition. |
| **40** | Curcumin (Turmeric Derivative) | Curcumin is a bioactive compound in turmeric with anti-inflammatory properties. Although preliminary studies suggest it may have general health benefits, there is no proof that it affects autism treatment. |
| **41** | Cysteine (N-Acetylcysteine, NAC) | NAC is an antioxidant used for some psychiatric conditions, but there is no proof that it "cures" autism. |
| **42** | Dairy Detox | The removal of all dairy products from the diet as an "autism treatment" follows the same logic as casein detox, without any scientific evidence of effectiveness. |
| **43** | Dandelion Leaf | Dandelion is used in alternative medicine to "purify the liver", but there is no evidence that it has any effect on autism. |
| **44** | Detox Protocols | "Detox protocols" include practices such as chelation, aggressive enemas, restrictive diets, and consumption of dangerous substances like chlorine dioxide (MMS/CDS). These methods have no scientific basis and can be extremely harmful, causing poisoning, severe dehydration, and even liver failure. |
| **45** | Diatomaceous Earth | Diatomaceous earth is a substance used as a natural insecticide, but some groups claim its ingestion "eliminates parasites" and "cures autism". There is no scientific evidence for this, and consumption can cause severe irritation in the digestive and respiratory tracts. |
| **46** | Diet | While a healthy diet is essential for overall well-being, there is no "diet to cure autism". Restrictive diets without medical supervision can cause severe nutritional deficiencies. |
| **47** | Digestive Enzymes | Some diets promote the use of digestive enzymes to "improve nutrient absorption" in autistic individuals. While enzymes can be beneficial for people with specific gastrointestinal issues, there is no evidence that they have any impact on autism. |
| **48** | Dimethylglycine | This supplement is promoted as a "cognitive enhancer" for autism. However, there is no scientific evidence proving its effectiveness. |
| **49** | Dysphania Ambrosioides (Epazote/Paico) | This herb is traditionally used as a vermifuge, and some groups claim it can "cure autism by removing parasites". As previously discussed, ASD is not caused by parasites, and the improper use of this herb can be toxic to the liver. |
| **50** | EGCG (Green Tea Catechin) | EGCG, an antioxidant found in green tea, is promoted as a substance that "improves autism". While it is a healthy compound, there is no scientific evidence that it impacts ASD. |
| **51** | Enemas | Enemas are promoted by conspiracy groups as a way to "remove toxins" or "eliminate parasites that cause autism". Some practices include the use of coffee enemas, chlorine dioxide (MMS/CDS), baking soda, and aggressive herbs. Frequent and unsupervised use can cause severe intestinal irritation, dehydration, inflammation, bleeding, and rectal perforation, putting the child's life at risk. There is no scientific basis to justify using enemas to treat ASD. |
| **52** | Ethylenediaminetetraacetic Acid (EDTA) | EDTA is used in chelation therapy, a legitimate treatment for heavy metal poisoning, but it is dangerously promoted for autistic children. Chelation does not treat or cure autism and can be extremely dangerous, causing kidney damage, seizures, and even death. |
| **53** | Faith in God | Faith can be a positive element for many people, but autism is not a condition that can be "cured" through faith, prayer, or religious practices. This narrative is dangerous because it shifts focus away from evidence-based approaches and may lead to the rejection of appropriate treatments and support. |
| **54** | Fennel Tea | Fennel (anise) tea is used to relieve digestive issues, but there is no evidence that it has any effect on autism. |
| **55** | Folic Acid | Although folic acid is essential for fetal development, there is no evidence that its supplementation "cures" autism. Deficiency in this vitamin can cause neural tube defects, but it is not associated with ASD. |
| **56** | Gamma-Aminobutyric Acid (GABA) | This neurotransmitter is sold as a supplement to "calm" autistic children, but there is no scientific proof that GABA has any therapeutic effect on ASD. Additionally, neurotransmitter regulation in the brain is not as simple as taking a supplement. |
| **57** | Ginger | Ginger has antioxidant and anti-inflammatory properties, but there are no studies proving that it impacts ASD. |
| **58** | Gliadin | Gliadin is a protein found in gluten, and some groups claim it is linked to autism. While people with celiac disease may benefit from a gluten-free diet, there is no proof that removing gliadin improves or cures ASD. |
| **59** | Gluten Detox | Just like casein detox, removing gluten from the diet is promoted as a "cure for autism". Although some autistic individuals may have gluten sensitivity, there is no evidence that excluding gluten reverses or treats ASD. |
| **60** | Glycine | Glycine is an amino acid involved in neurotransmission, and some theories claim that its supplementation can "correct brain deficiencies" in autistic individuals. However, there is no scientific evidence proving this relationship. |
| **61** | Glyphosate | Conspiracy groups claim that glyphosate, an herbicide used in agriculture, is linked to autism. While the impact of pesticides on health is a relevant topic, there are no scientific studies proving that glyphosate causes ASD. |
| **62** | Hesperidin | Hesperidin, a flavonoid found in citrus fruits, is promoted as a "neuroprotector". However, there are no studies supporting its effectiveness in autism. |
| **63** | Homeopathy | Homeopathic treatments are widely promoted for various conditions, including autism. However, there is no scientific basis proving their effectiveness. Homeopathy is based on extreme dilution of substances, meaning homeopathic "remedies" do not contain detectable amounts of active ingredients. |
| **64** | Hydrogen Peroxide ($H_2O_2$) | The use of hydrogen peroxide to "cure autism" is one of the most dangerous and irresponsible therapies promoted by alternative groups. Some people encourage oral consumption or injection of this substance, claiming it kills parasites or eliminates "toxins". This is an extremely dangerous practice that can cause internal burns, severe digestive tract damage, and even death. |
| **65** | Hydroxychloroqui ne | After the COVID-19 pandemic, hydroxychloroquine was promoted as a supposed treatment for various conditions, including autism. There is no scientific evidence supporting this claim, and the indiscriminate use of this medication can cause cardiac arrhythmias, liver problems, and eye damage. |
| **66** | Hyperbaric Chamber (HBOT) | Hyperbaric oxygen therapy involves inhaling pure oxygen at high pressure and is used to treat certain conditions, such as carbon monoxide poisoning. However, there is no scientific evidence that HBOT treats or cures autism. Improper use can be dangerous, leading to lung damage, seizures, and oxygen toxicity. |
| **67** | Hyperbaric Oxygen Therapy (HBOT) | Hyperbaric oxygen therapy involves inhaling pure oxygen under high pressure and is promoted as a cure for autism. There is no scientific evidence proving its efficacy, and improper use can lead to seizures, oxygen toxicity, and lung damage. |
| **68** | Hypertonic Seawater | Supporters of this practice claim that hypertonic seawater "reprograms" the body and improves autism. However, no studies demonstrate any real benefit, and excessive consumption can lead to dehydration, sodium toxicity, and kidney problems. |
| **69** | Infrared Light | There are claims that infrared light therapy improves "brain functions" in autistic individuals. While infrared light has legitimate medical applications, there is no scientific evidence that it impacts ASD. |
| **70** | Inhaled Ibuprofen | Some theories suggest that inhaled ibuprofen can "reduce brain inflammation" and treat autism. There is no scientific evidence supporting this idea, and inhaling ibuprofen can be extremely toxic to the lungs. |
| **71** | Injectable Glutathione | Glutathione is a natural antioxidant, and some groups promote its injectable application as a "detox therapy" for autism. There is no scientific evidence proving this practice, and unnecessary glutathione injections can cause side effects, including allergic reactions and metabolic imbalances. |
| **72** | Ivermectin | Ivermectin, mistakenly promoted during the COVID-19 pandemic, is also suggested as a treatment for autism. Some groups claim that it eliminates "parasites that cause ASD", but this theory is completely false, as autism is not caused by parasitic infections. The improper use of ivermectin can lead to liver damage, seizures, and even coma. |
| **73** | Lactoferrin | Lactoferrin is a protein found in breast milk and some supplements. Some claims state that its supplementation could "regulate the immune system" and treat autism, but there is no scientific proof for this claim. |
| **74** | Lemon | Some narratives suggest that lemon consumption "detoxifies the body" and improves autism. While it is a healthy food, there is no evidence that lemon has any impact on ASD. |
| **75** | Lion's Mane Mushroom | This medicinal mushroom is promoted as a "neuronal regenerator" that could improve autism. While there are preliminary studies on its cognitive effects, there is no robust evidence that it has any effect on ASD. |
| **76** | Liquid Chlorophyll | Liquid chlorophyll is promoted as a "detoxifier" that removes toxic substances associated with autism. However, there is no scientific evidence that chlorophyll has any impact on ASD. Additionally, excessive consumption can cause photosensitivity and digestive issues. |
| **77** | Lithium Orotate | Lithium is used in psychiatric treatments for bipolar disorder, and some claims suggest that its orotate form could "correct brain imbalances" in autistic individuals. There is no scientific evidence supporting this claim, and improper use of lithium can cause severe toxicity, kidney damage, and heart problems. |
| **78** | Liver Cleanse | "Liver cleansing" is an alternative practice that claims to eliminate toxins that cause autism. This method typically involves consuming large amounts of oils, magnesium salts, and citrus juices, which can cause severe diarrhea, dehydration, and metabolic disturbances. There is no scientific evidence that autism is caused by liver toxins. |
| **79** | Macerated Garlic | A similar version of raw garlic, promoted to "cleanse" the body and eliminate supposed toxins that cause autism. There is no scientific evidence for this claim, and excessive use can cause digestive damage. |
| **80** | Magnesium | Magnesium is an essential mineral, and some claims suggest that its deficiency causes autism. There is no scientific proof that magnesium supplementation has any effect on ASD. Excessive magnesium intake can cause diarrhea, low blood pressure, and cardiac arrhythmias. |
| **81** | Maleic Acid | Supposedly promoted as a "detoxifier", maleic acid has no relation to autism, and its use can cause adverse gastrointestinal effects. |

[ english version / português abaixo ]

| | | |
|---|---|---|
| **82** | Mebendazole | Mebendazole is an anthelmintic drug, and some pseudoscientific communities claim that "eliminating parasites cures autism". This theory is completely false, as autism is not caused by parasitic infections. Improper use of mebendazole can cause liver problems and severe gastrointestinal disorders. |
| **83** | Methylene Blue | Methylene blue is a chemical dye with medical applications, but pseudoscientific groups claim that it improves "brain oxygenation" and treats autism. This substance can be toxic when ingested, causing neurological damage, high blood pressure, cardiac arrhythmias, and respiratory problems. Its unsupervised use can be extremely dangerous. |
| **84** | Microbiome / Fecal Transplant | Some research suggests that the gut microbiome may influence aspects of neurodevelopment, leading pseudoscientific groups to promote fecal transplants as a cure for autism. While fecal transplantation is a legitimate treatment for severe intestinal infections, there is no scientific evidence proving its efficacy for ASD. Additionally, this procedure can be extremely dangerous, exposing patients to infections and serious complications. |
| **85** | Milk Thistle | This herb is promoted as a "liver detoxifier", supposedly removing substances that cause autism. However, there is no scientific evidence that it has any impact on ASD. |
| **86** | Monoatomic Gold | This concept, popular in esoteric groups, claims that "special gold particles" can alter consciousness and improve cognition in autistic individuals. This has no scientific basis and is a fraud. |
| **87** | Monoatomic Vibrational Dioxide | This term has no basis in science and is promoted by sellers of false cures. There is no proof that this substance has any effect on autism. |
| **88** | Mustard Sprouts | Some alternative diets claim that mustard sprouts "detoxify the body" and improve autism. However, there is no scientific evidence supporting this claim. |
| **89** | NAC (N-Acetylcysteine) | N-acetylcysteine is an antioxidant studied for various psychiatric conditions, and some studies explore its potential for autism. However, there is no evidence that it is a cure for ASD, and its use should only be under medical supervision. |
| **90** | NAD+ IV | NAD+ (nicotinamide adenine dinucleotide) is promoted by alternative therapy clinics as a "cellular cure" for autism. However, there is no scientific evidence to support this claim. Furthermore, intravenous NAD+ infusions can cause severe side effects, such as low blood pressure and adverse reactions. |
| **91** | Neem (Indian Lilac) | Neem oil is promoted as a "body cleanser" and a "cure for autism", but there is no scientific evidence to support this claim. Excessive consumption of neem can be toxic to the liver and cause severe gastrointestinal disorders. |
| **92** | Neodymium Copper Bracelet | Some people believe that copper magnetic bracelets improve brain function and cure autism. This has no scientific basis and is merely a placebo with no real effect. |
| **93** | Nitric Oxide | Nitric oxide is a gas that helps regulate vascular and neurological functions, and some claims suggest that its supplementation could improve autism. However, no scientific studies support this theory, and excessive use may cause blood pressure changes and cardiovascular risks. |
| **94** | Oregano Oil | Oregano oil has antimicrobial and antifungal properties, but there is no evidence that it has any impact on autism. Excessive use may cause gastrointestinal problems and mucosal irritation. |

[ english version / português abaixo ]

| | | |
|---|---|---|
| **95** | Organic GHEE Butter | GHEE is a clarified fat used in cooking and Ayurvedic medicine. Some narratives claim that it "enhances the brain and treats autism", but there are no studies supporting this idea. |
| **96** | Orgonites (Crystals) | Orgonites are crystal and resin objects promoted as "energy regulators" to "cure autism". This has no scientific basis and is part of esoteric beliefs with no proven efficacy. |
| **97** | ORMUS (Crushed Sea Salt and Himalayan Salt Crystals) | ORMUS is a pseudoscientific concept suggesting that crushed sea salts contain "monoatomic elements" with quantum properties that could cure autism. This is a fraud with no scientific basis. There is no evidence that ORMUS has any effect on ASD. |
| **98** | Orthosilicic Acid ($SiOH_4$) | Some claims state that this compound helps "eliminate heavy metals" and, consequently, "treats" autism. However, there is no scientific evidence supporting this idea, and excessive ingestion can be toxic. |
| **99** | Ostrich Oil | Ostrich oil is promoted by some groups as an anti-inflammatory supplement for ASD. However, no studies prove its effectiveness in treating autism. |
| **100** | Ozone Therapy | Ozone therapy is promoted by pseudoscientific groups as a way to "purify the body", but there is no evidence that this works for autism. Ozone can be toxic when inhaled or injected, causing lung damage, severe inflammation, and embolism risk. |
| **101** | Pancho Clays | Conspiracy groups claim that ingesting or applying "purifying" clays on the body can detoxify and cure autism. This has no scientific basis, and some clays may contain heavy metals that are toxic when ingested. |
| **102** | Parasite Detox Protocols | "Parasite detox protocols" are based on the false premise that autism is caused by parasitic infections. These treatments include the use of dangerous anthelmintics, chlorine dioxide enemas, and extreme diets. Excessive use of antiparasitic drugs can cause liver damage, neurotoxicity, and severe imbalances in the body. |
| **103** | Pine Needle Tincture | This plant extract is promoted as a "detoxifier", but there is no evidence that it has any impact on autism. |
| **104** | Pine Needles | Some communities promote pine needle extracts as a way to "detoxify" the body and improve autism. However, there is no scientific proof of its effectiveness, and unsupervised consumption can be dangerous due to the presence of potentially toxic substances. |
| **105** | Pine Resin | Pine resin is promoted as a "natural detoxifier", but there is no scientific evidence that it has any impact on autism. Furthermore, improper consumption can be toxic and cause severe allergic reactions. |
| **106** | Polyporus | This medicinal mushroom is promoted by some communities as an immune system modulator for autistic individuals, but there is no scientific proof that it has any effect on ASD. |
| **107** | Potassium Citrate | Potassium citrate is used to treat kidney problems, but there is no evidence that it affects ASD. Excessive intake can cause cardiac arrhythmias and muscle problems. |
| **108** | Probiotics | While the gut microbiome may influence general health aspects, there is no evidence that probiotics cure autism. Some studies suggest they may help with gastrointestinal symptoms associated with ASD, but they do not alter the neurodevelopmental condition. |

[ english version / português abaixo ]

| | | |
|---|---|---|
| **109** | Prunella Vulgaris (Self-Heal Herb) | This medicinal plant is promoted as an "immune system modulator" for autistic individuals. However, there are no scientific studies supporting this claim. |
| **110** | Psyllium Husk | Psyllium is a fiber used to regulate bowel movements. While some autistic individuals have gastrointestinal issues, psyllium is not a cure for autism. Excessive use can cause bloating, gas, and intestinal blockages. |
| **111** | Pumpkin Seeds | Some claims suggest that pumpkin seeds eliminate "parasites causing autism". This assertion has no scientific basis, and consuming pumpkin seeds has no effect on ASD. |
| **112** | Pycnogenol | Pycnogenol is a natural antioxidant extracted from maritime pine bark. Although it has antioxidant properties, there is no scientific evidence proving its effect on autism. |
| **113** | Quercetin | Quercetin is an antioxidant flavonoid found in fruits and vegetables. Some claims suggest that its supplementation helps with autism, but there are no robust scientific studies proving this relationship. |
| **114** | Raw Garlic | Garlic is promoted as a "natural antibiotic" and "detoxifier", being suggested to eliminate "parasites that cause autism". As previously discussed, autism is not caused by parasites, and eating raw garlic does not alter neurodevelopment. Additionally, excessive consumption can cause severe gastrointestinal irritation. |
| **115** | Rosemary | Rosemary is an aromatic herb used in cooking and some natural remedies. Some claims state that it improves "brain function" and reverses autism, but no scientific studies support this idea. |
| **116** | Rotenone | Rotenone is a natural pesticide that has paradoxically been associated with an increased risk of neurodegenerative diseases. Some theories suggest that exposure to rotenone may be linked to autism symptoms, but there is no scientific evidence that it is a treatment for ASD. Rotenone can be highly toxic when ingested, causing irreversible neurological damage. |
| **117** | Rue (Ruta graveolens) | Rue is promoted as a natural remedy for various health problems, including autism. However, there is no scientific evidence proving its effectiveness. Additionally, rue can be toxic in large quantities, causing liver damage and severe gastrointestinal irritation. |
| **118** | Selenium | Selenium is an essential mineral for metabolism, but there is no evidence that its supplementation treats or cures autism. Excess selenium can be toxic, leading to hair loss, gastrointestinal problems, and neurological damage. |
| **119** | Serotonin | Some theories suggest that autism is related to a "serotonin imbalance" and that supplementing or manipulating this neurotransmitter could cure it. While medications that affect serotonin may be prescribed to treat specific symptoms, there is no evidence that serotonin cures autism. Improper administration of substances that alter serotonin levels can cause serotonin syndrome, leading to mental confusion, seizures, and cardiac arrhythmias. |
| **120** | Shikimic Acid | Pseudoscientific groups claim that this acid, found in some plants, helps "reverse autism". However, there is no scientific research supporting this claim, and the belief that natural compounds can "cure" ASD ignores its neurodevelopmental and genetic basis. |
| **121** | Silicon | Silicon is an essential mineral for bone formation, but there is no evidence that it has any effect on autism. |
| **122** | Spike Protein Detox | This anti-vaccine theory claims that people who received COVID-19 vaccines need to "eliminate" the spike protein from their bodies to prevent neurological damage, including ASD. There is no scientific basis for this theory, and the products sold for this "detoxification" can be harmful to health. |
| **123** | Star Anise Tea | Star anise contains substances that can be toxic in high doses. Some groups claim it "improves autism", but there is no scientific evidence for this claim. Excessive consumption can cause seizures, vomiting, and neurotoxic effects. |
| **124** | Sugar Detox | The claim that eliminating sugar from the diet "cures" autism has no scientific basis. While a balanced diet can improve overall health, there is no evidence that removing sugar treats ASD. |
| **125** | Sulforaphane | Sulforaphane, found in broccoli, has been studied for its potential neuroprotective effects, but there is not enough evidence to claim that it cures or treats autism. |
| **126** | Superoxide Dismutase (SOD) | SOD is an antioxidant enzyme promoted as a supplement to combat "oxidative stress" in autism. However, there is no scientific evidence that its supplementation has therapeutic effects on ASD. |
| **127** | Suramin | Suramin is an antiparasitic drug that has been investigated in preliminary studies for possible effects on autism. However, there is no robust scientific evidence proving it to be safe or effective for ASD. Improper use can cause severe side effects, including kidney toxicity, immune system suppression, and even death. |
| **128** | Sweat (Sauna Therapy) | Some alternative practices claim that "sweating out toxins" in saunas can cure autism. This has no scientific basis. Exposing children to saunas can be dangerous, leading to severe dehydration, hyperthermia, and kidney damage. |
| **129** | Taurine | Taurine is an amino acid found in energy supplements and some diets. Some theories suggest it helps with cognition, but there is no scientific evidence that taurine supplementation treats or cures autism. |
| **130** | Tesla Frequency Therapy | This treatment is based on the pseudoscientific idea that electromagnetic frequencies can "adjust the brain" of autistic individuals. There is no scientific evidence to support this claim. |
| **131** | Transdermal Melatonin | Melatonin is used to regulate sleep and can benefit autistic individuals with sleep difficulties. However, it is not a cure for ASD. Additionally, transdermal use (absorbed through the skin) has no proven scientific efficacy. |
| **132** | Trementina (Turpentine Oil) | Turpentine, a chemical solvent, is promoted as a "parasite eliminator" and "detoxifier". Its ingestion can cause severe poisoning, kidney failure, and death. |
| **133** | Turmeric Tincture | Turmeric has antioxidant properties, but there is no evidence that its tincture has any effect on autism. |
| **134** | Turpentine | Turpentine (oil derived from pine resin) is promoted as a "body purifier" for autistic individuals, but there is no scientific basis for this claim. Ingesting or inhaling turpentine can cause severe poisoning, liver failure, neurological damage, and even death. |
| **135** | Tyrosine | Tyrosine is an amino acid essential for neurotransmitter production, but there is no evidence that supplementing it alters ASD symptoms. |

[ english version / português abaixo ]

| | | |
|---|---|---|
| **136** | Vaccine Detox | "Vaccine detox" is a practice promoted by anti-vaccine groups claiming that vaccines contain toxins that need to be eliminated to "reverse autism". This practice has no scientific basis and can be highly dangerous, as it involves the use of toxic substances such as chlorine dioxide (MMS/CDS), activated charcoal, and aggressive chelators. Vaccines DO NOT cause autism, and any treatment claiming to "detoxify" vaccines is a dangerous scam. |
| **137** | Vitamin A | Although vitamin A is essential for various bodily functions, there is no scientific evidence that its supplementation cures or treats autism. Excess vitamin A can be toxic, causing liver problems and neurological damage. |
| **138** | Vitamin B1 | Some claims suggest that vitamin B1 deficiency is linked to autism. There is no scientific proof of this relationship, and excessive supplementation can cause adverse neurological effects. |
| **139** | Vitamin B12 | Vitamin B12 is essential for brain function, but there is no evidence that its supplementation cures autism. |
| **140** | Vitamin B2 | Riboflavin is important for metabolism, but there is no evidence that its supplementation affects ASD. |
| **141** | Vitamin B6 | While vitamin B6 is involved in neurotransmission, there is no proof that its supplementation alters autism. |
| **142** | Vitamin B8 | Biotin is essential for energy metabolism, but no studies prove any relationship between biotin deficiency and ASD. |
| **143** | Vitamin B9 | Folic acid is essential for fetal development, but there is no evidence that its supplementation after birth alters autism. |
| **144** | Vitamin C | Although vitamin C is important for the immune system, there is no scientific evidence that it treats or cures autism. |
| **145** | Vitamin D | While some studies investigate the relationship between vitamin D and neurological development, there is no proof that its supplementation cures autism. |
| **146** | Vitamin E | Vitamin E is an important antioxidant, but there are no studies proving its impact on ASD. |
| **147** | Walnut Tincture | Walnut tincture is suggested as an antiparasitic treatment for autism. As previously discussed, autism is not caused by parasites, and consuming this tincture without supervision can be toxic. |
| **148** | Wormwood (Losna) | Artemisia absinthium, known as wormwood, is promoted by some groups as a vermifuge "capable of curing autism". However, there is no scientific basis proving this claim. Additionally, wormwood contains thujone, a compound that can cause seizures, liver damage, and neurotoxic effects. |
| **149** | Zeolite | Zeolite is promoted as a "heavy metal eliminator" for autistic individuals, but there is no evidence that it works. |
| **150** | Zinc | Zinc is essential for metabolism, but there is no evidence that its supplementation treats autism. |

Source: Own elaboration (2025).

### 4.2. Network

The graphs (Figures 13 and 14) and tables (07 and 08) below present a perspective on the coexistence of users (authors) of publications about autism and autism-related content within conspiracy theory communities across different countries. The objective is to identify sharing patterns and connections between nations based on the repetition of user IDs and the dissemination of similar content. To achieve this, co-authorship and content similarity networks were generated, along with matrices that quantify these interactions.

Figure 13 represents the network of connections between countries based on the coexistence of user IDs that posted about autism within these communities. Each node represents a country, and the node size is proportional to the number of users who published on the topic and also appeared in other countries. The edges, or connections between nodes, indicate how frequently the same users were found posting about autism in multiple countries. The thicker the edge, the higher the number of users shared between nations.

What stands out in this network is that no country remains isolated — connections exist throughout the structure, demonstrating the transnational nature of these conspiratorial narratives. Communities categorized as Transnational, as well as Mexico, Colombia, and Argentina, have the largest nodes and the thickest connections, indicating that they concentrate a significant number of users who post about autism in communities from different countries. The Transnational category node, for instance, is the largest in the network, suggesting that this group includes users who are simultaneously active across multiple nations, serving as a convergence point for the dissemination of these narratives.

**Figure 13.** Network graph of coexisting user IDs between countries (authors).

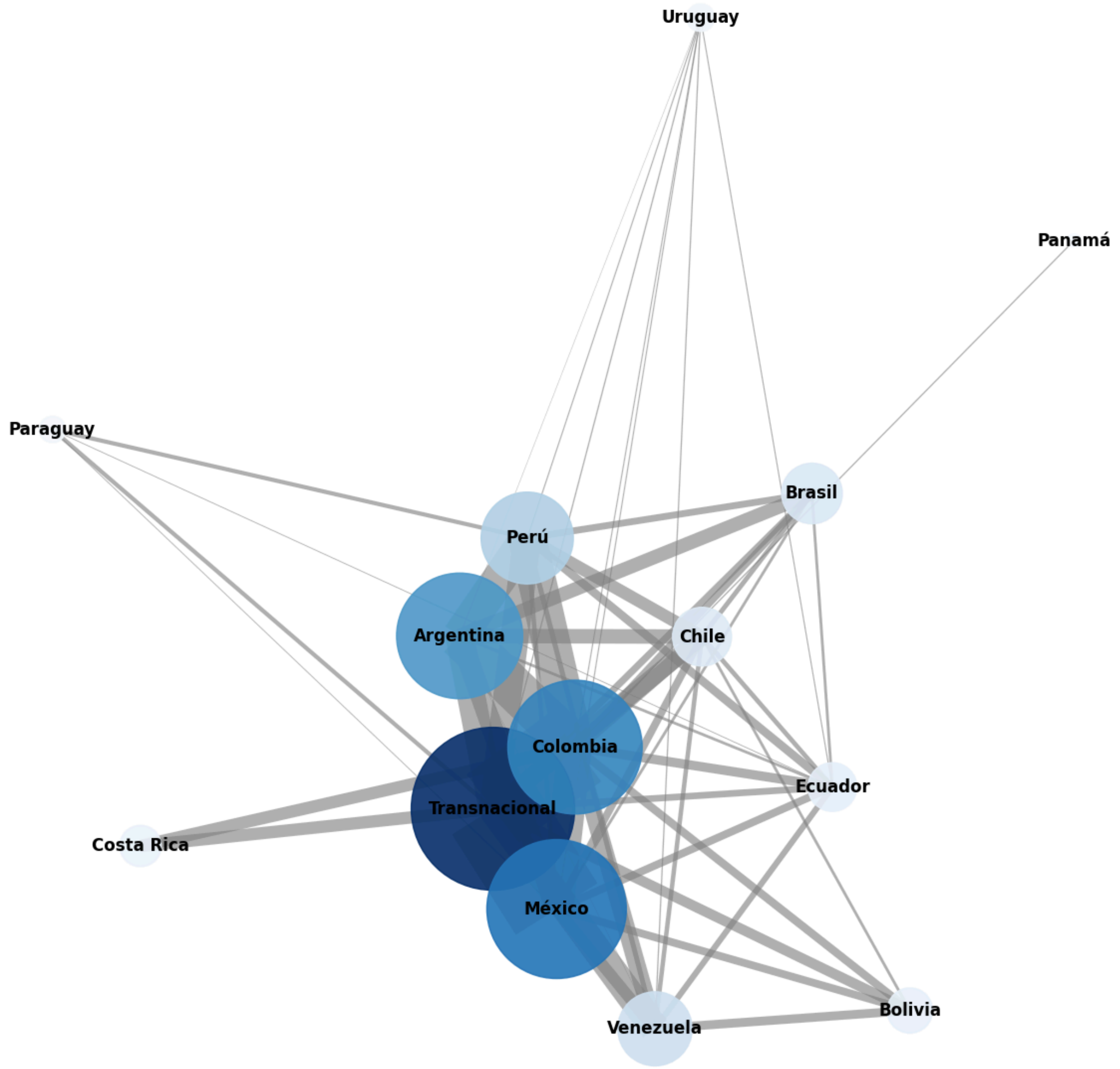

Source: Own elaboration (2025).

Table 07 details these connections, presenting a user coexistence matrix between countries. The main diagonal of the table shows how many times the same users posted about autism within their own country, while the off-diagonal values indicate the number of times these users also published about the topic in another country. This metric provides a better understanding of which nations have the highest volume of users engaged in disseminating these narratives and highlights the main interconnection flows between countries.

The numbers reveal intriguing patterns. Brazilian groups show 10,280 occurrences of user IDs posting about autism within their own groups, making it the largest hub in the network. Transnational communities (4,970), Mexico (1,129), and Colombia (1,490) also have high volumes within their respective communities. These data indicate that some groups act as aggregators of users, maintaining a fixed base while simultaneously exporting these narratives to other countries.

The coexistence of users between different countries follows a similar pattern. For example, Mexico and Transnational communities share 697 user IDs, while Colombia and Transnational communities have 623 IDs in common. Additionally, Mexico and Colombia share 244 IDs, suggesting a significant intersection in the narratives being spread. These numbers confirm that certain communities serve as amplifiers, with users navigating between different digital spaces and replicating the same publications across multiple platforms.

**Table 07.** Coexistence of the same users across countries (authors).

| | **Perú** | **Transnacional** | **México** | **Colombia** | **Panamá** | **Argentina** | **Bolivia** | **Venezuela** | **Chile** | **Ecuador** | **Brasil** | **Costa Rica** | **Uruguay** | **Paraguay** |
|---|---|---|---|---|---|---|---|---|---|---|---|---|---|---|
| **Perú** | 672 | 202 | 32 | 277 | 0 | 245 | 0 | 36 | 89 | 60 | 49 | 0 | 8 | 30 |
| **Transnacional** | 202 | 4,970 | 697 | 623 | 0 | 204 | 74 | 209 | 129 | 48 | 50 | 89 | 9 | 30 |
| **México** | 32 | 697 | 1,129 | 244 | 0 | 152 | 56 | 84 | 55 | 56 | 20 | 0 | 8 | 6 |
| **Colombia** | 277 | 623 | 244 | 1,490 | 9 | 317 | 58 | 126 | 109 | 67 | 50 | 89 | 8 | 0 |
| **Panamá** | 0 | 0 | 0 | 9 | 9 | 0 | 0 | 0 | 0 | 0 | 0 | 0 | 0 | 0 |
| **Argentina** | 245 | 204 | 152 | 317 | 0 | 1,488 | 0 | 3 | 105 | 20 | 100 | 0 | 3 | 0 |
| **Bolivia** | 0 | 74 | 56 | 58 | 0 | 0 | 184 | 64 | 20 | 0 | 0 | 0 | 0 | 0 |
| **Venezuela** | 36 | 209 | 84 | 126 | 0 | 3 | 64 | 261 | 32 | 43 | 0 | 0 | 9 | 0 |
| **Chile** | 89 | 129 | 55 | 109 | 0 | 105 | 20 | 32 | 580 | 33 | 34 | 0 | 0 | 0 |
| **Ecuador** | 60 | 48 | 56 | 67 | 0 | 20 | 0 | 43 | 33 | 119 | 20 | 0 | 8 | 6 |
| **Brasil** | 49 | 50 | 20 | 50 | 0 | 100 | 0 | 0 | 34 | 20 | 10,280 | 0 | 0 | 0 |
| **Costa Rica** | 0 | 89 | 0 | 89 | 0 | 0 | 0 | 0 | 0 | 0 | 0 | 159 | 0 | 0 |
| **Uruguay** | 8 | 9 | 8 | 8 | 0 | 3 | 0 | 9 | 0 | 8 | 0 | 0 | 56 | 0 |
| **Paraguay** | 30 | 30 | 6 | 0 | 0 | 0 | 0 | 0 | 0 | 6 | 0 | 0 | 0 | 73 |

Source: Own elaboration (2025).

Figure 14 provides a complementary perspective by analyzing the circulation of autism-related content within these networks. In this case, each node represents a country, and its size reflects the volume of internationally repeated content. The edges indicate the frequency with which the same content was shared between countries, with varying thickness depending on the intensity of the exchange of posts. Unlike the user analysis, which measures the authority and influence of those who post, this graph highlights the dissemination patterns of narratives, that is, how certain texts, images, and videos circulate transnationally, appearing in different communities with the same content.

Once again, it is evident that no country is isolated, suggesting that the same pieces of disinformation appear simultaneously in different communities around the world. Transnational communities, along with countries such as Colombia and Mexico, emerge as central hubs in content dissemination, with particularly strong connections to Chile, Peru, and Brazil. The Brazilian node, for example, shows a significant concentration of shared content within the country (10,591 posts) but also actively participates in cross-border exchanges of publications with other nations.

**Figure 14.** Network graph of coexisting content IDs across countries.

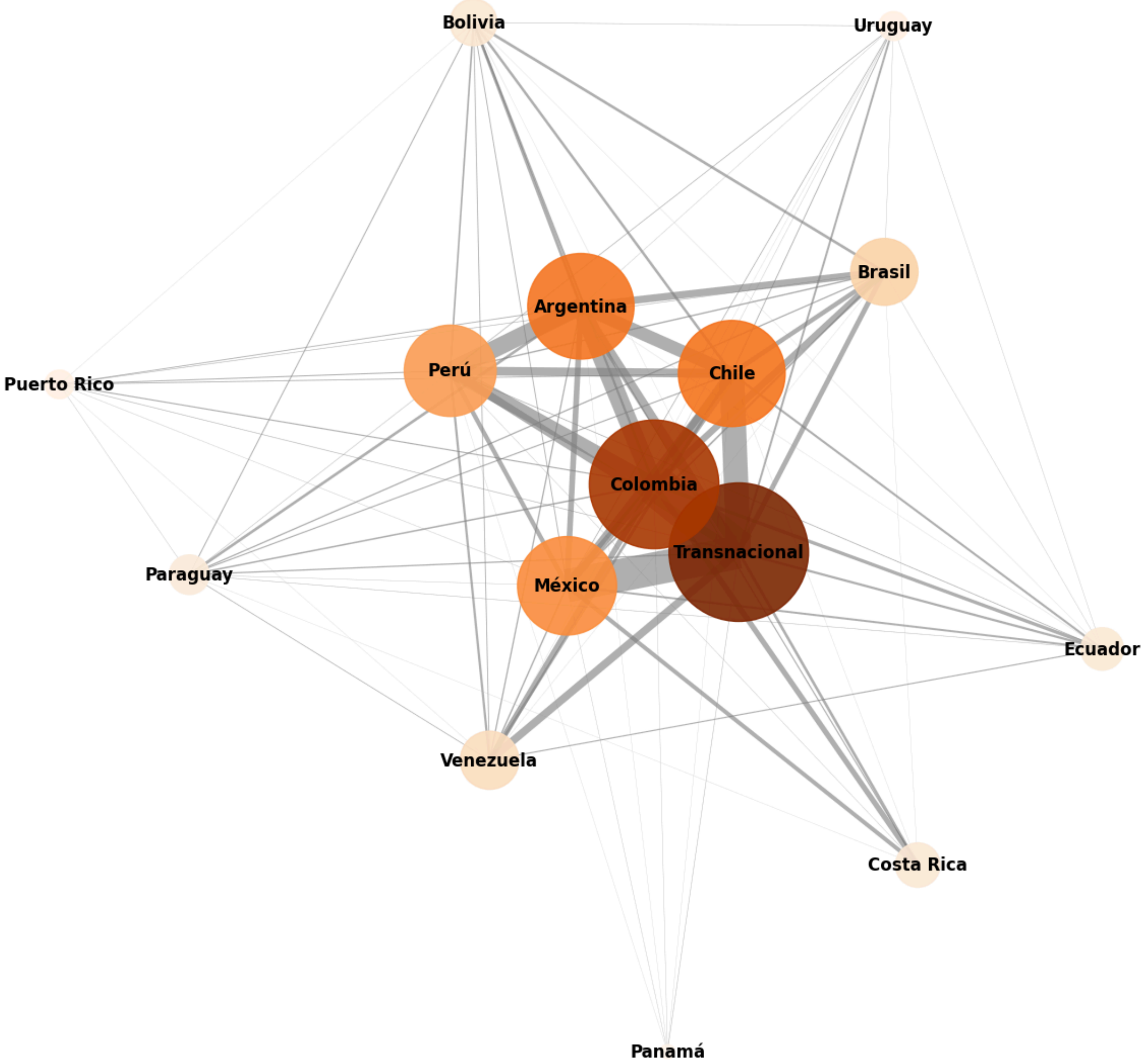

Source: Own elaboration (2025).

Table 08 complements this visualization by quantifying the coexistence of content across countries. Similar to the user matrix, the main diagonal presents the number of repeated posts within the same country, while the values outside the diagonal indicate how many times the same content appeared in different nations.

The data reveal that Brazil and Transnational groups are the main hubs for content dissemination. Brazil records 10,591 internal repetitions of content, while Transnational communities account for 5,039. In terms of international connections, Colombia and Mexico share 567 identical pieces of content, while Argentina and Transnational groups have 64 replicated posts between them. The volume of these connections indicates a structured dynamic of content circulation, where certain narratives are reused and adapted for audiences.

Countries that might be considered secondary in this ecosystem, such as Costa Rica, Paraguay, and Puerto Rico, also appear connected to the network, demonstrating that these narratives are not limited to major population centers or nations within the same geopolitical axis. This suggests the presence of an active network of content translation and adaptation, allowing disinformation to overcome language barriers and reach diverse audiences.

**Table 08.** Coexistence of identical content across countries.

| | Argentina | Transnacional | Colombia | México | Chile | Perú | Brasil | Venezuela | Bolivia | Costa Rica | Ecuador | Paraguay | Uruguay | Puerto Rico | Panamá |
|---|---|---|---|---|---|---|---|---|---|---|---|---|---|---|---|
| **Argentina** | 1.517 | 64 | 130 | 40 | 91 | 122 | 52 | 9 | 20 | 8 | 1 | 19 | 2 | 6 | 1 |
| **Transnacional** | 64 | 5.039 | 291 | 234 | 173 | 55 | 35 | 55 | 1 | 19 | 16 | 9 | 14 | 4 | 3 |
| **Colombia** | 130 | 291 | 1.567 | 50 | 85 | 96 | 47 | 40 | 16 | 38 | 24 | 11 | 2 | 8 | 2 |
| **México** | 40 | 234 | 50 | 1.178 | 54 | 31 | 12 | 10 | 6 | 29 | 14 | 2 | 5 | 2 | 2 |
| **Chile** | 91 | 173 | 85 | 54 | 582 | 62 | 32 | 19 | 18 | 1 | 15 | 8 | 6 | 6 | 1 |
| **Perú** | 122 | 55 | 96 | 31 | 62 | 675 | 10 | 16 | 12 | 2 | 5 | 1 | 3 | 7 | 1 |
| **Brasil** | 52 | 35 | 47 | 12 | 32 | 10 | 10.591 | 1 | 19 | 1 | 2 | 9 | 2 | 4 | 0 |
| **Venezuela** | 9 | 55 | 40 | 10 | 19 | 16 | 1 | 262 | 6 | 0 | 8 | 5 | 1 | 1 | 0 |
| **Bolivia** | 20 | 1 | 16 | 6 | 18 | 12 | 19 | 6 | 184 | 0 | 1 | 7 | 3 | 1 | 0 |
| **Costa Rica** | 8 | 19 | 38 | 29 | 1 | 2 | 1 | 0 | 0 | 160 | 0 | 1 | 0 | 0 | 0 |
| **Ecuador** | 1 | 16 | 24 | 14 | 15 | 5 | 2 | 8 | 1 | 0 | 126 | 3 | 2 | 0 | 0 |
| **Paraguay** | 19 | 9 | 11 | 2 | 8 | 1 | 9 | 5 | 7 | 1 | 3 | 73 | 2 | 2 | 0 |
| **Uruguay** | 2 | 14 | 2 | 5 | 6 | 3 | 2 | 1 | 3 | 0 | 2 | 2 | 56 | 0 | 0 |
| **Puerto Rico** | 6 | 4 | 8 | 2 | 6 | 7 | 4 | 1 | 1 | 0 | 0 | 2 | 0 | 23 | 0 |
| **Panamá** | 1 | 3 | 2 | 2 | 1 | 1 | 0 | 0 | 0 | 0 | 0 | 0 | 0 | 0 | 9 |

Source: Own elaboration (2025).

When analyzing the combined data, it becomes evident that autism-related disinformation is not confined to a single country or language but is instead part of an interconnected transnational ecosystem, where engaged users and replicated content contribute to the consolidation of false narratives. The fact that no country is isolated in these graphs and matrices indicates that these networks are highly organized and efficient in spreading conspiracy theories, leveraging the dynamics of digital platforms to reach diverse audiences. This transnational structure underscores the importance of coordinated containment strategies between countries, aiming to mitigate the impact of such disinformation and protect vulnerable populations from harmful beliefs and practices associated with autism.

### 4.3. Time series

As shown in Figure 15, the volume of autism-related publications within conspiracy communities has experienced exponential growth in recent years. In January 2019, there were only four monthly posts on the topic, a number that increased to 35 posts in January 2020, marking a 775% rise in just one year. With the onset of the COVID-19 pandemic, the growth became even more pronounced: in January 2021, the number of posts reached 260, representing a 635% increase compared to the previous year. This surge reflects the intensification of autism-related narratives during the pandemic, often linked to anti-vaccine discourse and fraudulent alternative treatments.

The upward trend continued in the following years, reaching 438 posts in January 2022 and 365 posts in January 2023, demonstrating that autism disinformation remained relevant even after the pandemic's critical phase. By 2024, the number of posts had risen to 587, and in 2025, it hit a record 611 monthly posts, marking a total increase of over 15,000% (x151) since 2019. These figures highlight how autism has become a central theme within these communities, exploited to fuel disinformation and conspiracy theories.

**Figure 15.** Time series graph of autism-related content (total).

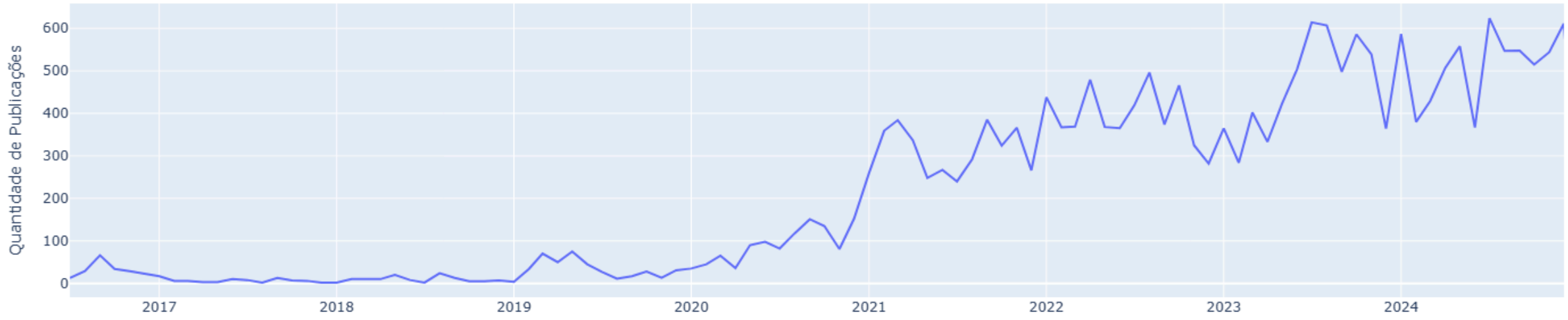

Source: Own elaboration (2025).

With Figure 16, we can observe content segmentation by country. The analysis reveals that Brazil leads in the volume of autism-related posts within these conspiracy communities, with monthly peaks exceeding 300 posts from 2023 onwards. Brazil stands out as the largest disseminator of content related to the topic, both due to its active online population and the historical influence of anti-vaccine movements in the country. Another key finding is the role of Transnational groups, which appear as a major category in autism-related discussions, indicating a high degree of content intersection between different countries, detached from a specific national context.

Countries such as Mexico, Colombia, and Argentina also show significant increases in autism-related posts over the years, although at a smaller scale. From 2021 onwards, there is a steady rise in the volume of publications in these countries, with peaks aligning with strategic moments, such as vaccination campaigns or the release of new scientific research on autism. The presence of smaller countries within the network, such as Costa Rica, Paraguay, and Venezuela, suggests that autism disinformation has managed to penetrate various spheres, going beyond major population centers and reaching audiences previously less exposed to these narratives.

**Figure 16.** Time series graph of autism-related content (by country).

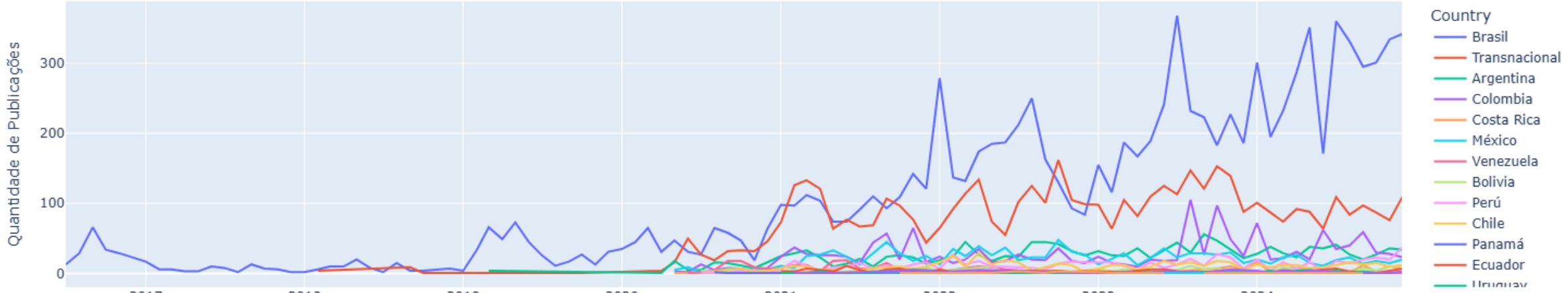

Source: Own elaboration (2025).

Figure 17 further segments these posts by category, helping to identify which specific narratives have driven the rise in autism-related disinformation in recent years. Among the analyzed categories, the ones with the most significant growth include: Off Label and Quackery, Antivax and QAnon. These categories are directly linked to conspiracy narratives that exploit fear and uncertainty among parents of autistic children. In 2019, the Off Label and Quackery category was virtually nonexistent, but by 2021, it exceeded 200 monthly posts, marking a sharp surge in this type of content.

The Antivax category also saw notable growth, with monthly peaks surpassing 150 posts, particularly during periods of high circulation of anti-vaccine campaigns. Despite being widely debunked by the scientific community, the claim that vaccines cause autism remains a core pillar of conspiracy narratives in these communities. Also, the presence of QAnon — a conspiracy theory originally focused on politics and global control — indicates that autism has been integrated into broader narratives about the manipulation of health and science.

Other categories, such as flat Earth theories, climate change denial, and alien-related conspiracies, appear at lower frequencies but demonstrate an intersection between various conspiracy theories. This suggests that autism disinformation is not an isolated phenomenon but rather part of a larger ecosystem of adaptable disinformation that shifts according to global events and crises. The growth of these categories over time reinforces how autism narratives are being opportunistically leveraged to amplify misleading content.

**Figure 17.** Time series graph of autism-related content (by category).

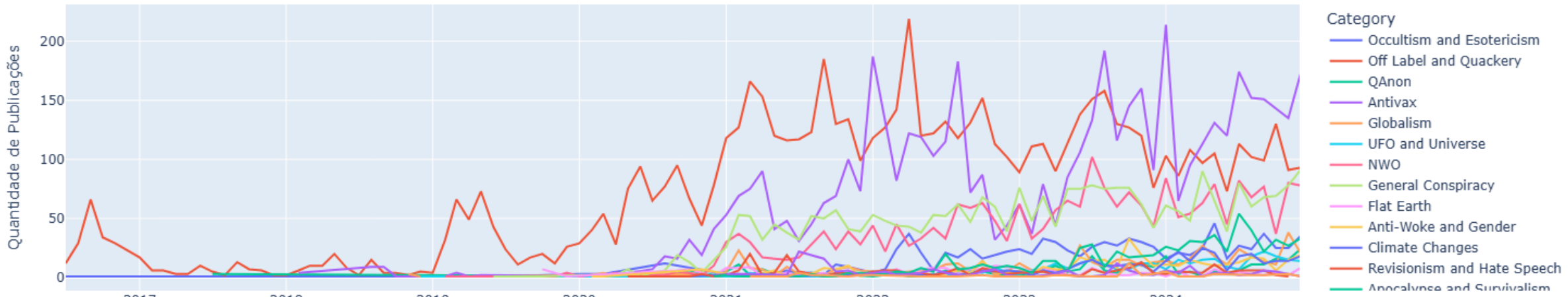

Source: Own elaboration (2025).

The graphs highlight the exponential rise of autism-related disinformation within conspiracy communities, especially after the COVID-19 pandemic. Brazil emerges as the largest hub for disseminating these narratives, with a monthly volume surpassing all other countries, followed by transnational networks and Latin American countries such as Mexico, Colombia, and Argentina. The most widespread narratives are linked to unauthorized treatments, anti-vaccine conspiracies, and globalist agendas, demonstrating how autism has been instrumentalized by various disinformation actors.

### 4.4. Content analysis

Having already analyzed the temporal and geographical progression of autism-related posts, this content analysis aims to complement the findings by exploring the most frequent words in the collected corpus. As shown in the word cloud (Figure 18), the predominance of terms such as “autismo” (26,845 mentions), “vacinas” (7,677), “vacuna” (9,530), “autista” (4,153), and “crianças” (4,532) highlights how discussions are primarily centered on narratives related to the autism spectrum and its connection to immunization. The recurring use of these words in multiple languages suggests that these narratives circulate widely across multilingual contexts, reinforcing their transnational nature. A noteworthy aspect is the high frequency of links, indicated by terms like “https” (23,763 mentions) and “www” (6,878). This suggests a dissemination strategy heavily reliant on external sources, many of which are likely associated with pseudoscience, conspiracy forums, and alternative health websites.

Another key finding is the correlation between autism and chemical substances, particularly the prominence of terms such as “alumínio” (3,300), “cloro” (2,068), “mercúrio” (1,266), and “timerosal” (1,133). These substances are frequently cited in conspiracy theories about autism, particularly in claims that attribute the condition to alleged toxic agents in vaccines or the environment — a claim thoroughly debunked by scientific studies. The term “dióxido” (1,626), often linked to chlorine dioxide (MMS, or Miracle Mineral Solution), further reinforces the presence of discourses promoting fraudulent and dangerous treatments that falsely claim to "cure" autism.

The analysis also reveals how autism is embedded within broader conspiracy narratives related to population control and distrust in medical institutions. Terms such as “CDC” (2,551), “gov” (1,780), “NIH” (1,427), and “comusav” (2,190) suggest a strong anti-institutional discourse, frequently questioning public health agencies and spreading disinformation about healthcare policies. Additionally, the appearance of “Bill Gates” (while not among the most frequent terms, it is visible in the word cloud) suggests intersections with conspiracy theories that attribute the rise in autism cases to global vaccination agendas supposedly orchestrated by powerful elites. This interconnection with broader conspiracy narratives underscores how autism is being strategically instrumentalized within the larger disinformation ecosystem.

Finally, the lexical analysis indicates the use of alarmist rhetoric, reinforced by terms such as “problemas” (3,999), “risco” (while not in the top list, it is highlighted in the word cloud), “causar” (2,191), and “doenças” (2,743). The prevalence of fear-inducing language suggests an engagement strategy based on exploiting negative emotions, which can make new users more susceptible to conspiracy narratives. Additionally, terms such as “cura” (1,447), “tratamento”, and “verdade” (1,439) indicate that these groups also position themselves as alternatives to conventional medicine, promoting pseudoscientific solutions for autism.

In combination, these patterns reinforce the urgent need for continuous monitoring and coordinated responses to counteract the spread of autism-related disinformation. The misuse of medical and scientific terminology, combined with fear-based engagement tactics, not only distorts public perception of autism but can also lead to harmful decisions regarding treatment and inclusion policies. Addressing this issue requires interdisciplinary efforts across fact-checking, policy-making, and digital literacy initiatives to mitigate the impact of these misleading narratives.

**Figure 18.** Word cloud of the most frequent terms in autism-related posts.

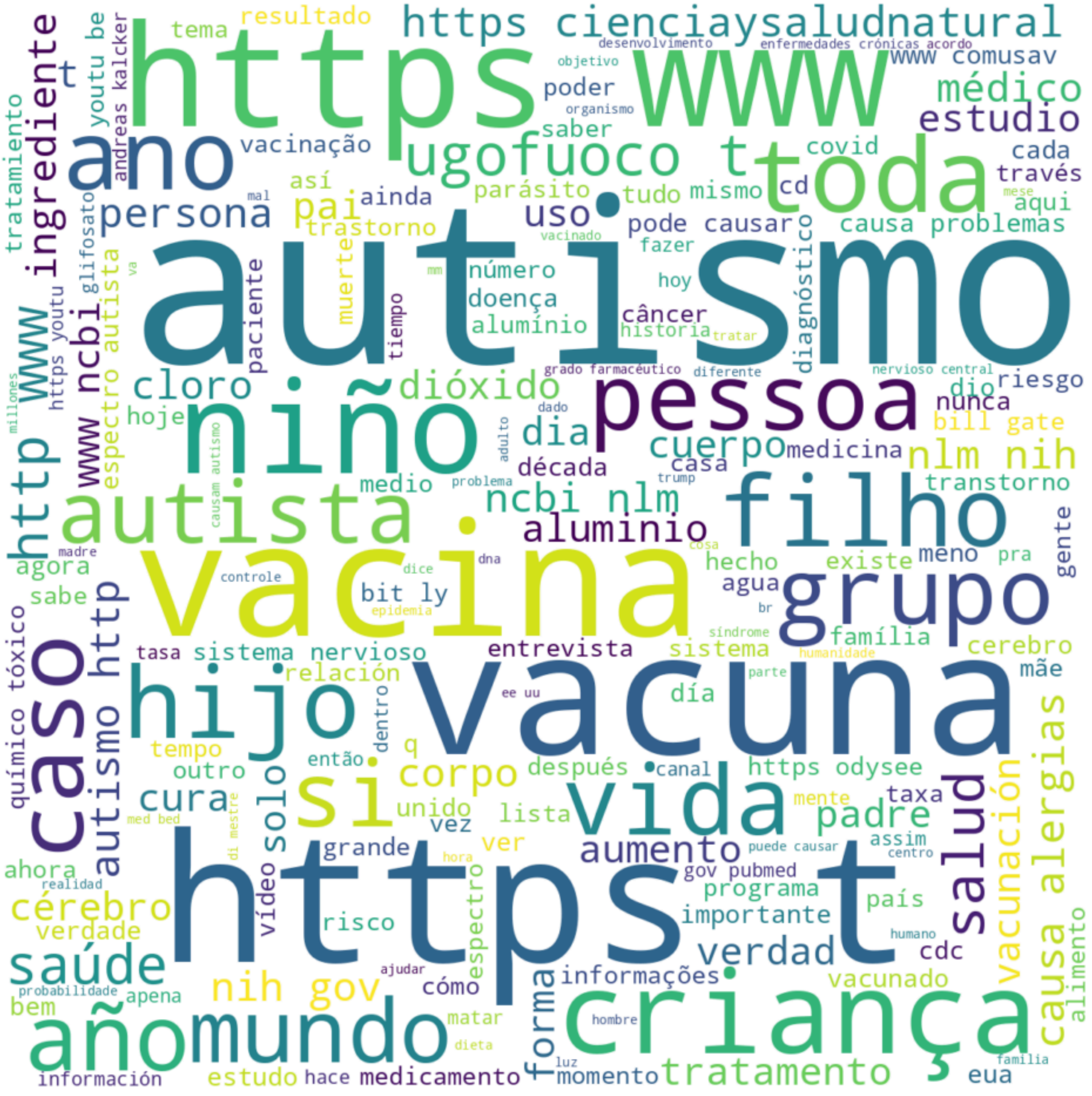

Source: Own elaboration (2025).

## 5. Reflections and future works

To address the research question — ***"How do conspiracy theory communities across Latin America and the Caribbean structure, articulate, and sustain the spread of*** ***disinformation about autism?"*** — this study adopted a methodological approach consistent with a broader series of investigations into conspiracy theories on Telegram. Following an extensive investigation, a total of **1,649 conspiracy theory communities** across Latin America and the Caribbean were identified, encompassing a wide array of disinformation themes. Within these communities, **58,637,137 contents** were published between **December 2015** (earliest records) and **January 2025** (study period), with a total of **5,345,332 users**.

To systematically analyze the dynamics of these communities, four main methodological approaches were applied: **(i) Descriptive Analysis of Autism-Related Claims** – A structured classification was conducted on 150 alleged causes and 150 alleged cures for autism, categorizing them by thematic origin and type of proposed treatment to identify narrative patterns; **(ii) Network Analysis** – A proprietary algorithm was developed to map interconnections between communities by tracking Telegram invitations (*t.me/* links) shared among groups and channels, revealing how communities reinforce internal narratives and cross-promote different conspiracy theories; **(iii) Time Series Analysis** – Using the *Pandas* library (McKinney, 2010) for data structuring and the *Plotly* library (Plotly Technologies Inc., 2015) for visualization, temporal patterns were analyzed to examine content production trends and engagement fluctuations over time; **(iv) Content Analysis** – Text analysis techniques were applied to examine word frequencies and thematic variations, offering insights into the persistence and evolution of autism-related narratives.

The following sections present the key findings of this study, offering an assessment of the structural characteristics and behavioral patterns of conspiracy theory communities in Latin America and the Caribbean, followed by recommendations for future research.

### 5.1. Main reflections

The market of autism influencers and coaches who promote a pathologizing view of the condition is closely linked to the rise of conspiracy communities operating in the lucrative "autism cure" industry. These influencers often portray autism as a tragic and debilitating condition, reinforcing stigma and prejudice while creating fertile ground for the spread of miraculous and pseudoscientific solutions. By reinforcing the idea that autism is a disease to be fought or cured, these influencers open space for dangerous practices such as chlorine dioxide (MMS) consumption, fecal transplants, and even spiritual approaches promising to "free" individuals from autism through exorcism. These so-called treatments lack scientific backing and pose significant risks to the health and well-being of autistic individuals.

The pathologizing discourse also fuels disinformation, contributing to the formation of conspiracy communities that distrust evidence-based medicine and reject approaches that promote neurodiversity acceptance. These groups frequently share alarmist narratives about vaccines, environmental toxins, and alleged pharmaceutical industry conspiracies, amplifying conspiracy theories that gain traction on social media. Malicious influencers and users who profit from the desperation and disinformation of families with autistic individuals often position themselves as saviors or self-proclaimed experts, selling courses, mentorships, and alternative therapies that promise a cure. By creating a sense of urgency and fear, they attract a vulnerable audience willing to pay for any promise of normalcy.

By pathologizing autism and presenting easy solutions aimed at its "cure" or "eradication", these discourses perpetuate the idea that neurodivergent traits are defects to be corrected, echoing the eugenicist logic of exclusion and conformity to a normative ideal. This dehumanizes autistic people by treating them as biological errors, legitimizing invasive and potentially harmful interventions. This scenario not only reinforces prejudice against autistic individuals but also diverts attention from public policies and human rights-based approaches that prioritize social inclusion, respect for neurodiversity, and access to appropriate support. Therefore, the pathologization of autism not only fuels profitable markets and pseudoscientific conspiracies but also sustains a eugenicist social project aimed at normalizing bodies and minds, erasing neurocognitive differences in favor of a capacitive social uniformity.

An important point worth emphasizing is how the market for miracle cures and alternative remedies operates without borders, functioning in a coordinated manner to promote ready-to-deliver networks of chemical products, with suppliers distributed across the entire continent. Some screenshots from these communities help illustrate this phenomenon (Figure 19), as they not only provide direct contact with sellers but even feature a child demonstrating how to prepare "immunotherapy eye drops" as part of one of the so-called "autism cures." In the same video where a child is shown preparing the chemical, there is a link for purchasing the preparation kit.

[ english version / português abaixo ]

**Figure 19.** Examples of the commercialization of miracle cures for autism.

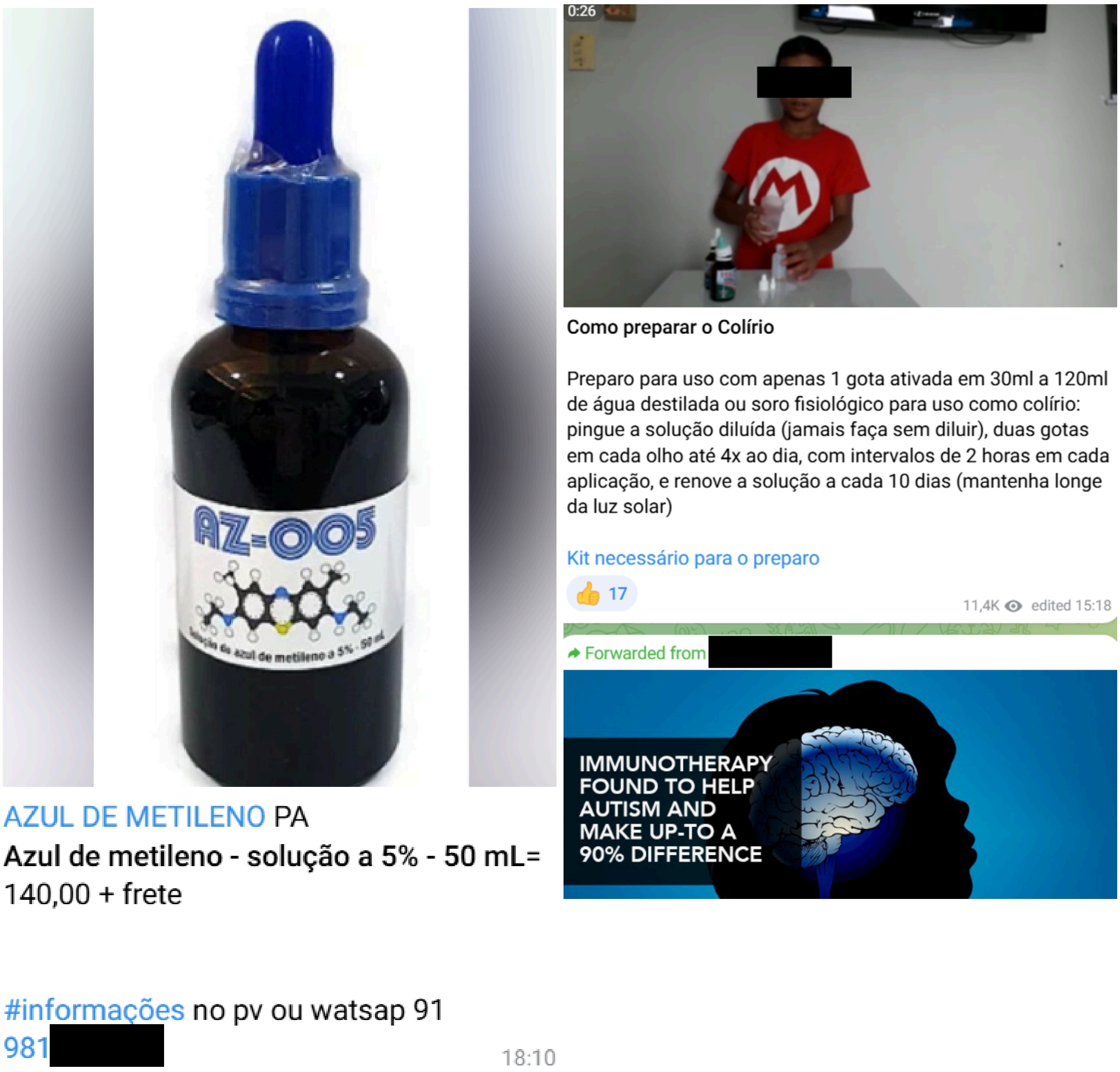

Source: Own elaboration (2025).

In this context, the main findings are:

**Approximately 100 million views and 4 million users:** Conspiracy theories about autism reached at least 4,186,031 users in Latin America and the Caribbean between 2015 and 2025, totaling 99,318,993 views, 107,880 reactions in 47,261 mapped and categorized posts across the continent.

**Brazil ranks first in the continent:** Brazilian conspiracy communities account for 46% of autism-related content in the region, totaling 22,007 posts, reaching up to 1,726,364 users and 13,944,477 views. Following Brazil, Argentina, Mexico, Venezuela, and Colombia also stand out as the countries producing the most conspiracy content on autism.

**150 false causes of autism, from parasites to Doritos:** Among the mapped explanations for the "cause" of autism are theories ranging from serotonin deficiency and aluminum exposure to claims linking it to eating Doritos, Earth's magnetic field inversion, and chemtrails. Other theories resort to moral panic and scientific denialism, blaming autism on 5G, Wi-Fi, microwaves, and even vaccines.

**150 false autism cures, from chlorine dioxide to Tesla electroshock:** The false promise of an autism cure has become a lucrative business, driven by disinformation and opportunism. Among the 150 “cures” identified, dangerous practices stand out, such as consuming chlorine dioxide (CDS), known as "MMS", a toxic substance promoted as a miracle solution. Additionally, absurd methods like ozone therapy, Tesla electroshock therapy, and even ingesting colloidal silver and methylene blue are sold as supposedly effective treatments. Many of these products and practices are openly marketed by groups that exploit desperate families, profiting from monetized falsehoods while endangering lives.

**15,000% increase in autism-related conspiracies post-pandemic:** The COVID-19 pandemic was the gateway to the explosion of autism-related disinformation across the continent. Between 2019 and 2024 (five years), the volume of misleading content grew by over 15,000% (x151), with a striking 635% (x7,35) increase during the pandemic period (2020-2021) alone. This accelerated growth demonstrates how the health crisis provided an opening for conspiracy narratives that continued expanding in the following years.

### 5.2. Future works

Despite the broad scope of this research, some limitations deserve attention and may be addressed in future studies:

**Expansion to other digital platforms:** This study focused exclusively on Telegram, but future investigations could analyze the spread of autism-related disinformation on other social networks such as X (formerly Twitter), Facebook, Instagram, YouTube, and TikTok to assess differences in strategies and the reach of these narratives.

**Impact of disinformation on family decision-making:** While this study mapped the dissemination of disinformation, there is still a gap in understanding how these beliefs influence medical and therapeutic decisions made by parents and caregivers of autistic children. Future research could explore the direct impacts of these narratives on access to evidence-based treatments.

**Monetization strategies and profits from disinformation:** The commercialization of false autism cures is a critical aspect of disinformation. Future studies could investigate the economics of disinformation, identifying who benefits financially from selling fraudulent products and how these practices are structured.

**Strategies to combat disinformation:** Understanding how regulatory agencies, healthcare professionals, and the scientific community respond to these narratives could help develop more effective disinformation counteraction campaigns. Studies could explore communication approaches and the effectiveness of interventions in online spaces.

**Regional variations in the adoption of conspiracy narratives:** This research revealed that different Latin American and Caribbean countries exhibit unique dynamics in the dissemination of autism disinformation, even while remaining interconnected. Future studies could delve deeper into regional differences, exploring sociopolitical and cultural factors that either encourage or inhibit these beliefs in different national contexts.

Finally, this work seeks to contribute objective reflections to the Family Health Strategy (ESF), the National Immunization Program (PNI), and the National Humanization Policy (PNH) of the Ministry of Health (MS). In this regard, it presents reflections aimed at the Health with Science Program, the Committee for Combating Disinformation on the National Immunization Program and Public Health Policies of the Federal Government, the Brazilian Institute of Information in Science and Technology (IBICT), and the Parliamentary Front for Vaccine Advocacy in the Chamber of Deputies.

## 7. Authors biographies

**Ergon Cugler de Moraes Silva** is autistic. Holds a Master's degree in Public Administration and Government (FGV), a Postgraduate MBA in Data Science & Analytics (USP), and a Bachelor's degree in Public Policy Management (USP). He collaborates with the Interdisciplinary Observatory of Public Policies (OIPP USP), the Study Group on Technology and Innovations in Public Management (GETIP USP), the Monitor of Political Debate in the Digital Environment (Monitor USP), and the Working Group on Strategy, Data, and Sovereignty of the Study and Research Group on International Security at the Institute of International Relations of the University of Brasília (GEPSI UnB). He was previously a researcher at the Brazilian Institute of Information in Science and Technology (IBICT), where he worked on strategies against disinformation for the Federal Government. He is now affiliated with Getulio Vargas Foundation (FGV) and National Council for Scientific and Technological Development (CNPq) through the Laboratory for Studies on Information Disorder and Public Policies (DesinfoPop), with the Center for Studies in Public Administration and Government (CEAPG/FGV/EAESP). He is also the Director of Public Policy, Research, and Advocacy at the National Association for the Inclusion of Autistic People (Autistas Brasil). In addition, he is enrolled in the postgraduate program in Data Science for Social and Business Analytics at the University of Barcelona. Website: https://ergoncugler.com/. Contact: contato@ergoncugler.com.

**Arthur Ataide Ferreira Garcia** is autistic. He began his activism in defense of autistic people's rights at the age of 11, participating in and mobilizing actions, lectures, and events to promote inclusion across all sectors of society. He is a speaker and vice president of the National Association for the Inclusion of Autistic People (Autistas Brasil), an organization he co-founded. He is currently pursuing a degree in Medicine at the Universidade Metropolitana de Santos (UNIMES) and is the only young Brazilian to have spoken at the Stanford Neurodiversity Summit. He also authored and is the creator of state law 17.759/2023, the “Individualized Assessment Protocol.” Additionally, he collaborates with various initiatives of the Ministry of Education and the Ministry of Health. Website: www.autistas.org.br. Contact: arthur.afg.2003@gmail.com.

**Guilherme de Almeida** is autistic. He is a PhD candidate (CAPES Excellence Scholarship recipient) and holds a Master’s degree in Education from the School of Education at the University of Campinas (UNICAMP). He earned his Bachelor's degree in Law from the Pontifical Catholic University of Paraná (PUC/PR). Currently, he serves as the President of the National Association for the Inclusion of Autistic People (Autistas Brasil). He is the only Brazilian researcher who is a member of the Stanford Neurodiversity Project, where he works on the Committees for Inclusion in Higher Education and Workplace Inclusion. He is also a member of the PAIDEIA Research and Study Group at the School of Education at UNICAMP and the Judicial Committee on the Rights of Persons with Disabilities of the National Justice Council (CNJ). Website: www.autistas.org.br. Contact: g229669@dac.unicamp.br.

**Julie Ricard** is a researcher and Ph.D. candidate in Public Administration and Government at Fundação Getulio Vargas (FGV). She holds master's degrees in International Relations from Sciences Po (France) and in Gender Studies from Université Paris 7. Her research intersects disinformation, public policy, and gender, with a focus on public health, democracy, and technology. Currently, she is a fellow from the Brazilian National Council for Scientific and Technological Development (CNPq), working as a researcher at the Center for Studies in Public Administration and Government (CEAPG/FGV/EAESP) within the Laboratory for Studies on Information Disorder and Public Policy (DesinfoPop). She previously served as a consultant for UNESCO, supporting the Social Communication Secretariat of the Presidency of Brazil in efforts to counter public health disinformation, and as Director of the Technology and Democracy program at Data-Pop Alliance. Julie is also the founder of Eureka, a platform promoting spaces for political and cultural education through book and film clubs in partnership with civil society organizations and universities. Website: https://www.eureka.club. Contact: juliec.ricard@gmail.com.

[ versão português / english above ]

# **Desinformação sobre autismo na América Latina e no Caribe:** Mapeando 150 falsas causas e 150 falsas curas do TEA em comunidades de teorias da conspiração no Telegram

*Ergon Cugler de Moraes Silva* [1]

Fundação Getulio Vargas
São Paulo, São Paulo, Brasil

contato@ergoncugler.com
www.ergoncugler.com

*Arthur Ataide Ferreira Garcia* [2]

Univ. Metropolitana de Santos
Santos, São Paulo, Brasil

arthur.afg.2003@gmail.com
www.autistas.org.br

*Guilherme de Almeida* [3]

Univ. Estadual de Campinas
Campinas, São Paulo, Brasil

g229669@dac.unicamp.br
www.autistas.org.br

*Julie Ricard* [4]

Fundação Getulio Vargas
São Paulo, São Paulo, Brasil

juliec.ricard@gmail.com
www.eureka.club

**Resumo**

Como as comunidades de teorias da conspiração na América Latina e no Caribe estruturam, articulam e sustentam a disseminação da desinformação sobre o autismo? Para responder à questão, esta pesquisa investiga a estruturação, articulação e sustentação da disseminação da desinformação sobre o autismo em comunidades de teorias da conspiração na América Latina e no Caribe. A partir da análise das publicações de 1.659 comunidades no Telegram durante 10 anos (2015-2025) e observados mais de 58 milhões de conteúdos compartilhados por aproximadamente 5,3 milhões de usuários, o estudo examina como narrativas falsas sobre o autismo são promovidas, incluindo alegações infundadas sobre suas causas e promessas de curas milagrosas. A metodologia adotada combina análise de redes, séries temporais, agrupamento temático e análise de conteúdo, permitindo identificar padrões de disseminação, principais influenciadores e interconexões com outras teorias da conspiração. Entre os principais achados, destaca-se que as comunidades brasileiras lideram a produção e circulação dessas narrativas na região, sendo responsáveis por 46% dos conteúdos analisados. Além disso, observou-se um crescimento exponencial de 15.000% (x151) no volume de desinformação sobre autismo desde a Pandemia da COVID-19 na América Latina e no Caribe, evidenciando a relação entre crises sanitárias e o aumento da crença em teorias conspiratórias. A pesquisa revela ainda que falsas curas, como o uso de dióxido de cloro (CDS), ozonioterapia e dietas extremas, são amplamente promovidas nessas comunidades e comercializadas, muitas vezes explorando o desespero de famílias em troca de dinheiro. Ao responder à pergunta de pesquisa, este estudo busca contribuir para o entendimento do ecossistema de desinformação e sugere provocações ao enfrentamento a essas narrativas nocivas.

**Principais descobertas:**

- **Cerca de 100 milhões de visualizações e 4 milhões de usuários:** Teorias da conspiração envolvendo autismo alcançaram pelo menos 4.186.031 usuários na América Latina e no Caribe entre 2015 e 2025, totalizando 99.318.993 visualizações, 107.880 reações em 47.261 publicações mapeadas e categorizadas no continente.
- **Brasil em primeiro lugar no Continente:** As comunidades brasileiras de teorias da conspiração somam 46% dos conteúdos sobre autismo no continente, totalizando 22.007 publicações, alcançando até 1.726.364 usuários e 13.944.477 visualizações. Seguido do Brasil, Argentina, México, Venezuela e Colômbia também ocupam posição de destaque dentre os países que mais produzem conteúdos conspiracionistas sobre autismo.
- **150 falsas causas do autismo, de parasitas a Doritos:** Entre as versões mapeadas, encontram-se desde explicações para "causa" do autismo como deficiência de serotonina e exposição a alumínio, até alegações como consumo de Doritos, inversão do campo magnético da Terra e influência de *chemtrails*. Outras teorias recorrem ao pânico moral e ao negacionismo científico, atribuindo o autismo ao 5G, Wi-Fi, microondas e até às vacinas.
- **150 falsas curas do autismo, de dióxido de cloro a eletrochoque de Tesla:** A falsa promessa de cura para o autismo se tornou um negócio lucrativo, impulsionado por desinformação e oportunismo. Entre as 150 "curas" identificadas, destacam-se práticas perigosas, como o consumo de dióxido de cloro (CDS), conhecido como "MMS", uma substância tóxica promovida como solução milagrosa. Além disso, métodos absurdos como ozonioterapia, terapia de eletrochoque de Tesla e até a ingestão de prata coloidal e azul de metileno são vendidos como tratamentos supostamente eficazes. Muitos desses produtos e práticas são comercializados abertamente por grupos que exploram o desespero de familiares, lucrando com a monetização da mentira e colocando vidas em risco.
- **Crescimento de 15.000% no conspiracionismo sobre autismo pós-Pandemia:** A Pandemia da COVID-19 foi a porta de entrada para a explosão da desinformação sobre autismo no Continente. Entre 2019 e 2024 (cinco anos), o volume de conteúdos enganosos cresceu mais de 15.000% (x151), com um aumento expressivo de 635% (x7,35) apenas durante o período pandêmico (2020-2021). Esse crescimento acelerado demonstra como a crise sanitária abriu espaço para narrativas conspiratórias que continuaram se expandindo nos anos seguintes.

## 1. Introdução

Nos últimos anos, as comunidades de teorias da conspiração no Telegram tornaram-se um vetor significativo para a disseminação da desinformação em toda a América Latina e no Caribe. Com base em pesquisas anteriores focadas exclusivamente no Brasil, este estudo amplia o escopo da análise para uma perspectiva regional mais ampla, cobrindo múltiplos países e variações linguísticas. Ao navegar sistematicamente por **1.659 comunidades de teorias da conspiração** e analisar **58.521.152 conteúdos** compartilhados por **5.310.728 usuários**, esta pesquisa busca mapear, caracterizar e compreender a estrutura e a dinâmica das redes de desinformação relacionadas ao autismo.

Diferentemente de estudos anteriores, que se concentraram na hesitação vacinal e em *off-label*, esta investigação examina especificamente a disseminação de mentiras sobre o autismo. Em diversos ambientes digitais, as narrativas conspiratórias sobre o autismo variam desde alegações infundadas sobre suas causas — como supostas ligações com vacinas, metais pesados ou fatores dietéticos — até falsas "curas" ou "tratamentos" enganosos que promovem terapias prejudiciais, como quelação, câmaras hiperbáricas e o consumo de dióxido de cloro (conhecido como *"MMS"* ou *"CDS"*). Ao analisar a evolução, as interconexões e as sobreposições ideológicas dessas narrativas, este estudo busca fornecer uma visão abrangente sobre como a desinformação relacionada ao autismo se espalha, quem são seus principais promotores e quais estratégias utilizam para ganhar credibilidade entre diferentes públicos.

Para isso, aplicamos uma metodologia sistemática consistente com uma série de estudos, garantindo a comparabilidade e a reprodutibilidade dos achados. Essa abordagem envolve análise de redes, avaliação de séries temporais, agrupamento temático e classificação de conteúdo, permitindo determinar como as comunidades conspiratórias interagem e reforçam a desinformação sobre o autismo. Além disso, o estudo destaca especificidades regionais, considerando nuances sociopolíticas e culturais que moldam a adoção de teorias da conspiração em diferentes contextos latino-americanos e caribenhos.

O conjunto de dados inclui grupos e usuários de diversos países, como **Brasil, Uruguai, Colômbia, Peru, México, Venezuela, Equador, Paraguai, Argentina, Chile, Bolívia, Costa Rica, Honduras, Panamá, República Dominicana, Porto Rico, Cuba, Guatemala e Jamaica**. Enquanto alguns grupos operam dentro de limites nacionais, outros funcionam como núcleos transnacionais, facilitando a disseminação rápida da desinformação por fronteiras linguísticas e geopolíticas. Essa dimensão transnacional ressalta a necessidade de uma análise comparativa e de uma compreensão mais profunda sobre como as infraestruturas digitais de desinformação transcendem os limites dos países individuais.

Um aspecto crítico deste estudo é seu compromisso ético com a proteção de dados e a privacidade dos usuários, em conformidade com legislações regionais, como a *Lei Geral de Proteção de Dados Pessoais* (LGPD – Lei nº 13.709/2018) no Brasil e marcos regulatórios equivalentes em outros países da América Latina e do Caribe. O estudo não busca expor indivíduos, mas analisar padrões, redes e estratégias utilizadas para propagar a desinformação.

Diante das amplas consequências da desinformação sobre o autismo — incluindo seus impactos na saúde pública, na tomada de decisão de pais e responsáveis e na vida de pessoas neurodivergentes — este estudo levanta uma questão central: **como as comunidades de teorias da conspiração na América Latina e no Caribe estruturam, articulam e sustentam a disseminação da desinformação sobre o autismo?** Ao abordar essa questão, a pesquisa contribui para o combate à desinformação e auxilia na formulação de políticas e iniciativas de alfabetização digital para mitigar os danos causados por essas narrativas.

## 2. Revisão da literatura e justificativa

O pensamento conspiratório tornou-se uma força disseminada no discurso público contemporâneo, moldando atitudes não apenas em relação à política e às instituições, mas também à ciência e à saúde. Pesquisadores mostraram que esse tipo de pensamento é frequentemente alimentado por baixa confiança institucional, percepção de perda de controle e um ambiente midiático polarizado. Em particular, Enders (2019) e Albertson & Guiler (2020) identificaram que a crença em teorias da conspiração se correlaciona fortemente com partidarismo, extremismo ideológico e desconfiança dos processos democráticos. A Pandemia de COVID-19 acelerou essa tendência, prosperando a partir do pensamento conspiratório e alimentando a desconfiança em relação aos processos e instituições científicas.

Estudos na literatura sobre teorias da conspiração enfatizam como essas crenças podem fomentar atitudes autoritárias, desconfiança eleitoral e rejeição ao pluralismo (Stoeckel, 2023; Jiang, 2021). Por exemplo, Pickel et al. descobriram que as conspirações relacionadas à COVID estavam associadas ao aumento do apoio a formas autoritárias de governo, enquanto Czech (2022) examinou como o nacionalismo católico polonês integra a identidade religiosa com lógicas conspiratórias que se opõem às normas democráticas seculares. De maneira semelhante, Yendell & Herbert (2022) demonstram que o fundamentalismo religioso é frequentemente um preditor do pensamento conspiratório, especialmente quando enquadrado como uma guerra espiritual ou uma luta cósmica entre o bem e o mal. Essas dimensões religiosas e políticas são especialmente relevantes na América Latina e entre as diásporas latinas, onde Cortina & Rottinghaus (2020) encontraram altos níveis de pensamento conspiratório ligados à marginalização institucional, baixa confiança cívica e discurso político etnicizado.

Crucialmente, o pensamento conspiratório frequentemente oferece terreno fértil para a adoção de crenças pseudocientíficas sobre saúde. Esses dois sistemas epistêmicos estão profundamente interconectados: ambos desafiam a autoridade dominante, baseiam-se em raciocínio anedótico ou intuitivo e prosperam em ambientes digitais. Essa relação é evidente no contexto de teorias conspiratórias sobre saúde, que frequentemente se nutrem da desconfiança em processos e instituições científicas. Esse tipo de pensamento pode levar à aceitação de alegações pseudocientíficas, como se viu durante a Pandemia de COVID-19, quando desinformação e teorias da conspiração sobre o vírus e as vacinas se proliferaram.

Alegações pseudocientíficas são frequentemente apresentadas como narrativas envolventes, oferecendo histórias emocionalmente satisfatórias de curas milagrosas, verdades ocultas ou encobrimentos por elites. Subjacente a essas crenças, geralmente há uma percepção de relações de causa e efeito que parecem intuitivamente verdadeiras, mas carecem de base empírica. Chow et al. (2021) descobriram que indivíduos frequentemente endossam mitos sobre saúde com base na contingência percebida — a crença de que uma ação causa determinado resultado de saúde, mesmo quando não há evidência ou esta é contraditória. Essas percepções não são puramente cognitivas, mas frequentemente estão ligadas ao pensamento mágico ou paranormal, revelando um desejo mais profundo de controle, significado ou esperança em situações de incerteza, como no cuidado de uma criança autista.

No caso da desinformação relacionada ao autismo, esses sistemas convergem em narrativas poderosas que oferecem falsas causas e curas, às custas da evidência científica e da saúde pública — como será evidenciado adiante. Tais crenças podem afetar perigosamente o comportamento relacionado à saúde e a saúde pública em geral. Como se viu no caso das vacinas, crenças conspiratórias moldaram comportamentos de saúde durante a Pandemia (Kowalska-Duplaga & Duplaga, 2023) e reduziram a adesão a medidas preventivas — como a vacinação — (Prooijen, 2022).

## 3. Metodologia

A metodologia deste estudo está organizada em três subseções, sendo: **2.1. Extração de dados**, que descreve o processo e as ferramentas utilizadas para coletar as informações das comunidades no Telegram; **2.2. Tratamento de dados**, onde são abordados os critérios e métodos aplicados para classificar e anonimizar os dados coletados; e **2.3. Abordagens para análise de dados**, que detalha as técnicas utilizadas para investigar as conexões, séries temporais, conteúdos e sobreposições temáticas das comunidades de teorias da conspiração.

### 3.1. Extração de dados

Este projeto teve início em fevereiro de 2023 com a publicação da primeira versão do TelegramScrap (Silva, 2023), uma ferramenta proprietária, gratuita e de código aberto que utiliza a Application Programming Interface (API) do Telegram por meio da biblioteca Telethon, organizando ciclos de extração de dados de grupos e canais abertos na plataforma. Ao longo dos meses, o banco de dados foi expandido e aprimorado com a aplicação de quatro abordagens para a identificação de comunidades de teorias da conspiração:

**(i) Uso de palavras chave:** No início do projeto, foi elaborada uma lista de palavras-chave para facilitar a identificação direta de grupos e canais de teorias da conspiração no Telegram em toda a América Latina e no Caribe. Essas palavras-chave incluíam apocalipse, sobrevivencialismo, mudanças climáticas, terra plana, teoria da conspiração, globalismo, nova ordem mundial, ocultismo, esoterismo, curas alternativas, QAnon, reptilianos, revisionismo, alienígenas, entre outras. A busca foi conduzida tanto em português quanto em espanhol, garantindo cobertura regional. Inicialmente, essa abordagem possibilitou a identificação de comunidades cujos títulos e/ou descrições faziam referência explícita a teorias da conspiração. No entanto, com o tempo, tornou-se evidente que muitas comunidades utilizavam variações ortográficas, caracteres especiais ou linguagem codificada para evitar a detecção. Para contornar esse desafio, a lista de palavras-chave foi refinada iterativamente com base nos padrões observados em diferentes países, expandindo o conjunto de dados além dos grupos brasileiros para incluir comunidades do Uruguai, Colômbia, Peru, México, Venezuela, Equador, Paraguai, Argentina, Chile, Bolívia, Costa Rica, Honduras, Panamá, República Dominicana, Porto Rico, Cuba, Guatemala e Jamaica.

**(ii) Mecanismo de recomendação de canais do Telegram:** Um dos principais achados durante a investigação foi o uso do sistema de recomendação de canais do Telegram (mas não de grupos). Quando um usuário acessa um canal específico, o Telegram sugere automaticamente até dez canais semelhantes, com base no conteúdo do canal original. Esse mecanismo foi essencial para identificar comunidades adicionais de teorias da conspiração na América Latina e no Caribe, que não haviam sido detectadas por meio da busca por palavras-chave. Ao explorar essas recomendações, foi possível descobrir uma rede mais ampla de comunidades interconectadas, muitas das quais não possuíam referências explícitas a teorias da conspiração em suas descrições, mas estavam alinhadas tematicamente.

**(iii) Abordagem de bola de neve para identificação de convites:** Após a identificação de um conjunto inicial de comunidades para extração, foi desenvolvido um algoritmo proprietário para detectar e analisar URLs contendo o prefixo "t.me/", utilizado para convites a grupos e canais no Telegram. Esse processo permitiu a acumulação de centenas de milhares de links, que foram classificados por frequência e analisados para identificar novos grupos. A abordagem "bola de neve" foi crucial para revelar comunidades adicionais de teorias da conspiração em toda a América Latina e no Caribe, pois muitos dos convites circulavam internamente dentro dos grupos já identificados. A repetição iterativa desse método garantiu a descoberta contínua e a ampliação do conjunto de dados.

**(iv) Ampliação para tweets publicados no X que mencionassem convites:** Para diversificar ainda mais as fontes de identificação das comunidades de teorias da conspiração, foi desenvolvida uma consulta de busca personalizada para identificar convites para o Telegram compartilhados no X (antigo Twitter). A busca se concentrou em palavras-chave associadas a teorias da conspiração combinadas com o prefixo "t.me/", utilizando a sintaxe para busca: https://x.com/search?q=lang%3Aes%20%22t.me%2F%22%20SEARCH-TERM e https://x.com/search?q=lang%3Apt%20%22t.me%2F%22%20SEARCH-TERM. Esse método se mostrou particularmente eficaz para capturar links de grupos transnacionais de teorias da conspiração, que utilizam o X como plataforma de disseminação para atrair novos membros.

Ao implementar essas múltiplas estratégias de identificação — desenvolvidas ao longo de meses de refinamento metodológico — foi possível construir um banco de dados abrangente com **1.649 comunidades de teorias da conspiração** em toda a América Latina e no Caribe. Coletivamente, essas comunidades **publicaram 58.637.137 conteúdos** entre **dezembro de 2015 e janeiro de 2025**, com um total combinado de **5.345.332 usuários**. É importante destacar que esse número envolve duas considerações: primeiro, o total de usuários é variável, uma vez que membros entram e saem das comunidades com frequência, o que significa que o total registrado representa um recorte temporal do período de extração dos dados; segundo, um mesmo usuário pode fazer parte de vários grupos, o que pode gerar duplicações na contagem geral. Embora o número real de indivíduos únicos engajados com o conteúdo conspiratório possa ser menor devido a essas sobreposições, a escala de participação permanece significativa no ecossistema digital latino-americano e caribenho.

### 3.2. Tratamento de dados

Com todos os grupos e canais de teorias da conspiração no Telegram extraídos de toda a América Latina e do Caribe, foi realizada uma classificação manual com base no título e na descrição de cada comunidade. Se houvesse uma menção explícita no título ou na descrição referindo-se a um tema específico, a comunidade foi categorizada em uma das seguintes classificações: (i) "Anticiência"; (ii) "Anti-Woke e Gênero"; (iii) "Antivax"; (iv) "Apocalipse e Sobrevivencialismo"; (v) "Mudanças Climáticas"; (vi) Terra Plana; (vii) "Globalismo"; (viii) "Nova Ordem Mundial"; (ix) "Ocultismo e Esoterismo"; (x) "Off Label e

Charlatanismo”; (xi) “QAnon”; (xii) “Reptilianos e Criaturas”; (xiii) “Revisionismo e Discurso de Ódio”; (xiv) “OVNI e Universo”. Se nenhuma referência explícita a esses temas fosse encontrada no título ou na descrição, a comunidade foi classificada como (xv) Conspiração geral. Dado o escopo ampliado deste estudo, o processo de classificação considerou variações linguísticas e terminológicas em diferentes países da América Latina e do Caribe, especialmente em comunidades de língua portuguesa e espanhola. Além disso, alguns temas apresentaram nuances regionais, como a predominância de narrativas conspiratórias em países. Nas tabelas a seguir, apresentamos as métricas relacionadas à classificação das comunidades de teorias da conspiração na América Latina e no Caribe.

**Tabela 01.** Países das comunidades de teorias da conspiração (métricas até jan. de 2025).

| País | Grupos | Usuários | Conteúdos | Comentários | Total |
|---|---|---|---|---|---|
| Argentina | 62 | 545.594 | 1.459.065 | 4.796.166 | 6.255.231 |
| Bolivia | 09 | 4.622 | 96.010 | 2.347 | 98.357 |
| Brasil | 1.000 | 2.537.760 | 15.779.699 | 16.110.578 | 31.890.277 |
| Chile | 43 | 76.375 | 469.916 | 431.154 | 901.070 |
| Colombia | 71 | 152.946 | 1.121.331 | 2.075.572 | 3.196.903 |
| Costa Rica | 13 | 7.728 | 128.306 | 11.095 | 139.401 |
| Cuba | 01 | 71 | 439 | 00 | 439 |
| Ecuador | 29 | 15.559 | 111.280 | 662.661 | 773.941 |
| Guatemala | 03 | 97 | 159 | 22 | 181 |
| Honduras | 02 | 299 | 1.551 | 76 | 1.627 |
| Jamaica | 01 | 25 | 22 | 00 | 22 |
| México | 59 | 318.740 | 692.847 | 2.171.015 | 2.863.862 |
| Panamá | 07 | 4.343 | 15.704 | 3.695 | 19.399 |
| Paraguay | 12 | 9.430 | 49.317 | 16.210 | 65.527 |
| Perú | 35 | 47.341 | 1.149.962 | 305.918 | 1.455.880 |
| Puerto Rico | 03 | 2.026 | 3.080 | 626 | 3.706 |
| República Dominicana | 01 | 12 | 29 | 10 | 39 |
| Transnacional | 291 | 1.555.733 | 3.648.008 | 6.289.729 | 9.937.737 |
| Uruguay | 08 | 14.987 | 115.012 | 223.401 | 338.413 |
| Venezuela | 09 | 17.040 | 73.699 | 505.441 | 579.140 |
| **Total** | **1.659** | **5.310.728** | **24.915.436** | **33.605.716** | **58.521.152** |

Fonte: Elaboração própria (2025).

**Tabela 02.** Categorias das comunidades de teorias da conspiração (métricas até jan. de 2025).

| Categoria | Grupos | Usuários | Conteúdos | Comentários | Total |
|---|---|---|---|---|---|
| Anticiência | 31 | 102.163 | 323.012 | 982.414 | 1.305.426 |
| Anti-*Woke* e Gênero | 57 | 181.257 | 636.734 | 1.981.732 | 2.618.466 |
| Antivacinas (*Antivax*) | 280 | 985.438 | 3.562.816 | 3.775.987 | 7.338.803 |
| Apocalipse e Sobrevivência | 44 | 169.568 | 1.503.161 | 583.290 | 2.086.451 |
| Mudanças Climáticas | 43 | 46.154 | 504.505 | 171.285 | 675.790 |
| Terraplanismo | 52 | 48.725 | 556.063 | 1.337.079 | 1.893.142 |
| Conspirações Gerais | 219 | 1.016.301 | 5.102.585 | 7.595.481 | 12.698.066 |
| Globalismo | 62 | 504.759 | 1.246.500 | 1.097.291 | 2.343.791 |
| Nova Ordem Mundial (NOM) | 195 | 607.731 | 3.985.405 | 5.888.434 | 9.873.839 |
| Ocultismo e Esoterismo | 83 | 208.393 | 1.801.951 | 2.285.810 | 4.087.761 |
| Medicamentos *off label* | 343 | 966.010 | 3.013.077 | 6.839.001 | 9.852.078 |
| QAnon | 42 | 126.342 | 893.353 | 273.350 | 1.166.703 |
| Reptilianos e Criaturas | 28 | 128.543 | 180.786 | 70.342 | 251.128 |
| Revisionismo e Ódio | 108 | 111.628 | 495.319 | 265.194 | 760.513 |
| OVNI e Universo | 72 | 107.716 | 1.110.169 | 459.026 | 1.569.195 |
| **Total** | **1.659** | **5.310.728** | **24.915.436** | **33.605.716** | **58.521.152** |

Fonte: Elaboração própria (2025).

Além disso, é importante enfatizar que apenas comunidades abertas foram extraídas para este estudo. Esses são grupos e canais que não apenas são publicamente identificáveis, mas também permitem acesso irrestrito ao seu conteúdo, ou seja, qualquer usuário do Telegram pode entrar e visualizar as discussões sem a necessidade de aprovação ou convite. Além disso, em conformidade com as regulamentações regionais de proteção de dados, particularmente a Lei Geral de Proteção de Dados Pessoais (LGPD – Lei nº 13.709/2018) do Brasil, todos os dados extraídos foram totalmente anonimizados para garantir a privacidade e o cumprimento de padrões éticos de pesquisa. Esse processo de anonimização não se limita aos dados dos usuários, mas também inclui a identificação das comunidades, garantindo que nenhum grupo, canal ou participante possa ser rastreado por meio deste estudo. Assim, embora os dados analisados sejam provenientes de conteúdo publicamente disponível, camadas de proteção à privacidade foram implementadas para impedir qualquer forma de atribuição direta, reforçando o compromisso ético com a confidencialidade dos usuários.

### 3.3. Abordagens para análise de dados

**(i) Análise descritiva das alegações relacionadas ao autismo:** Este estudo examina sistematicamente 150 supostas causas e 150 supostas curas para o autismo, conforme propagadas dentro das comunidades de teorias da conspiração no Telegram. Por meio de um modelo estruturado de classificação, cada alegação é categorizada com base em sua origem temática (por exemplo, fatores ambientais, vacinas, influências dietéticas, teorias genéticas) e na natureza da suposta cura (por exemplo, medicina alternativa, protocolos de desintoxicação, alegações farmacêuticas, intervenções pseudocientíficas). Ao mapear essas narrativas, a análise busca identificar padrões, motivações ideológicas e mecanismos de inter-referência que sustentam e amplificam a desinformação sobre o autismo. Essa abordagem descritiva fornece uma base essencial para compreender como a desinformação é estruturada dentro dessas comunidades e como ela evolui ao longo do tempo.

**(ii) Análise de redes:** Esse modelo metodológico possibilita avaliar se essas comunidades funcionam como fontes autorreferenciais de legitimação ou se promovem ativamente outras teorias conspiratórias, expandindo o envolvimento dos usuários com redes de desinformação. Além disso, este estudo se baseia na abordagem de análise de redes de Rocha et al. (2024), que aplicaram uma técnica semelhante para identificar similaridades de conteúdo em comunidades do Telegram, utilizando IDs únicos para cada mensagem.

**(iii) Análise de séries temporais:** Para estruturar e analisar tendências temporais, a biblioteca Pandas (McKinney, 2010) foi utilizada na organização dos dataframes, permitindo a observação de: (a) o volume de publicações ao longo do tempo; (b) a dinâmica de engajamento dos usuários, incluindo metadados de visualizações, reações e comentários extraídos durante a coleta. Além da análise volumétrica, a biblioteca Plotly (Plotly Technologies Inc., 2015) foi empregada para gerar gráficos das variações temporais, facilitando a compreensão das flutuações na produção de conteúdo e interação dos usuários.

**(iv) Análise de conteúdo:** Além da análise geral de frequência de palavras, foram aplicadas séries temporais para examinar a variação dos termos mais frequentes ao longo dos semestres. O conjunto de dados abrange o período de julho de 2017 (primeiras publicações sobre autismo) até janeiro de 2025 (período de estudo). Utilizando as bibliotecas Pandas (McKinney, 2010) e WordCloud (Mueller, 2020), os resultados são apresentados tanto volumetricamente quanto graficamente, fornecendo insights sobre mudanças temáticas e a persistência de narrativas específicas ao longo do tempo.

A metodologia aplicada abrangeu todo o processo de pesquisa, desde a extração de dados — realizada com a ferramenta proprietária TelegramScrap (Silva, 2023) — até o processamento e análise dos dados coletados. Um conjunto diversificado de abordagens analíticas foi empregado para identificar, classificar e examinar sistematicamente as comunidades de teorias da conspiração no Telegram em toda a América Latina e Caribe. Cada etapa do estudo foi projetada para garantir a integridade dos dados e assegurar a rigorosa conformidade com as regulamentações regionais de proteção de dados, particularmente a Lei Geral de Proteção de Dados Pessoais (LGPD – Lei nº 13.709/2018) do Brasil. As seções a seguir apresentam os resultados do estudo, oferecendo uma análise aprofundada da dinâmica estrutural, das tendências temáticas e das interconexões dentro das comunidades analisadas.

## 4. Resultados

A distribuição geográfica das publicações mapeadas sobre autismo revela uma concentração significativa de grupos e conteúdos no Brasil, seguido por um número expressivo de interações em comunidades transnacionais. Países como Colômbia, Argentina e

México também apresentam um volume considerável de publicações e discussões sobre o tema, demonstrando o alcance e a relevância das narrativas sobre autismo em diferentes contextos nacionais. O elevado número de usuários e conteúdos registrados nos grupos brasileiros indica uma intensa participação e engajamento na disseminação de informações – ou desinformações – relacionadas ao autismo, tornando essencial a análise qualitativa dessas interações para compreender suas implicações.

**Tabela 03.** Publicações mapeadas sobre autismo por países (métricas até jan. de 2025).

| País | Grupos | Usuários | Conteúdos | Comentários | Total |
|---|---|---|---|---|---|
| Brasil | 487 | 1.726.364 | 10.591 | 11.416 | 22.007 |
| Transnacional | 194 | 1.385.420 | 5.039 | 3.126 | 8.165 |
| Colombia | 57 | 112.453 | 1.567 | 1.179 | 2.746 |
| Argentina | 45 | 528.190 | 1.517 | 4.299 | 5.816 |
| México | 42 | 261.407 | 1.178 | 1.770 | 2.948 |
| Perú | 21 | 43.543 | 675 | 331 | 1.006 |
| Chile | 26 | 62.086 | 582 | 558 | 1.140 |
| Venezuela | 06 | 16.869 | 262 | 2.276 | 2.538 |
| Bolivia | 04 | 4.295 | 184 | 06 | 190 |
| Costa Rica | 06 | 6.085 | 160 | 06 | 166 |
| Ecuador | 17 | 13.339 | 126 | 51 | 177 |
| Paraguay | 05 | 4.780 | 73 | 19 | 92 |
| Uruguay | 04 | 14.857 | 56 | 174 | 230 |
| Puerto Rico | 01 | 1.869 | 23 | 04 | 27 |
| Panamá | 06 | 4.339 | 09 | 01 | 10 |
| Honduras | 01 | 135 | 02 | 01 | 03 |
| **Total** | **922** | **4.186.031** | **22.044** | **25.217** | **47.261** |

Fonte: Elaboração própria (2025).

Além da análise por país, a classificação por categorias temáticas permite uma visão mais detalhada das principais abordagens dentro das discussões sobre autismo. Como se observa na Tabela 04, há uma interseção significativa entre esse tema e pautas ligadas ao movimento antivacinas, ao globalismo e a teorias conspiratórias mais amplas, como a Nova Ordem Mundial (NOM). O alto volume de conteúdos e interações dentro dessas categorias aponta que o autismo frequentemente aparece em debates marcados por desinformação e narrativas controversas, reforçando a necessidade de monitoramento e estratégias eficazes de enfrentamento à propagação de discursos que possam impactar negativamente a sociedade.

**Tabela 04.** Publicações mapeadas sobre autismo por categorias (métricas até jan. de 2025).

| Categoria | Grupos | Usuários | Conteúdos | Comentários | Total |
|---|---|---|---|---|---|
| Anticiência | 17 | 53.846 | 213 | 1.117 | 1.330 |
| Anti-*Woke* e Gênero | 30 | 105.511 | 425 | 960 | 1.385 |
| Antivacinas (*Antivax*) | 183 | 877.305 | 5.140 | 3.915 | 9.055 |
| Apocalipse e Sobrevivência | 24 | 138.720 | 683 | 230 | 913 |
| Mudanças Climáticas | 28 | 30.957 | 358 | 23 | 381 |
| Terraplanismo | 17 | 25.827 | 238 | 632 | 870 |
| Conspirações Gerais | 129 | 849.959 | 2.827 | 2.150 | 4.977 |
| Globalismo | 43 | 363.933 | 549 | 609 | 1.158 |
| Nova Ordem Mundial (NOM) | 119 | 456.193 | 2.414 | 4.293 | 6.707 |
| Ocultismo e Esoterismo | 41 | 138.320 | 929 | 2.984 | 3.913 |
| Medicamentos *off label* | 200 | 803.150 | 7.321 | 7.805 | 15.126 |
| QAnon | 28 | 115.696 | 301 | 133 | 434 |
| Reptilianos e Criaturas | 08 | 97.836 | 66 | 13 | 79 |
| Revisionismo e Ódio | 36 | 61.551 | 231 | 205 | 436 |
| OVNI e Universo | 19 | 67.227 | 349 | 148 | 497 |
| **Total** | **922** | **4.186.031** | **22.044** | **25.217** | **47.261** |

Fonte: Elaboração própria (2025).

A seguir, os resultados são detalhados na ordem prevista na metodologia.

## 4.1. Descritivo

### 4.1.1. Falsas causas do autismo

Uma das narrativas mais disseminadas associa o autismo à radiação eletromagnética, à inversão do campo magnético da Terra e ao uso de agrotóxicos. Essas alegações sugerem que a modernização tecnológica e o uso intensivo de produtos químicos estariam diretamente ligados ao aumento de diagnósticos de TEA. A radiação eletromagnética emitida por celulares, redes Wi-Fi e antenas 5G não é ionizante, ou seja, não tem capacidade de modificar o DNA humano ou causar alterações cerebrais. Estudos científicos demonstram que essa forma de radiação não representa risco ao desenvolvimento neurológico.

Já a ideia de que a inversão do campo magnético terrestre causaria autismo carece de qualquer evidência. Esse fenômeno ocorre em ciclos geológicos e não há qualquer registro histórico de impactos neurológicos em seres humanos. Os agrotóxicos, por sua vez, são frequentemente apontados como uma das causas ambientais do autismo. Embora a exposição prolongada a pesticidas possa ter impactos na saúde, não há estudos que comprovem uma relação causal direta entre o uso de agrotóxicos e o desenvolvimento do TEA.

Além disso, uma vertente preocupante dessa desinformação sugere que os parasitas seriam responsáveis pelo autismo, promovendo assim protocolos de desparasitação. Esse tipo de prática é perigosa, pois induz pais a administrarem substâncias tóxicas, como dióxido de cloro (MMS), além de outros protocolos nocivos de desparasitação, iludindo responsáveis de crianças autistas à esperança de uma cura milagrosa. O autismo não é causado por parasitas e não há qualquer base científica que justifique tratamentos antiparasitários para essa condição.

**Figura 01.** Exemplos de conspirações sobre causas do autismo.

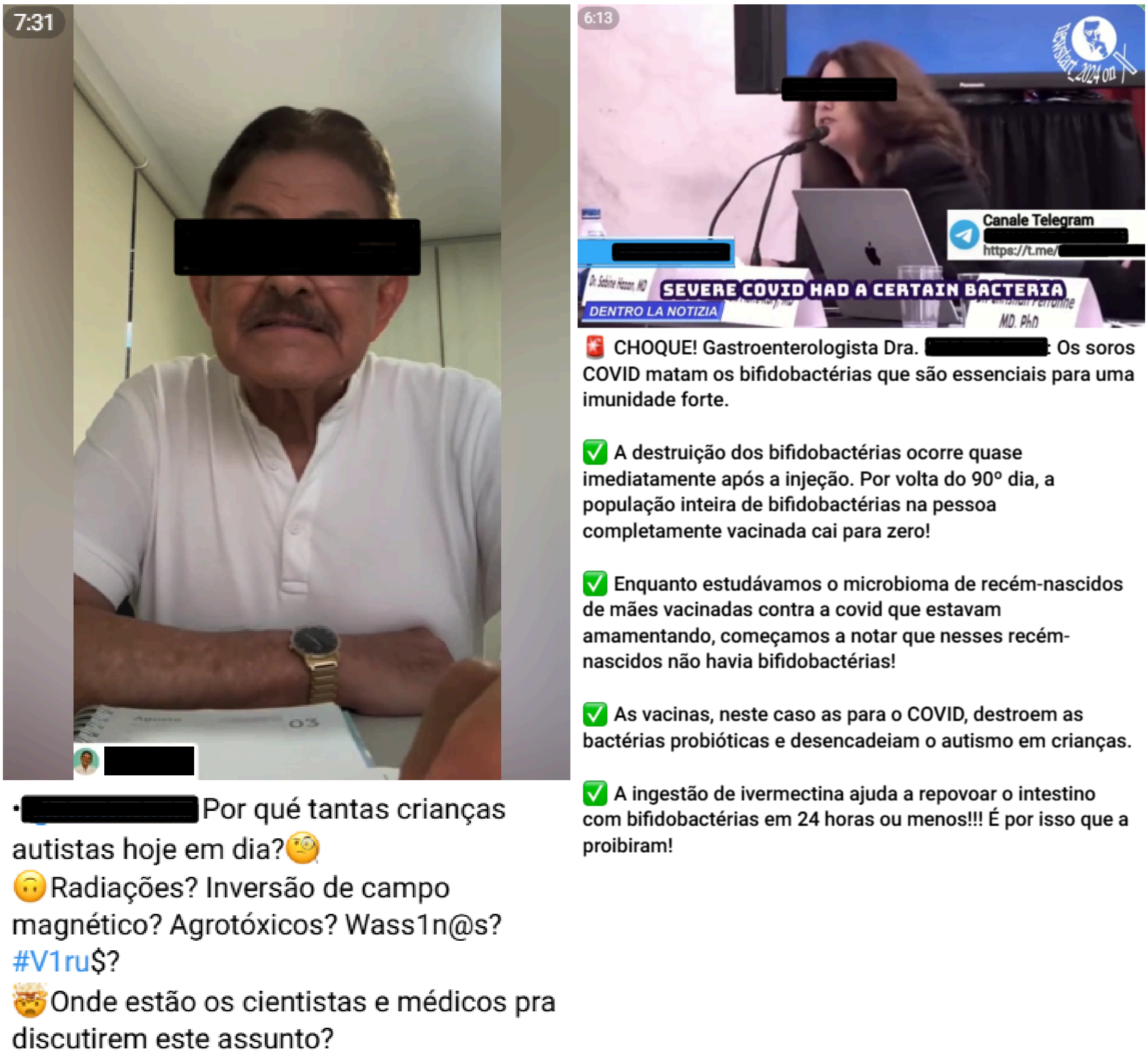

Fonte: Captura de tela (2025).

Outra narrativa amplamente divulgada em comunidades conspiratórias sugere que óleos vegetais e alimentos ultraprocessados, como Doritos, poderiam causar autismo. O argumento baseia-se na alegação de que os ácidos graxos ômega-6, presentes em óleos como soja, canola e girassol, atuariam como um "veneno metabólico", promovendo inflamações cerebrais. Embora um equilíbrio entre ômega-3 e ômega-6 seja importante para a saúde geral, não há qualquer estudo que relacione diretamente o consumo desses óleos ao autismo. Da mesma forma, a presença de ingredientes geneticamente modificados em produtos industrializados, como o milho transgênico em salgadinhos, não tem qualquer correlação científica com o desenvolvimento do TEA. Essa narrativa se insere em uma tendência maior de desinformação alimentar, que demoniza determinados grupos de alimentos sem base científica, desviando o foco de discussões reais sobre nutrição e saúde.

**Figura 02.** Exemplos de conspirações sobre causas do autismo.

**Este óleo vegetal está vinculado a doenças crônicas**

**Óleo de soja pode causar mudanças irreversíveis em seu cérebro**

**Resumo da matéria**:
- Óleos vegetais e óleos de sementes como soja, canola, girassol, semente de uva, milho, cártamo, amendoim e óleo de farelo de arroz são carregados com ácido linoleico ômega-6 (LA), que atua como um veneno metabólico quando consumido em excesso. Qualquer coisa acima de 10 gramas por dia causará problemas a longo prazo;
- Os óleos de sementes são incrivelmente pró-inflamatórios e promovem a oxidação em seu corpo. Essa oxidação, por sua vez, desencadeia a disfunção mitocondrial que então impulsiona o processo da doença;
- O óleo de soja demonstrou causar alterações genéticas irreversíveis no cérebro de camundongos. Isso foi verdade tanto para o óleo de soja não modificado quanto para o óleo de soja modificado para ser baixo em AL.; • Ambos produziram efeitos pronunciados no hipotálamo, que regula o metabolismo e as respostas ao estresse;
- Vários genes nos camundongos que foram alimentados com óleo de soja não estavam funcionando corretamente, incluindo um gene que produz ocitocina, o "hormônio do amor". Cerca de 100 outros genes também foram afetados. Essas mudanças podem ter ramificações para o metabolismo energético, função cerebral adequada e doenças como autismo e doença de Parkinson;

Por que você não deveria comer Doritos de qualquer maneira

1. Doritos contém ingredientes OGM, como o primeiro ingrediente, milho

Os OGM usados para fazer Doritos são considerados cancerígenos que têm sido associados ao câncer de mama, autismo, alergias ao glúten, diabetes, inflamação e distúrbios que afetam os sistemas digestivo e reprodutivo.

2. É feito com uma grande quantidade de óleos hidrogenados processados comercialmente

Isso pode levar a um aumento de radicais livres no corpo. Eles também são geneticamente modificados e carregados com gorduras trans, que podem causar inflamação, imunidade comprometida, aumento da circulação de estrogênio ruim e falta de nutrientes.

3. É feito com corantes, incluindo corante amarelo 5, corante amarelo 6 e corante vermelho 40.

Os efeitos colaterais de longo prazo da ingestão de corantes incluem distúrbios imunológicos, A.D.D. e TDAH, especialmente em crianças.

Fonte: Captura de tela (2025).

Outra linha de desinformação atribui o autismo à presença de alumínio em vacinas, ao suposto despejo de substâncias químicas na atmosfera por aviões (chemtrails) e ao projeto HAARP, um programa de pesquisa atmosférica dos Estados Unidos. O alumínio é um dos elementos mais abundantes na natureza e está presente naturalmente em alimentos, na água e em produtos cosméticos. A pequena quantidade utilizada como adjuvante em vacinas não apresenta risco à saúde humana, sendo rapidamente eliminada pelo organismo.

Os chemtrails, por sua vez, são uma teoria da conspiração sem qualquer comprovação científica. As trilhas deixadas por aviões são apenas condensação de vapor d'água e não têm qualquer relação com a saúde humana. Já o HAARP (High-Frequency Active Auroral Research Program), frequentemente citado em teorias conspiratórias, é um programa de pesquisa sobre a ionosfera terrestre. Não há qualquer evidência de que ele tenha capacidade de manipular mentes ou influenciar condições neurológicas como o autismo.

**Figura 03.** Exemplos de conspirações sobre causas do autismo.

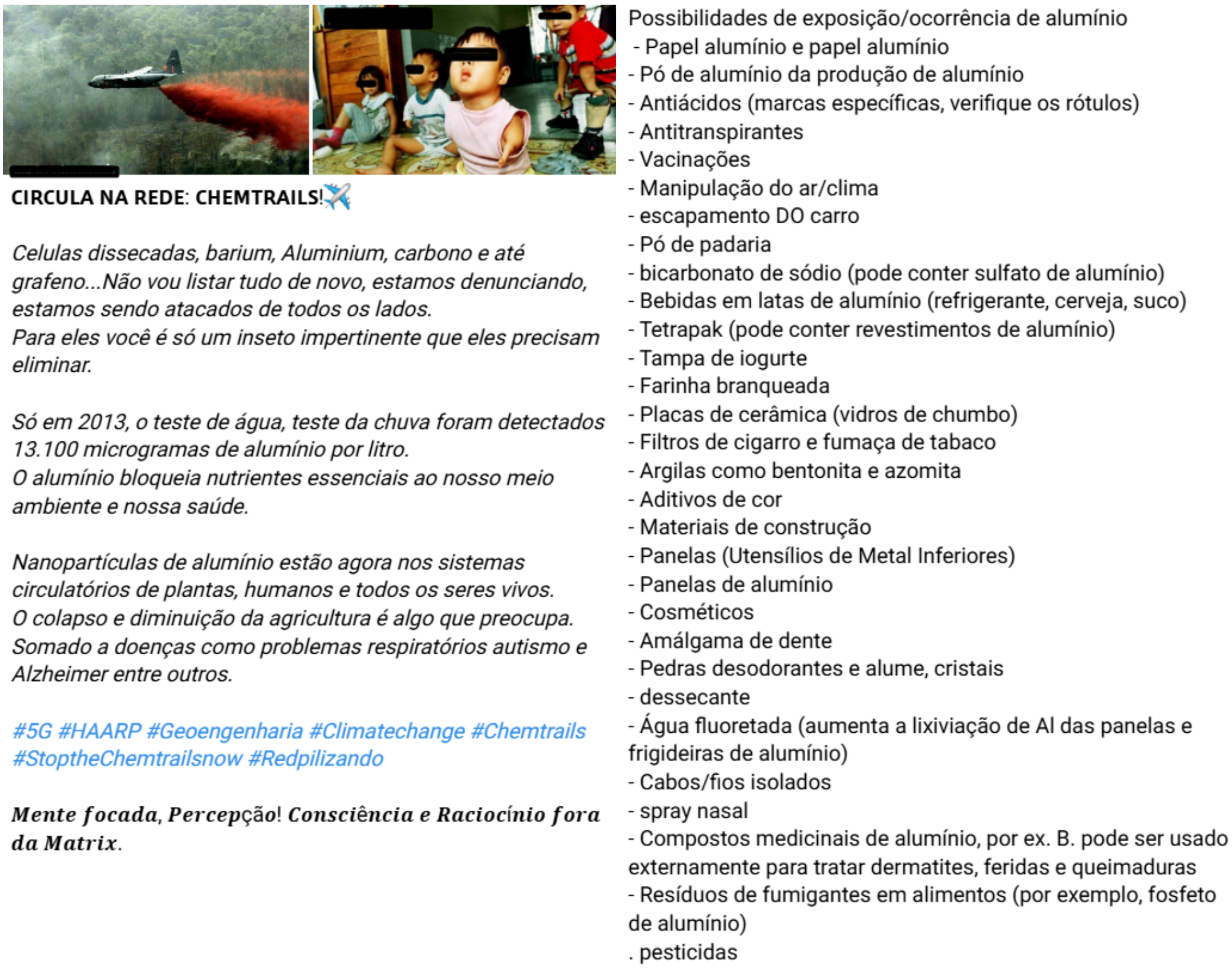

Fonte: Captura de tela (2025).

Além dos elementos já citados, algumas comunidades conspiratórias divulgam uma lista extensa de substâncias químicas que supostamente causariam autismo, incluindo:

- Arsênio, N-Hexano, Organoclorados, Organofosforados, Tolueno e Tricloroetileno – Esses compostos químicos podem ser tóxicos em altas doses, mas não há nenhuma comprovação de que causem autismo.
- Álcool, Bicarbonato de Sódio, Bórax e Cafeína – Algumas dessas substâncias podem afetar o sistema nervoso, mas não há nenhuma relação causal com o TEA.
- Chemtrails, Enxofre, Petróleo, Wireless e WPO – São frequentemente citados como agentes que alterariam o cérebro humano, mas não há qualquer base científica que corrobore essa hipótese.

A disseminação dessas alegações cria um cenário de paranoia e desinformação, afastando familiares e cuidadores de informações médicas confiáveis.

**Figura 04.** Exemplos de conspirações sobre causas do autismo.

❤️❤️**EPIGENÉTICA Y ENFERMEDADES CEREBRALES: PAPEL DE LA EXPOSICIÓN TEMPRANA A CONTAMINANTES.**❤️❤️

La zeolita es un anti-veneno, un quelador de metales pesados y pesticidas. Se sospecha que son los causantes de numerosas enfermedades. Podemos mencionar las enfermedades neurológicas porque ciertos metales como el aluminio y el mercurio "aman" el cerebro. Los científicos han descubierto que la exposición temprana a los contaminantes, es decir, en el feto y en los niños pequeños, favorece la aparición de lesiones de Alzheimer y Parkinson. Aquí es donde entra la noción de epigenética, una nueva ciencia que se define como el silenciamiento de un gen a lo largo de la vida o de varias generaciones. Esto se debe a las modificaciones genéticas adquiridas a través de los venenos. Así que hay "interruptores" epigenéticos, ¿cómo evitarlos? ¿Es útil un quelante de arcilla o zeolita para las mujeres embarazadas?

❤️**La epigenética en la aparición de enfermedades neurológicas**

Hay pruebas de la exposición temprana a los contaminantes y la aparición de enfermedades neurológicas. Estos factores ambientales alteran la genética del individuo; estamos entrando en el campo de la epigenética, una ciencia que ha cobrado impulso. Todo lo que hacemos, comemos, absorbemos como alimento o veneno puede influir en el ADN. Dependiendo de las alteraciones y la predisposición de los individuos; tendremos cánceres, enfermedades neurológicas y muchas otras dolencias.

- Transmissão deliberada de parasitas através do ar, vacinação e alimentos - Morgellons, autismo (verme), vermes, tênias, vermes, fungos como candida e muitos mais.
- Radiação, celular, wireless - altas frequências discordantes, criam desequilíbrio
- Digitalização humana - chips RFID, nanobots, parasitas sintéticos, realidade virtual
- Microondas – especialmente o popular micro-ondas na cozinha, destrói completamente a estrutura dos alimentos, deixando apenas o enchimento tóxico.
- Chemtrails - Morgellons, escurecimento global, manipulação do clima, pulverização de toxinas em geral.
- Medicamentos que contêm aditivos nocivos e/ou foram modificados quimicamente - Muitas substâncias são benéficas mesmo em sua forma pura ou por que mais foram colocadas sob o BTMG e o álcool está disponível gratuitamente?
- Limpadores de corpo antigos e comprovados, como petróleo, WPO, bórax, bicarbonato de sódio, enxofre, terebintina, são fornecidos com avisos de perigo e caveiras - porque sabem o quão forte é o efeito de limpeza
- Bactérias (microrganismos) são nossos "inimigos" - quando na verdade são nossos "amigos"
- Drogas pesadas como açúcar, cafeína e álcool estão disponíveis gratuitamente - uma verdadeira planta medicinal (cânhamo), por outro lado, foi transformada em uma droga "ruim" e criminalizada, que foi massivamente geneticamente modificada

Fonte: Captura de tela (2025).

Outra teoria recorrente sugere que o autismo seria causado por uma deficiência no microbioma intestinal ou pela insuficiência de vitamina B12. Essa narrativa tem levado muitas pessoas a acreditarem que o uso de probióticos, dietas restritivas e suplementação intensa de vitamina B12 poderiam "reverter" o TEA. Embora estudos demonstrem que o microbioma pode desempenhar um papel na saúde neurológica, não há qualquer evidência de que um desequilíbrio intestinal cause autismo. Algumas crianças autistas podem apresentar diferenças na microbiota, mas isso pode ser um efeito da condição, e não uma causa. A deficiência de vitamina B12 pode, de fato, causar problemas neurológicos, mas não há relação científica entre deficiência de B12 e o TEA. A suplementação excessiva pode levar a efeitos adversos, incluindo desequilíbrios metabólicos.

**Figura 05.** Exemplos de conspirações sobre causas do autismo.

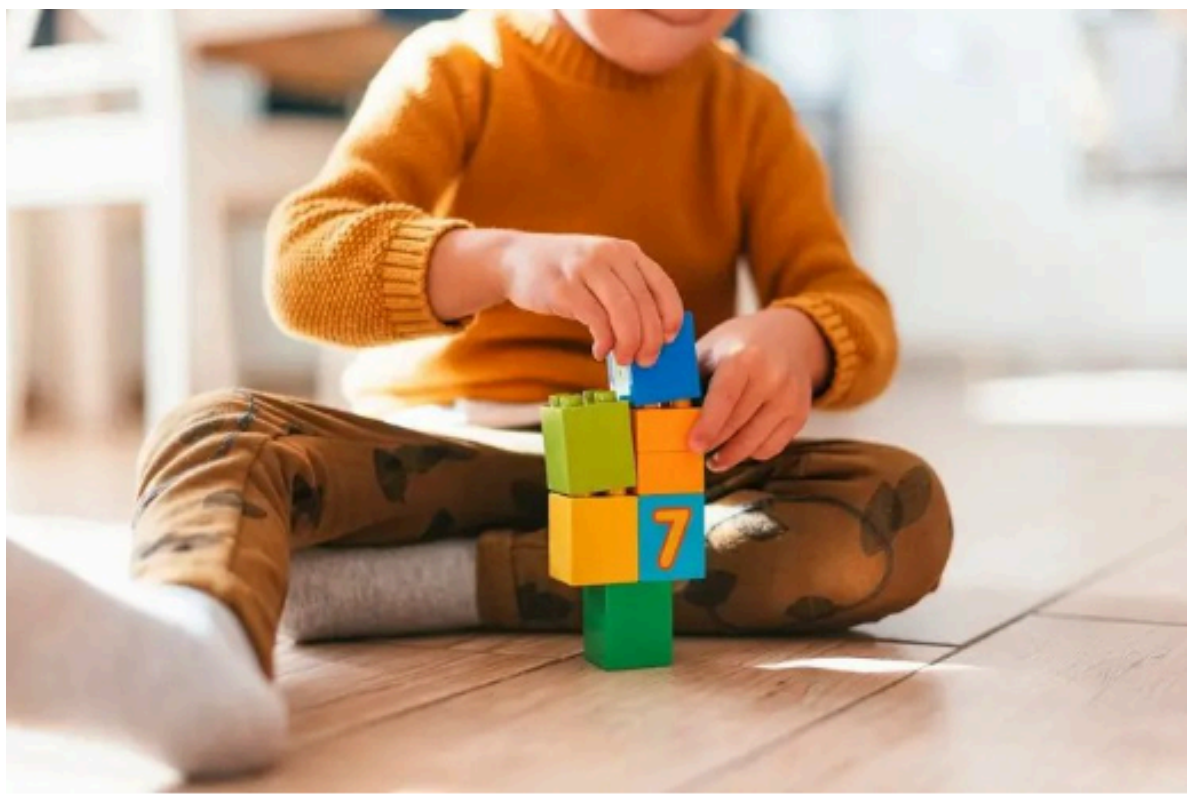

O AUTISMO ESTÁ RELACIONADO COM A ALIMENTAÇÃO

Existem muitas causas para o autismo, daí o nome de perturbação do espectro do autismo. Se existe um espectro de perturbações, é provável que exista um espetro de problemas subjacentes. Um deles demonstrou ser uma deficiência de carnitina, que é…

Cloridrato de betaína disbiose do intestino falta de b12 e autismo
O cloridrato de betaína que faz subir a acidez no estômago ... e assim tratar a baixa acidez que faz com que não haja uma digestão das proteínas animais... e assim não se produzem os peptideos e os aminoácidos para quebrar assim as proteínas animais no estômago... Para não chegar pedaços de carne (peixe, ovo... etc) inteiros no intestino e assim apodrecer no intestino e provocando lesões na parede do intestino... fazendo disbiose e intestino permeável... levando à não produção dos neurotransmissores serotonina (90% é produzida no intestino) e dopamina e produzindo assim sintomas de autismo ...
Falta de uma boa digestão também para a produção de vitamina B12 ...
Cloridrato de betaína HCL ajuda a uma boa digestão... (mas dizem para não usar em pessoas menores de idade...¿¿¿)
Má digestão vai provocar disbiose intestinal... permeabilidade intestinal... e consequentemente inflamação cerebral... e falta de nutrientes no cérebro... levando à um estado de sintomas de autismo
15:03

Fonte: Captura de tela (2025).

Com o avanço das vacinas contra COVID-19, algumas comunidades conspiratórias passaram a associar a proteína spike gerada pelas vacinas ao desenvolvimento do autismo. Essa alegação, além de infundada, alimenta um discurso antivacinação já amplamente refutado pela ciência. A proteína spike presente nas vacinas de mRNA é temporária e eliminada pelo organismo em poucos dias, não havendo qualquer evidência de que cause danos neurológicos ou alterações genéticas. Além disso, a teoria de que vacinas causam autismo surgiu de um estudo fraudulento publicado em 1998 e posteriormente desmentido. Diversos estudos epidemiológicos de grande escala comprovaram que não há qualquer relação entre vacinação e o TEA. Infelizmente, essas narrativas incentivam o movimento antivacinação, colocando em risco não apenas crianças autistas, mas toda a população, ao aumentar a disseminação de doenças preveníveis.

**Figura 06.** Exemplos de conspirações sobre causas do autismo.

**DESINTOXICAR LA PROTEINA DE LA ESPÍCULA**

Aquí tenemos un buen recopilatorio del impacto de la proteína S en la salud, tanto en los procesos celulares como en los distintos órganos del cuerpo, así como dónde tiende a acumularse.

También incluye algunas de las herramientas terapéuticas para inhibir su impacto (especialmente importante si se sigue produciendo).

**Ivermectina** (con prescripción médica)
**Suramina** (con prescripción médica) También eficaz para autismo.
**NAC** (N-acetilcisteína)
**EGCG** (catequina del té verde)
**Curcumina** (derivado de la cúrcuma)
**Prunella Vulgaris**
**Agujas de Pino**
**Extracto de Hoja de Diente de León**
**Extracto de Margosa de la India** (NEEM Extract)

Se puede ver más información en The World Council for Health (WCH):
https://worldcouncilforhealth.org/resources/spike-protein-detox-guide/

35,1K [redacted] 23:14

La FDA ha publicado pruebas concluyentes en su sitio web de que la vacuna DTap puede causar autismo. De acuerdo con el documento en línea de Vacunas de Sangre Biológicas de la FDA, un fabricante de vacunas admite en su paquete que su vacunación puede causar autismo como una de las muchas reacciones adversas.

- Estos eventos adversos informados durante el uso posterior a la aprobación de la vacuna Tripedia incluyen púrpura trombocitopénica idiopática, SIDS, reacción anafiláctica, celulitis, autismo, convulsión / convulsión de gran mal, encefalopatía, hipotonía, neuropatía, somnolencia y apnea. Los eventos se incluyeron en esta lista debido a la gravedad o frecuencia de los informes. Debido a que estos eventos se informan voluntariamente de una población de tamaño incierto, no siempre es posible estimar de manera confiable sus frecuencias o establecer una relación causal con los componentes de la vacuna Tripedia.

Fonte: Captura de tela (2025).

Embora essas comunidades aleguem ter descoberto as verdadeiras causas do autismo, a ampla e muitas vezes contraditória variedade de explicações revela a falta de coerência dessas narrativas. Se o autismo supostamente é causado por tudo — desde radiação eletromagnética, alimentos industrializados e vacinas até deficiências vitamínicas e exposição a metais pesados — , isso demonstra não apenas a ausência de embasamento científico, mas também a inconsistência interna desses discursos. A seguir, apresenta-se, em formato de quadro, a íntegra das 150 supostas causas do autismo que circulam nessas comunidades conspiratórias, evidenciando como a desinformação se adapta e se expande para reforçar diferentes temores e crenças pseudocientíficas.

**Quadro 01.** Lista de falsas causas do autismo mapeadas nas comunidades (A-Z).

| | Suposta causa | Exemplo |
|---|---|---|
| **1** | 5G | Algumas teorias da conspiração alegam que a radiação das redes 5G interfere no desenvolvimento neurológico e causa autismo. No entanto, não há nenhuma evidência científica que sustente essa afirmação. A radiação emitida pelo 5G é não ionizante, o que significa que não tem energia suficiente para causar alterações no DNA ou no cérebro humano. A disseminação dessa teoria reflete um padrão mais amplo de desinformação sobre tecnologia e saúde, sem qualquer embasamento científico. |
| **2** | Acetaminofeno | Algumas comunidades alegam que o uso de paracetamol causa autismo ao "intoxicar" o cérebro infantil. No entanto, não há evidências científicas que relacionem o medicamento ao transtorno, sendo seguro quando usado corretamente. |

[ versão português / english above ]

| | | |
|---|---|---|
| **3** | Ácido Ascórbico | A vitamina C é equivocadamente apontada como um fator no desenvolvimento do autismo por supostamente alterar processos metabólicos. No entanto, a deficiência dessa vitamina pode causar escorbuto, mas não há ligação com TEA. |
| **4** | Ácido Linoleico Ômega-6 (LA) | Algumas narrativas sugerem que o consumo excessivo desse ácido graxo promove inflamação cerebral e leva ao autismo. No entanto, o equilíbrio entre ômega-6 e ômega-3 é importante para a saúde, sem evidências de relação com TEA. |
| **5** | Ácido Palmítico | Alega-se que esse ácido graxo, presente no leite materno e em óleos vegetais, estaria ligado ao autismo por alterar a função cerebral. No entanto, ele é um componente essencial da nutrição infantil. |
| **6** | Açúcares | Algumas comunidades afirmam que o consumo excessivo de açúcar causa inflamação no cérebro e leva ao autismo. No entanto, não há nenhuma evidência científica de que o açúcar influencie no desenvolvimento do TEA. |
| **7** | Água da Torneira | Alega-se que substâncias como flúor ou metais pesados presentes na água potável causam autismo. No entanto, a regulação da qualidade da água impede níveis prejudiciais desses compostos. |
| **8** | Álcool | Algumas narrativas associam o consumo de álcool na gestação ao autismo, mas isso é uma confusão com a Síndrome Alcoólica Fetal, que é um distúrbio diferente e não relacionado ao TEA. |
| **9** | Alergia IgG a Alimentos | Defensores da dieta sem glúten/lactose afirmam que alergias alimentares aumentam o risco de autismo. No entanto, a reação IgG não tem relação causal com TEA e muitas dessas intolerâncias não estão cientificamente validadas. |
| **10** | Alto uso de telas | Algumas teorias afirmam que o uso de telas na infância pode causar autismo, apesar do uso intenso de telas poder prejudicar o desenvolvimento na infância, não existem evidências significativas de que possa causar autismo. |
| **11** | Alumínio | Comunidades antivacinas alegam que o alumínio presente em adjuvantes de vacinas provoca autismo. No entanto, não há evidências de que o alumínio em quantidades seguras cause qualquer alteração neurológica. |
| **12** | Amálgama de Dente | Há o mito de que o mercúrio dos amálgamas dentários leva ao autismo. No entanto, estudos mostram que a exposição ao mercúrio em obturações não altera o desenvolvimento neurológico. |
| **13** | Amoxicilina | Algumas teorias afirmam que antibióticos alteram a microbiota intestinal e desencadeiam autismo. Embora antibióticos possam impactar a flora intestinal, não há qualquer evidência de que causem TEA. |
| **14** | Antiácidos | Sugere-se que o uso de antiácidos na gravidez interfere na absorção de nutrientes essenciais e leva ao autismo. No entanto, não há base científica para essa alegação. |
| **15** | Antitranspirantes | Alegam que o alumínio presente em desodorantes obstrui glândulas linfáticas e afeta o desenvolvimento neurológico. No entanto, a absorção dérmica desse alumínio é insignificante. |
| **16** | Arsênio | Alguns estudos associam a exposição excessiva ao arsênio a distúrbios neurológicos, mas não há comprovação de que seja uma causa direta do autismo. |
| **17** | Aspartame | Adoçantes artificiais são frequentemente alvo de desinformação e são erroneamente ligados ao TEA por supostamente alterarem neurotransmissores. No entanto, estudos mostram que seu consumo moderado é seguro. |

[ versão português / english above ]

| | | |
|---|---|---|
| **18** | Azomita | Algumas comunidades pseudocientíficas afirmam que a exposição a minerais raros da azomita pode influenciar no desenvolvimento do autismo. No entanto, não há qualquer base científica para essa alegação. |
| **19** | Bactéria Intestinal | Há alegações de que desequilíbrios na microbiota intestinal causam autismo. Embora a flora intestinal possa influenciar a saúde geral, não há evidências de que seja um fator causal do TEA. |
| **20** | Bentonita | Argilas como a bentonita são promovidas como “desintoxicantes” para prevenir o autismo, mas não há evidências científicas de que eliminação de toxinas previna ou reverta o transtorno. |
| **21** | Bicarbonato de sódio | Defensores da "cura natural" alegam que o bicarbonato reverte o autismo ao alcalinizar o corpo, mas essa teoria não tem embasamento científico. |
| **22** | Biotina (em Etanol) | Alega-se que a biotina em certas formas químicas causaria TEA por interferir no metabolismo. No entanto, a biotina é essencial para várias funções do organismo e não há qualquer relação com o autismo. |
| **23** | Bisfenol | O bisfenol (BPA), presente em plásticos, já foi estudado por possíveis efeitos endócrinos, mas não há evidência de que cause autismo. |
| **24** | Bórax | Algumas comunidades pseudocientíficas promovem o bórax como "desintoxicante", sugerindo que sua ausência levaria ao TEA. No entanto, o bórax pode ser tóxico e não tem relação com o autismo. |
| **25** | Cafeína | Algumas teorias afirmam que o consumo de cafeína na gestação altera o desenvolvimento neurológico e causa TEA, mas os estudos existentes não indicam essa relação. |
| **26** | Cândida | Alega-se que infecções por cândida no intestino causam autismo ao liberar toxinas no cérebro. No entanto, não há evidências que sustentem essa hipótese. |
| **27** | Caseína | Algumas dietas restritivas alegam que a caseína, proteína do leite, piora sintomas do autismo. Embora possa haver intolerâncias, não há relação causal com o TEA. |
| **28** | Chemtrails | A teoria da conspiração afirma que rastros químicos de aviões alteram o desenvolvimento neurológico e causam autismo, mas não há qualquer base científica para essa alegação. |
| **29** | Chumbo | A exposição ao chumbo pode ter efeitos neurotóxicos, mas não há comprovação de que seja um fator direto no desenvolvimento do TEA. Não há base sólida de evidência, muito menos consenso na comunidade científica de que metais pesados, como mercúrio ou chumbo, sejam responsáveis pelo autismo. |
| **30** | Cloreto de Cálcio | Algumas alegações pseudocientíficas sugerem que desequilíbrios minerais causam autismo, mas isso não é respaldado por evidências. |
| **31** | Cloreto de Magnésio | Se por um lado afirmam que cloreto de magnésio causaria autismo, por outro lado suplementos de magnésio são promovidos como “cura” do autismo por supostamente corrigirem deficiências, mas não há base científica para essa afirmação. |
| **32** | Cloreto de Potássio | Algumas narrativas alegam que desequilíbrios eletrolíticos, como excesso ou deficiência de potássio, causam autismo. No entanto, o cloreto de potássio é um nutriente essencial para funções celulares e não tem qualquer relação causal com o TEA. |

[ versão português / english above ]

| | | |
|---|---|---|
| **33** | Cloreto de Sódio | Há alegações de que o sal de cozinha altera neurotransmissores e leva ao autismo. No entanto, o sódio é essencial para o funcionamento do sistema nervoso e não há evidências científicas que sustentem essa hipótese. |
| **34** | Cloridrato de Uracila | Algumas teorias pseudocientíficas afirmam que compostos químicos desconhecidos do público podem causar autismo. No entanto, a uracila é uma base nitrogenada essencial no RNA, e não há estudos que sugiram qualquer relação com o TEA. |
| **35** | Conservante | Há uma crença popular de que conservantes artificiais, especialmente os usados em alimentos industrializados, alteram a função cerebral e causam autismo. No entanto, nenhum estudo confiável estabeleceu essa relação. |
| **36** | Corantes | Algumas comunidades alegam que corantes artificiais, como o tartrazina, afetam o neurodesenvolvimento e levam ao TEA. Embora alguns corantes possam causar hiperatividade em crianças sensíveis, não há evidências científicas que os associem ao autismo. |
| **37** | Cosméticos | Há teorias de que substâncias químicas presentes em produtos cosméticos, como parabenos e ftalatos, interferem no desenvolvimento neurológico. Embora esses compostos sejam regulados, não há qualquer estudo que comprove sua relação com o autismo. |
| **38** | COVID | Algumas narrativas sugerem que a infecção pelo vírus SARS-CoV-2 pode causar autismo, principalmente em gestantes. No entanto, o TEA tem origem predominantemente genética e não há evidências de que infecções virais estejam associadas ao seu desenvolvimento. |
| **39** | Deficiência de Aminoácidos | Algumas comunidades afirmam que a falta de aminoácidos essenciais interfere na formação cerebral e leva ao autismo. No entanto, enquanto a nutrição inadequada pode impactar o desenvolvimento geral, não há estudos que associem a deficiência de aminoácidos ao TEA. |
| **40** | Deficiência de B12 | Algumas teorias afirmam que a falta de vitamina B12 na gravidez ou infância leva ao autismo. Embora a B12 seja essencial para o desenvolvimento neurológico, sua deficiência pode causar outros problemas, como anemia, mas não está ligada ao TEA. |
| **41** | Deficiência de Cálcio | Narrativas pseudocientíficas sugerem que a deficiência de cálcio prejudica a comunicação neuronal e causa autismo. No entanto, enquanto o cálcio é fundamental para a saúde óssea e muscular, não há evidências de que sua ausência esteja relacionada ao TEA. |
| **42** | Deficiência de Carnitina | Algumas comunidades afirmam que baixos níveis de carnitina, um composto envolvido no metabolismo energético, causam autismo. No entanto, estudos não demonstram qualquer relação causal entre a carnitina e o TEA. |
| **43** | Deficiência de Glifosato | Essa alegação falsa surge de grupos conspiratórios que acreditam que o glifosato, um herbicida, seria essencial para processos metabólicos e que sua ausência levaria ao autismo. Não há qualquer base científica para essa afirmação. |
| **44** | Deficiência de Minerais | Algumas dietas alternativas alegam que a falta de minerais como zinco, ferro e magnésio está relacionada ao autismo. Embora esses minerais sejam importantes para diversas funções do organismo, a causa do TEA está ligada a fatores genéticos e não à deficiência mineral. |
| **45** | Deficiência de Serotonina | Há alegações de que níveis baixos de serotonina no cérebro causam autismo. Embora haja diferenças nos níveis de serotonina em algumas pessoas com TEA, isso é um efeito da condição e não a sua causa. |
| **46** | Deficiência de Sulfato de Colesterol | Algumas teorias sugerem que a falta desse composto no cérebro interfere na mielinização dos neurônios e leva ao autismo. No entanto, não há comprovação científica de que a deficiência de sulfato de colesterol seja um fator causal do TEA. |
| **47** | Deficiência de Vitamina A | Narrativas afirmam que baixos níveis de vitamina A durante a gestação causam autismo. Embora essa vitamina seja essencial para o desenvolvimento fetal, sua deficiência pode causar cegueira noturna e outros problemas, mas não está ligada ao TEA. |
| **48** | Deficiência de Vitamina B1 | Algumas teorias afirmam que a falta de vitamina B1 (tiamina) prejudica o desenvolvimento neurológico e causa autismo. No entanto, a deficiência de B1 pode causar beribéri, mas não há evidências de que leve ao TEA. |
| **49** | Deficiência de Vitamina B2 | Há alegações de que a deficiência de B2 (riboflavina) está associada ao autismo. No entanto, essa vitamina tem papel na produção de energia celular e sua ausência pode causar fadiga e problemas dermatológicos, mas não está ligada ao TEA. |
| **50** | Deficiência de Vitamina B6 | Algumas comunidades afirmam que baixos níveis de B6 afetam neurotransmissores e levam ao autismo. Embora a B6 seja importante para a função cerebral, sua deficiência pode causar irritabilidade e neuropatia, mas não tem relação com o TEA. |
| **51** | Deficiência de Vitamina B8 | A biotina (B8) é essencial para o metabolismo celular, mas não há qualquer evidência de que sua deficiência cause autismo. |
| **52** | Deficiência de Vitamina B9 | Algumas teorias alegam que a deficiência de B9 (ácido fólico) leva ao TEA. Embora a suplementação de ácido fólico na gestação seja importante para prevenir defeitos do tubo neural, não há evidências de que sua ausência cause autismo. |
| **53** | Deficiência de Vitamina C | Alguns grupos alegam que a falta de vitamina C afeta o cérebro e leva ao autismo. No entanto, sua deficiência causa escorbuto, mas não está associada ao TEA. |
| **54** | Deficiência de Vitamina D | Narrativas afirmam que baixos níveis de vitamina D na gravidez ou na infância causam autismo. Embora essa vitamina seja importante para a saúde óssea e imunológica, não há evidências científicas que sustentem essa afirmação. |
| **55** | Deficiência de Vitamina E | Algumas teorias sugerem que a falta de vitamina E afeta o sistema nervoso e leva ao autismo. Embora essa vitamina tenha propriedades antioxidantes, sua deficiência pode causar danos neurológicos, mas não está ligada ao TEA. |
| **56** | Disfunção Mitocondrial | Algumas teorias sugerem que falhas na produção de energia das células levam ao autismo. Embora certas disfunções mitocondriais possam ser encontradas em indivíduos com TEA, elas são consequências do transtorno em alguns casos, e não sua causa. |
| **57** | Doença Autoimune | Há alegações de que reações autoimunes da mãe durante a gestação atacam o cérebro do feto e causam TEA. Embora doenças autoimunes possam influenciar o desenvolvimento fetal, não há evidências científicas de que sejam a causa do autismo. |
| **58** | Doritos | Algumas teorias afirmam que o milho transgênico e os conservantes presentes em salgadinhos como Doritos levam ao autismo. No entanto, não há qualquer comprovação científica de que alimentos processados ou transgênicos causem TEA. |

[ versão português / english above ]

| | | |
|---|---|---|
| **59** | Enxofre | Há alegações de que o enxofre presente em certos alimentos e suplementos interfere no desenvolvimento neurológico e causa autismo. O enxofre é um mineral essencial para diversas funções do corpo, não havendo qualquer relação com TEA. |
| **60** | Epigene P450 | Algumas comunidades pseudocientíficas alegam que alterações no sistema de enzimas P450, que metaboliza substâncias no fígado, estariam ligadas ao autismo. No entanto, não há evidências científicas que sustentem essa hipótese. |
| **61** | Escapamento do Carro | Algumas teorias sugerem que a exposição a poluentes dos gases de escape dos veículos afeta o neurodesenvolvimento e leva ao autismo. Embora a poluição do ar tenha impactos negativos na saúde, não há comprovação de que seja uma causa do TEA. |
| **62** | Extrato de Boi | Algumas dietas afirmam que compostos derivados da carne bovina interferem na função cerebral e levam ao autismo. No entanto, não há qualquer base científica para essa alegação. |
| **63** | Farinha branqueada | Defensores de dietas naturais alegam que agentes branqueadores usados na farinha de trigo afetam o cérebro e causam autismo. No entanto, não há estudos que comprovem essa relação. |
| **64** | Feldspato | Algumas comunidades alegam que a exposição ao feldspato, um mineral comum na crosta terrestre, causa autismo. No entanto, não há qualquer base científica para essa afirmação. |
| **65** | Fluoreto | Há teorias conspiratórias que afirmam que o flúor na água e em produtos odontológicos causa autismo ao afetar a função cerebral. No entanto, estudos científicos demonstram que o flúor é seguro e essencial para a saúde bucal. |
| **66** | Formaldeído | Algumas teorias afirmam que a exposição ao formaldeído em móveis, cosméticos e vacinas causa autismo. Embora o formaldeído em grandes quantidades seja tóxico, as doses presentes no ambiente e em vacinas são extremamente baixas e não apresentam risco de causar TEA. |
| **67** | Fosfato de Sódio Dibásico | Esse composto químico, usado em alimentos e medicamentos, é apontado como causa do autismo por supostamente afetar a química do cérebro. No entanto, não há nenhuma evidência científica que relacione o fosfato de sódio ao TEA. |
| **68** | Fosfato Monopotássico | Alegações afirmam que esse aditivo alimentar prejudica o sistema nervoso e leva ao autismo. No entanto, não há qualquer base científica que sustente essa afirmação. |
| **69** | Fumaça de Cigarro | Algumas narrativas sugerem que a exposição ao cigarro durante a gravidez causa autismo. Embora o tabagismo na gestação esteja associado a diversos problemas de saúde, não há evidências de que seja um fator causal para o TEA. |
| **70** | Fungos | Algumas teorias afirmam que infecções fúngicas no intestino ou em outras partes do corpo afetam o cérebro e levam ao autismo. No entanto, não há nenhuma comprovação científica que relacione infecções fúngicas ao TEA. |
| **71** | Gelatina | Algumas alegações afirmam que a gelatina, devido à presença de colágeno animal e aditivos, pode causar autismo. No entanto, não há qualquer evidência científica que relacione gelatina ao TEA. |
| **72** | Glutamato Monossódico (MSG) | O MSG, um realçador de sabor, é frequentemente alvo de desinformação, incluindo a alegação de que prejudica o cérebro e causa autismo. No entanto, estudos científicos mostram que seu consumo em níveis normais é seguro e não tem relação com TEA. |
| **73** | Glutationa | Algumas comunidades afirmam que baixos níveis de glutationa, um antioxidante natural, causam autismo. Embora a glutationa tenha funções importantes no organismo, não há evidência de que sua deficiência seja um fator causal do TEA. |
| **74** | Glúten | Muitas dietas alternativas defendem que o consumo de glúten pode causar ou agravar o autismo, mas não há evidências científicas que sustentem essa relação. O glúten pode ser problemático apenas para pessoas com doença celíaca ou sensibilidade ao glúten. |
| **75** | Grafeno | Teorias conspiratórias afirmam que partículas de grafeno em vacinas ou no ambiente causam autismo. No entanto, não há qualquer comprovação científica de que o grafeno seja prejudicial ao cérebro ou esteja presente em doses relevantes para causar danos. |
| **76** | HAARP | O programa de pesquisa de alta frequência da ionosfera (HAARP) é alvo de teorias da conspiração que alegam que ele altera a mente humana e provoca doenças, incluindo o autismo. Não há nenhuma base científica para essa teoria, sendo um exemplo clássico de desinformação. |
| **77** | Heptaidrato de Sulfato Ferroso | Algumas narrativas pseudocientíficas sugerem que este composto químico, usado em suplementação de ferro, poderia causar TEA. No entanto, ele é seguro e essencial para prevenir anemia, sem qualquer relação com autismo. |
| **78** | Hidrocloridrato de Cisteína | Há alegações de que esse aminoácido, usado em suplementos e alimentos, afeta a função cerebral e causa autismo. No entanto, a cisteína é necessária para a síntese de proteínas e não há evidências de que cause TEA. |
| **79** | Hidrocloridrato de Piridoxina (em Etanol) | Algumas teorias afirmam que a vitamina B6 na forma de piridoxina, especialmente quando diluída em etanol, pode causar autismo. No entanto, a vitamina B6 é essencial para funções neurológicas e sua deficiência, e não seu consumo, pode causar problemas neurológicos. |
| **80** | Hidrocloridrato de Tirosina | Algumas alegações afirmam que esse aminoácido essencial afeta neurotransmissores e causa TEA. No entanto, a tirosina é fundamental para a produção de dopamina e outros neurotransmissores, e não há relação com o autismo. |
| **81** | Hidróxido de Sódio | Também conhecido como soda cáustica, é usado na indústria alimentícia e farmacêutica em quantidades mínimas e seguras. Não há qualquer evidência de que cause TEA. |
| **82** | Inflamação Cerebral | Algumas teorias sugerem que a inflamação no cérebro é uma causa direta do autismo. Embora processos inflamatórios possam ocorrer em algumas condições neurológicas, não há evidências de que sejam a origem do TEA, que tem base genética. |
| **83** | Inflamação Intestinal | Algumas comunidades afirmam que distúrbios intestinais são a raiz do autismo. Embora problemas gastrointestinais sejam comuns em algumas pessoas com TEA, eles são condições associadas, e não causas do transtorno. |
| **84** | Inflamação por Metais | Teorias conspiratórias sugerem que a "toxicidade de metais pesados" causa inflamação cerebral e leva ao autismo. Embora metais pesados possam ser tóxicos em altas doses, não há evidências de que sejam a causa do TEA. |
| **85** | Infusão de Coração de Bezerro | Algumas narrativas pseudocientíficas afirmam que esse tipo de infusão pode alterar funções neurológicas e levar ao autismo. No entanto, não há qualquer base científica para essa alegação. |
| **86** | Inversão de Campo Magnético | Algumas teorias afirmam que mudanças no campo magnético da Terra influenciam o neurodesenvolvimento e causam TEA. No entanto, não há nenhuma evidência de que o magnetismo terrestre tenha efeito sobre o autismo. |
| **87** | Iodo | Algumas alegações afirmam que deficiência ou excesso de iodo durante a gestação causam autismo. Embora o iodo seja essencial para o desenvolvimento da tireoide, não há comprovação de que esteja envolvido no TEA. |
| **88** | Lactose | Há afirmações de que a lactose, presente no leite, pode causar ou piorar o autismo. No entanto, não há qualquer relação comprovada entre o consumo de lactose e o TEA. |
| **89** | Lantanídeos | Alguns conspiracionistas sugerem que esses metais raros, encontrados em eletrônicos, interferem na função cerebral e causam autismo. No entanto, não há qualquer evidência científica que sustente essa hipótese. |
| **90** | Lidocaína | Algumas teorias afirmam que a lidocaína, um anestésico comum, causa autismo quando usada em procedimentos médicos. No entanto, não há nenhum estudo que relacione a lidocaína ao TEA. |
| **91** | Materiais de Construção | Algumas alegações afirmam que substâncias presentes em tintas, cimento e madeira tratada causam autismo. Embora certos produtos químicos possam ser tóxicos, não há comprovação de que sejam fatores causais do TEA. |
| **92** | Mercúrio | A teoria de que o mercúrio causa autismo vem de desinformações sobre vacinas e amálgamas dentários. Estudos científicos mostram que o mercúrio presente em vacinas (timerosal) nunca esteve associado ao TEA. |
| **93** | Merenda Escolar | Algumas narrativas afirmam que os alimentos servidos em escolas contêm aditivos que causam autismo. Sustentam uma teoria da conspiração de que o Governo estaria transformando as crianças em autistas. No entanto, não há qualquer base científica para essa afirmação. |
| **94** | Mica | Algumas comunidades sugerem que a exposição a mica, um mineral usado em cosméticos e tintas, causa autismo. No entanto, não há nenhuma evidência que relacione mica ao TEA. |
| **95** | Microondas | Há teorias que afirmam que a radiação de fornos micro-ondas altera a estrutura celular e causa autismo. No entanto, a radiação dos micro-ondas é não ionizante e não tem efeitos no DNA ou no cérebro humano. |
| **96** | Milho | Algumas narrativas sugerem que o milho transgênico contém toxinas que afetam o cérebro e causam TEA. No entanto, não há qualquer estudo que comprove essa relação, sendo essa uma das muitas desinformações sobre alimentos geneticamente modificados. |
| **97** | N-Hexano | Algumas teorias afirmam que esse solvente, presente em adesivos e tintas, pode causar autismo ao afetar o sistema nervoso. Embora o N-hexano possa ser neurotóxico em exposições prolongadas e em altas doses, não há qualquer evidência científica que o relacione ao TEA. |
| **98** | Nanopartículas de Dióxido de Titânio | Alegações conspiratórias sugerem que essas nanopartículas, usadas em cosméticos e alimentos, atravessam a barreira hematoencefálica e afetam o cérebro. No entanto, não há estudos que associem a exposição ao dióxido de titânio ao autismo. |
| **99** | Nanopartículas de Sílica | Algumas comunidades afirmam que a sílica em nanopartículas, usada em suplementos e alimentos, pode interferir no desenvolvimento neurológico e causar TEA. No entanto, não há comprovação científica para essa alegação. |

[ versão português / english above ]

| | | |
|---|---|---|
| **100** | Niacina | Algumas narrativas afirmam que a deficiência ou excesso de niacina (vitamina B3) pode desencadear autismo. Embora a niacina seja essencial para diversas funções metabólicas, não há qualquer relação causal entre seus níveis e o TEA. |
| **101** | Óleo de Canola | Algumas teorias afirmam que o consumo de óleo de canola, devido à sua origem transgênica e processamento industrial, leva ao autismo. No entanto, não há evidências científicas de que óleos vegetais influenciem no desenvolvimento do TEA. |
| **102** | Óleo de Farelo de Arroz | Algumas alegações afirmam que esse óleo, por conter certos antioxidantes, pode interferir no desenvolvimento neurológico. No entanto, não há qualquer evidência científica que associe o óleo de farelo de arroz ao autismo. |
| **103** | Óleo de Girassol | Há alegações de que o consumo desse óleo, rico em ômega-6, pode aumentar inflamações e desencadear autismo. No entanto, não há estudos científicos que comprovem essa hipótese. |
| **104** | Óleo de Soja | Algumas teorias conspiratórias afirmam que o óleo de soja contém fitoestrogênios que afetam o cérebro em desenvolvimento e causam TEA. Embora a soja contenha compostos bioativos, não há qualquer evidência de que seu consumo cause autismo. |
| **105** | Opióides | Algumas narrativas afirmam que o uso de opioides durante a gestação pode causar autismo. Embora o uso de opioides na gravidez possa estar relacionado a efeitos adversos no desenvolvimento fetal, não há estudos conclusivos que os vinculem diretamente ao TEA. |
| **106** | Organoclorados | Algumas alegações sugerem que pesticidas organoclorados, como o DDT, são responsáveis pelo aumento dos casos de autismo. Embora esses compostos tenham efeitos tóxicos para o sistema nervoso em altas exposições, não há comprovação científica de que sejam uma causa direta do TEA. |
| **107** | Organofosforados | Há teorias que afirmam que pesticidas organofosforados, usados na agricultura, causam autismo ao afetar o sistema nervoso. Embora a exposição excessiva possa ter impactos neurológicos, não há evidência de que cause TEA. |
| **108** | Ovos | Algumas comunidades afirmam que o consumo de ovos, devido ao seu teor de colesterol ou resíduos de antibióticos, pode levar ao autismo. No entanto, não há qualquer evidência científica que relacione o consumo de ovos ao TEA. |
| **109** | Pantotenato de Cálcio (em Etanol) | Algumas alegações sugerem que a presença de etanol em suplementos de vitamina B5 pode causar autismo. No entanto, não há base científica para essa afirmação, pois o pantotenato de cálcio é um nutriente essencial para o metabolismo celular. |
| **110** | Paracetamol | Algumas teorias afirmam que o uso de paracetamol durante a gestação ou na infância pode desencadear TEA. No entanto, estudos científicos não demonstram qualquer relação causal entre o uso desse medicamento e o autismo. |

[ versão português / english above ]

| | | |
|---|---|---|
| **111** | Parasita | Uma das teorias mais perigosas e amplamente disseminadas em comunidades conspiratórias afirma que o autismo é causado por infecções parasitárias no organismo e que, por isso, poderia ser "curado" através de protocolos de desparasitação. Essa narrativa se baseia na ideia falsa de que parasitas, supostamente alojados no intestino ou no cérebro, liberariam toxinas que alteram o desenvolvimento neurológico da criança. Essas alegações levaram ao surgimento de práticas extremamente prejudiciais, como o uso de substâncias tóxicas para a chamada "limpeza parasitária". Entre os métodos mais perigosos, destacam-se o uso de dióxido de cloro (MMS/CDS), ivermectina em doses inadequadas, óleo de rícino, terapias com ervas altamente tóxicas, uso abusivo de enemas, e até mesmo protocolos baseados em turpentina (solvente industrial). Esses métodos não apenas não têm qualquer base científica, como podem levar a danos irreversíveis na saúde, incluindo intoxicação severa, lesões gastrointestinais, falência hepática e até óbito. |
| **112** | Pasta de Dentes | Há alegações de que substâncias presentes em cremes dentais, como flúor e triclosan, causam autismo. No entanto, o flúor é seguro nas quantidades recomendadas e essencial para a saúde bucal, sem qualquer relação com o TEA. |
| **113** | Pedras Desodorantes | Algumas narrativas sugerem que desodorantes naturais à base de alúmen de potássio podem liberar alumínio no corpo e causar TEA. No entanto, não há evidências científicas que sustentem essa hipótese. |
| **114** | Petróleo | Algumas teorias afirmam que a exposição a derivados do petróleo em plásticos e combustíveis causa autismo. Embora certos compostos químicos possam ter efeitos na saúde, não há comprovação de que o petróleo ou seus derivados causem TEA. |
| **115** | Placas de Cerâmica | Algumas narrativas sugerem que a inalação de partículas de cerâmica ou seus componentes químicos pode afetar o cérebro e levar ao autismo. No entanto, não há qualquer estudo que relacione a exposição a cerâmica ao TEA. |
| **116** | Plástico / Microplástico | Há alegações de que a ingestão ou exposição a microplásticos interfere no desenvolvimento neurológico e causa autismo. Embora os microplásticos sejam um tema de preocupação ambiental, não há estudos que comprovem que sejam uma causa do TEA. |
| **117** | Pó de Padaria | Algumas comunidades pseudocientíficas afirmam que ingredientes utilizados no panificação, como fermentos químicos, podem afetar o cérebro e causar autismo. No entanto, não há nenhuma base científica para essa afirmação. |
| **118** | Polietilenoglicol (PEG) | Algumas alegações afirmam que o PEG, utilizado como excipiente em medicamentos e vacinas, afeta a barreira hematoencefálica e causa autismo. No entanto, não há qualquer evidência científica que sustente essa relação. |
| **119** | Polisorbato | O polisorbato, um emulsificante presente em alimentos e vacinas, é frequentemente citado por conspiracionistas como uma substância que causa autismo. No entanto, estudos mostram que ele é seguro e não tem qualquer ligação com TEA. |
| **120** | Prolina | Algumas teorias afirmam que a prolina, um aminoácido não essencial, afeta neurotransmissores e leva ao autismo. No entanto, não há nenhuma evidência científica que comprove essa alegação. |
| **121** | Proteína Spike | Narrativas antivacinas afirmam que a proteína spike, produzida pelo organismo após a vacinação contra COVID-19, pode causar autismo. No entanto, não há qualquer evidência que ligue a vacina contra COVID-19 ao TEA. |

[ versão português / english above ]

| | | |
|---|---|---|
| **122** | Raio-X | Algumas teorias afirmam que a exposição a radiação de raio-X durante a gravidez pode causar autismo. Embora doses elevadas de radiação sejam prejudiciais, os exames médicos utilizam níveis seguros e não há comprovação de que causem TEA. |
| **123** | Refrigerante | Algumas alegações afirmam que o consumo de refrigerantes, especialmente os que contêm adoçantes artificiais, causa autismo. No entanto, não há qualquer evidência científica que sustente essa hipótese. |
| **124** | Riboflavina (em Etanol) | Algumas teorias afirmam que a vitamina B2 (riboflavina), quando diluída em etanol, pode ter efeitos no neurodesenvolvimento e causar autismo. No entanto, a riboflavina é essencial para o metabolismo celular e não há evidências de que cause TEA. |
| **125** | Sais Minerais | Algumas comunidades afirmam que desequilíbrios em minerais como sódio, potássio e magnésio levam ao autismo. Embora esses minerais sejam importantes para a saúde, não há comprovação de que estejam ligados ao TEA. |
| **126** | Semente de Amendoim | Algumas alegações afirmam que substâncias presentes no amendoim, como aflatoxinas, causam autismo. Embora aflatoxinas possam ser tóxicas em grandes quantidades, não há evidências de que sejam responsáveis pelo TEA. |
| **127** | Semente de Cártamo | Algumas narrativas afirmam que o óleo ou extrato da semente de cártamo interfere nos neurotransmissores e causa autismo. No entanto, não há qualquer base científica para essa afirmação. |
| **128** | Semente de Uva | Algumas teorias sugerem que substâncias antioxidantes da semente de uva afetam o cérebro e podem causar TEA. No entanto, não há nenhuma evidência científica que sustente essa alegação. |
| **129** | Sessecante | Algumas alegações afirmam que substâncias químicas utilizadas para remover umidade de alimentos e produtos podem causar autismo. No entanto, não há comprovação científica de que sessecantes tenham qualquer impacto no neurodesenvolvimento. |
| **130** | Soja | Algumas narrativas afirmam que os fitoestrogênios presentes na soja podem afetar o neurodesenvolvimento e causar autismo. No entanto, não há nenhuma evidência científica que relacione o consumo de soja ao TEA. Os fitoestrogênios são compostos naturais que não possuem impacto negativo no desenvolvimento neurológico. |
| **131** | Spray Nasal | Algumas teorias sugerem que descongestionantes nasais e outros sprays poderiam conter substâncias que afetam o cérebro e levam ao autismo. No entanto, não há estudos que demonstrem qualquer relação entre o uso de sprays nasais e o TEA. |
| **132** | Sulfato de Alumínio | Há alegações de que esse composto químico, usado no tratamento de água potável, pode causar autismo ao afetar o cérebro. No entanto, não há evidências científicas que sustentem essa relação, e os níveis de alumínio na água são rigidamente controlados para garantir segurança. |
| **133** | Sulfato de Amônia | Algumas comunidades afirmam que o sulfato de amônia, usado como fertilizante, pode estar presente em alimentos e desencadear autismo. No entanto, não há nenhuma comprovação científica que ligue esse composto ao TEA. |
| **134** | Sulfato de Magnésio Hidratado | Algumas alegações afirmam que esse composto, utilizado em suplementos e medicamentos, pode interferir no desenvolvimento cerebral. No entanto, ele é seguro e não há qualquer evidência de que cause TEA. |
| **135** | Sulfato de Potássio | Algumas teorias conspiratórias sugerem que fertilizantes contendo sulfato de potássio alteram a função cerebral e causam autismo. No entanto, não há nenhuma comprovação científica que sustente essa alegação. |
| **136** | Sulfato Ferroso | Há afirmações de que o sulfato ferroso, usado para tratar anemia, pode interferir no neurodesenvolvimento e causar TEA. No entanto, ele é essencial para a saúde e não há qualquer relação comprovada com o autismo. |
| **137** | Testosterona | Algumas teorias afirmam que níveis alterados de testosterona durante a gravidez aumentam o risco de autismo. Embora hormônios influenciem o desenvolvimento fetal, não há evidências conclusivas de que a testosterona seja um fator causal do TEA. |
| **138** | Tetrapak | Há alegações de que produtos embalados em Tetrapak contêm compostos químicos que causam autismo. No entanto, não há qualquer evidência científica que comprove essa relação. |
| **139** | Tiamina (em Etanol) | Algumas narrativas afirmam que a tiamina (vitamina B1) em solução de etanol pode afetar o cérebro e causar TEA. No entanto, a tiamina é essencial para o metabolismo energético, e não há evidências científicas que associem sua forma química ao autismo. |
| **140** | Timerosal (Derivado do Mercúrio, Etilmercúrio) | Essa é uma das alegações mais disseminadas entre movimentos antivacinas, sugerindo que o timerosal, um conservante utilizado em algumas vacinas no passado, causaria autismo. No entanto, estudos científicos extensivos demonstraram que não há qualquer relação entre o timerosal e o TEA, e ele já foi removido de quase todas as vacinas infantis por precaução. |
| **141** | Tolueno | Algumas alegações sugerem que a exposição ao tolueno, um solvente químico, pode causar autismo. Embora a exposição prolongada a solventes possa ser prejudicial à saúde, não há nenhuma evidência científica que relacione o tolueno ao TEA. |
| **142** | Transgênicos | Há teorias conspiratórias que afirmam que alimentos geneticamente modificados (OGMs) afetam o cérebro e causam autismo. No entanto, estudos científicos demonstram que transgênicos não têm qualquer relação com o TEA. |
| **143** | Tricloroetileno | Algumas alegações sugerem que a exposição a esse solvente industrial pode desencadear TEA. Embora o tricloroetileno possa ser tóxico em exposições elevadas, não há evidências de que seja um fator causal do autismo. |
| **144** | Tris base | Algumas comunidades pseudocientíficas sugerem que esse composto químico, usado em biologia molecular, pode afetar o desenvolvimento neurológico. No entanto, não há qualquer estudo que comprove essa hipótese. |
| **145** | Tris cloridrato | Algumas teorias sugerem que esse tampão químico, utilizado em laboratórios, poderia estar presente em produtos de consumo e causar autismo. No entanto, não há nenhuma evidência científica para essa alegação. |
| **146** | Tylenol | Assim como o paracetamol, há alegações de que o Tylenol causa autismo quando utilizado durante a gestação. No entanto, estudos científicos não demonstram qualquer relação causal entre esse medicamento e o TEA. |
| **147** | Vacinas | A alegação de que vacinas causam autismo é uma das mais perigosas e amplamente refutadas pela ciência, mas continua sendo promovida por grupos antivacinas e propagadores de desinformação. Essa teoria surgiu em 1998, quando o ex-médico Andrew Wakefield publicou um estudo fraudulento na revista The Lancet, alegando uma relação entre a vacina tríplice viral (MMR – sarampo, caxumba e rubéola) e o autismo. Esse estudo foi completamente desmentido, Wakefield perdeu sua licença médica por fraude, e a publicação foi retratada. No entanto, a desinformação gerada por essa farsa persiste e tem levado a uma crise global de hesitação vacinal, impactando diretamente a saúde pública. Diversos estudos científicos de larga escala, envolvendo milhões de crianças ao redor do mundo, comprovaram que não há qualquer relação entre vacinas e autismo. Algumas das pesquisas mais robustas foram conduzidas na Dinamarca, nos Estados Unidos e no Reino Unido, acompanhando crianças vacinadas e não vacinadas ao longo dos anos e não encontrando nenhuma diferença na incidência de TEA entre os grupos. Mesmo diante dessas evidências, grupos conspiracionistas continuam propagando mitos sobre a relação entre vacinas e autismo, muitas vezes associando o suposto risco ao timerosal, um conservante à base de mercúrio que já foi utilizado em algumas vacinas. No entanto, o tipo de mercúrio presente no timerosal (etilmercúrio) é rapidamente eliminado pelo organismo e não se acumula no cérebro, sendo completamente seguro. Além disso, o timerosal foi removido da maioria das vacinas infantis há décadas, sem qualquer impacto na prevalência do autismo, reforçando que não há relação entre os dois. |
| **148** | Vírus | Algumas comunidades afirmam que infecções virais, especialmente durante a gravidez, podem causar autismo. Embora certas infecções possam ter efeitos na gestação, não há evidências de que vírus sejam uma causa direta do TEA. |
| **149** | Wireless | Teorias conspiratórias afirmam que sinais de Wi-Fi e radiação eletromagnética de dispositivos móveis alteram a função cerebral e causam autismo. No entanto, as ondas de Wi-Fi são de baixa frequência e não ionizantes, não tendo qualquer efeito no neurodesenvolvimento. |
| **150** | WPO | Algumas alegações sugerem que produtos de limpeza industrial WPO contêm substâncias que desencadeiam autismo. No entanto, não há nenhuma evidência científica que sustente essa relação. |

Fonte: Elaboração própria (2025).

### 4.1.2. Falsas curas e tratamentos do autismo

Muitas mensagens exploram a fé das famílias, sugerindo que o autismo poderia ser superado por meio de devoção religiosa e mudanças de estilo de vida. Essas narrativas propagam a ideia de que crianças autistas podem ser "curadas" apenas por acreditar mais em Deus, promovendo um discurso de culpa sobre pais e responsáveis. Além disso, essas mesmas mensagens frequentemente sugerem afastamento da medicina convencional, indicando mudanças nutricionais extremas, restrição alimentar sem orientação profissional e abandono de eletrodomésticos como micro-ondas. O incentivo ao uso de "detox" para limpar o organismo de supostas toxinas ligadas ao autismo é particularmente perigoso, pois muitas dessas práticas envolvem substâncias prejudiciais, especialmente para crianças. A homeopatia também aparece como solução milagrosa, promovendo "isoterápicos" — tratamentos que alegam reverter os efeitos de vacinas e substâncias químicas. A recorrente negação das vacinas e o incentivo a terapias sem evidência colocam vidas em risco.

**Figura 07.** Exemplos de conspirações sobre curas do autismo.

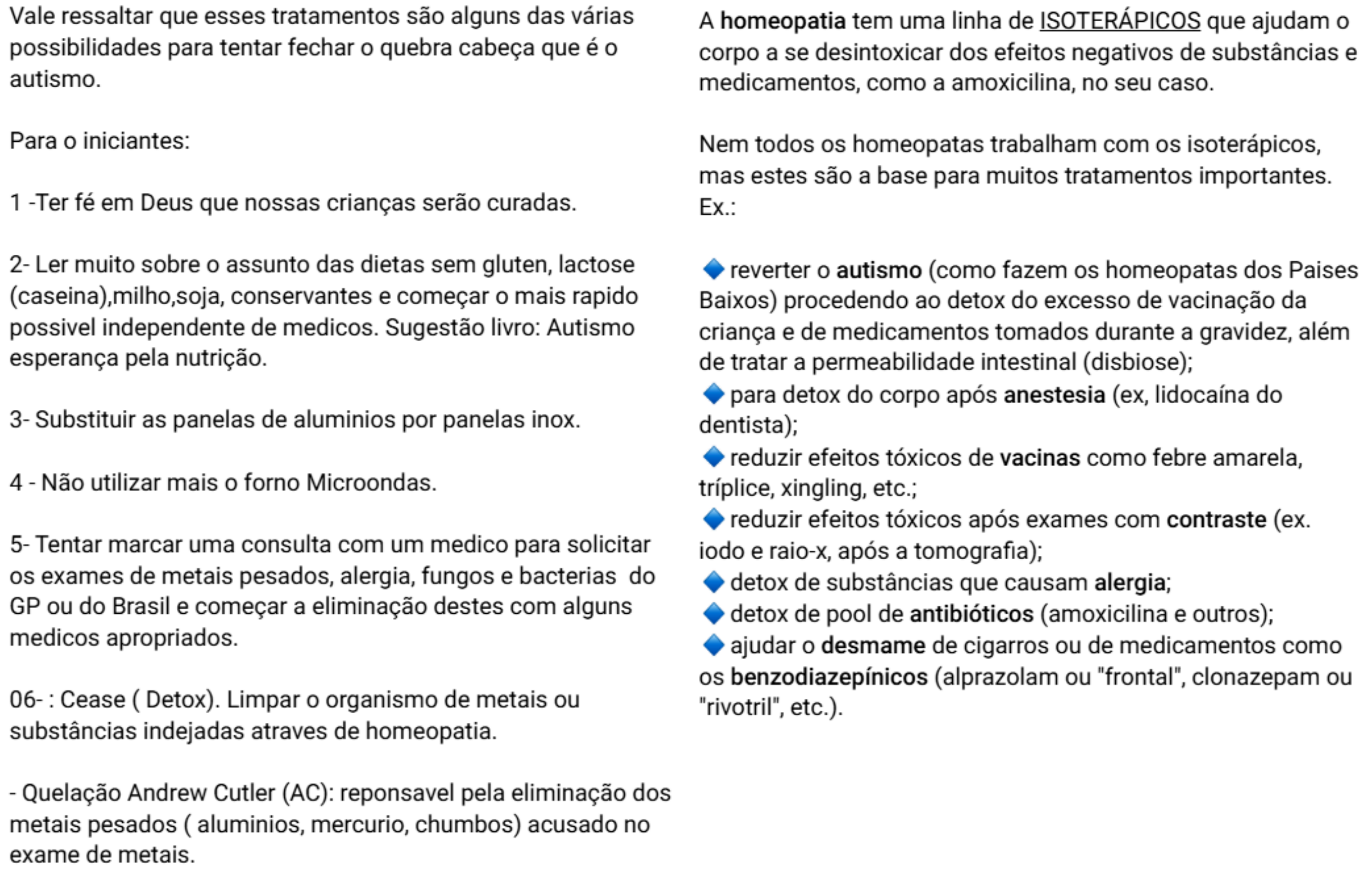

Vale ressaltar que esses tratamentos são alguns das várias possibilidades para tentar fechar o quebra cabeça que é o autismo.

Para o iniciantes:

1 -Ter fé em Deus que nossas crianças serão curadas.

2- Ler muito sobre o assunto das dietas sem gluten, lactose (caseina),milho,soja, conservantes e começar o mais rapido possivel independente de medicos. Sugestão livro: Autismo esperança pela nutrição.

3- Substituir as panelas de aluminios por panelas inox.

4 - Não utilizar mais o forno Microondas.

5- Tentar marcar uma consulta com um medico para solicitar os exames de metais pesados, alergia, fungos e bacterias do GP ou do Brasil e começar a eliminação destes com alguns medicos apropriados.

06- : Cease ( Detox). Limpar o organismo de metais ou substâncias indejadas atraves de homeopatia.

- Quelação Andrew Cutler (AC): reponsavel pela eliminação dos metais pesados ( aluminios, mercurio, chumbos) acusado no exame de metais.

A **homeopatia** tem uma linha de ISOTERÁPICOS que ajudam o corpo a se desintoxicar dos efeitos negativos de substâncias e medicamentos, como a amoxicilina, no seu caso.

Nem todos os homeopatas trabalham com os isoterápicos, mas estes são a base para muitos tratamentos importantes. Ex.:

🔹 reverter o **autismo** (como fazem os homeopatas dos Paises Baixos) procedendo ao detox do excesso de vacinação da criança e de medicamentos tomados durante a gravidez, além de tratar a permeabilidade intestinal (disbiose);
🔹 para detox do corpo após **anestesia** (ex, lidocaína do dentista);
🔹 reduzir efeitos tóxicos de **vacinas** como febre amarela, tríplice, xingling, etc.;
🔹 reduzir efeitos tóxicos após exames com **contraste** (ex. iodo e raio-x, após a tomografia);
🔹 detox de substâncias que causam **alergia**;
🔹 detox de pool de **antibióticos** (amoxicilina e outros);
🔹 ajudar o **desmame** de cigarros ou de medicamentos como os **benzodiazepínicos** (alprazolam ou "frontal", clonazepam ou "rivotril", etc.).

Fonte: Captura de tela (2025).

Outro eixo das falsas curas envolve a ideia de que o autismo seria causado pelo acúmulo de metais pesados e toxinas, que poderiam ser eliminadas por meio de "desintoxicação". Tratamentos com Ácido Etilenodiamino Tetra-Acético (EDTA) e Zeólita são promovidos como soluções eficazes para a remoção dessas substâncias do organismo, mesmo sem qualquer base científica para associá-las ao autismo. Além disso, esses grupos incentivam uma série de protocolos que incluem "detox de laticínios", consumo excessivo de aminoácidos, enzimas digestivas, infusões intravenosas de NAD+, uso de Orotato de Lítio e sessões com Oxigênio Hiperbárico, alegando que esses procedimentos reverteriam o autismo. Em muitos casos, esses produtos são vendidos diretamente pelas mesmas pessoas que propagam a desinformação, lucrando com o desespero das famílias. Uma substância particularmente perigosa promovida nesses círculos é a Suramina, um antiparasitário que, além de não ter qualquer comprovação científica, pode causar graves efeitos colaterais.

[ versão português / english above ]

**Figura 08.** Exemplos de conspirações sobre curas do autismo.

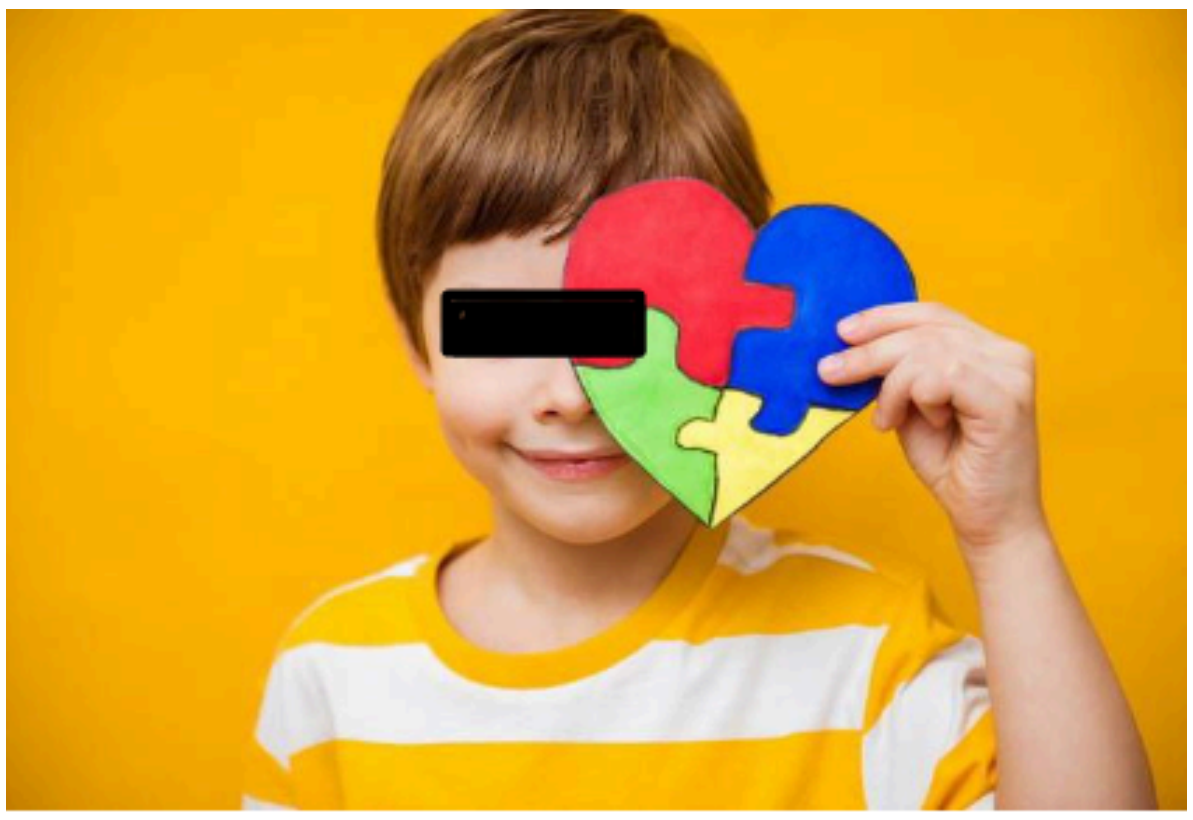

**Protocolo do doutor [redacted] para combater o autismo:**

- Desintoxicação (zeólita)
- Desintoxicação (EDTA)
- Tratamento com Oxigênio Hiperbárico
- NAD+ IV 500 mg
- Orotato de Lítio 20 mg
- Remover laticínios e glúten
- Fones de ouvido
- Mag Phos 1 mil
- Enzimas Digestivas
- Probióticos
- Aminoácidos
- CESSAR terapia
- NÃO VACINAR MAIS

SURAMINA

1 LITRO = 130,00+FRETE
#VENDAS NO MEU PRIVADO OU WATSAP [redacted]

**Pesquisadores testam droga que pode reverter sintomas do autismo**

**Os cientistas descobriram que medicamento corrige 17 tipos de anormalidades ligadas ao autismo, incluindo problemas de comportamento social**
**O que é esse medicamento?**
**A suramina é um fármaco antiparasitário desenvolvido nos laboratórios da Bayer em 1917 e utilizado para tratar infecções por tripanossoma no homem e em algumas espécies animais**
A Suramina, Ácido Shikimico, é um composto encontrado nas Agulhas de Pinheiro, o Antidoto para detox do corpo dos metais pesados

O ANTIDOTO DOS 100 ANOS -

O pinheiro é antioxidante, antidepressivo, antibacteriano, antiviral, antitumoral, anti-inflamatório, estimulador do sistema imunológico, protetor cardiovascular, triglicerídeo reduzindo e muito mais.
Suas substâncias também mata os parasitas e ajuda o autismo.

Fonte: Captura de tela (2025).

A promessa de cura do autismo por meio de produtos químicos como o Dióxido de Cloro (ClO2) — também conhecido como MMS (Miracle Mineral Solution) — é um dos casos mais graves de desinformação. Esse composto, promovido como uma panaceia para diversas doenças, na realidade, é um agente altamente tóxico, frequentemente comparado à ingestão de água sanitária. Em grupos conspiratórios, ele é vendido como uma solução para "limpar" o corpo, sendo aplicado em enemas ou administrado oralmente para crianças autistas, muitas vezes causando intoxicações severas. Além disso, outras substâncias, como cristais e quartzos, são comercializadas como auxiliares no tratamento do autismo, reforçando o aspecto mercadológico dessas comunidades. Os chamados "protocolos de desparasitação" também são preocupantes, pois baseiam-se na crença falsa de que o autismo seria causado por infecções parasitárias ocultas. Isso leva pais a submeterem seus filhos a tratamentos agressivos com antiparasitários sem necessidade, podendo resultar em sérios danos à saúde.

**Figura 09.** Exemplos de conspirações sobre curas do autismo.

**27 NIÑOS COLOMBIANOS CON AUTISMO, CURADOS CON CDS DIÓXIDO DE CLORO** 🙏🏽

Durante décadas hemos escuchado q el autismo es una enfermedad hereditaria e incurable, pero esto está demasiado alejado de la verdad

Desde q se inició con la vacunación en niños, esta enfermedad ha aumentado de manera exponencial

Hay muchos estudios q demuestran q estas "vacunas" son las q producen el autismo. La mal llamada ciencia es solo un negocio ☠️

El CDS llega a cualquier lugar del cuerpo por el torrente sanguíneo y produce la curación casi q de cualquier enfermedad

✅ **Tenemos Disponible:**

✔️DIÓXIDO DE CLORO A 3000PPM

✔️ZEOLITA MICRONIZADA CLINOPTILOLITA GRADO FARMACÉUTICO

✔️TIERRA DE DIATOMEAS GRADO ALIMENTICIO

✔️DMSO GRADO FARMACÉUTICO AL 99%

✔️AGUA DE MAR HIPERTONICA

✔️ ORMUS ORO MONOATÓMICO

✔️TREMENTINA RESINA DE PINO

✔️ GHEE ORGÁNICO

✔️ORGONITAS CONTRA ONDAS ELECTROMAGNÉTICAS, 5G

✔️CRISTALES CUARZOS

Fig. 31: Parásitos dentro de biofilm, también llamado de magma parasitario.
En la práctica en todos los niños afectados de autismo y en la mayoría de
las enfermedades crónicas, se ha podido ver una cantidad grande de mucosas, a
veces difícil de identificar ya que se asemeja a un Áscaris muerto o según dicen
algunos, a una mucosidad intestinal. Se encontraron mucosidades intestinales
por encima de 1 metro y por lo tanto es poco probable que sean mucosidades
del propio paciente. La Universidad de Bolonia, en Italia, afirma que es una
mucosidad propia del cuerpo. Sin embargo el Dr. Volinsky de la Universidad de
Florida ha podido hacer un análisis del ADN de la mucosidad, y opina que es
ajeno al cuerpo humano. Por lo tanto de momento, opino que es una forma de
'magma parasitario' no clasificado y por ello, tampoco aparece en los análisis de
los laboratorios. Las evidencias están dadas por los resultados.
Se han podido recuperar a más de 350 niños de autismo, basándose en este
protocolo, y todos expulsaron grandes cantidades de este plasma parasitario
(biofilm) y también otros parásitos. Después de cada expulsión, mejoraron
considerablemente. Lo mismo ocurre en muchas enfermedades crónicas,

Fonte: Captura de tela (2025).

O autismo também entra na lista de condições que supostamente seriam curadas pelas chamadas "Med Beds" — dispositivos fictícios que afirmam regenerar tecidos, curar doenças e até reverter a idade biológica das pessoas. A ideia dessas camas médicas é amplamente divulgada em grupos conspiratórios, prometendo eliminar doenças como câncer, Alzheimer e autismo por meio de tecnologias não comprovadas. Essas mensagens frequentemente apresentam um caráter quase messiânico, sugerindo que a "cura definitiva" do autismo já existe, mas estaria sendo ocultada por elites globais. Esse tipo de desinformação não apenas dá falsas esperanças a famílias vulneráveis, mas também cria um mercado lucrativo para venda de cursos, dispositivos e acessos a tratamentos inexistentes. Além disso, algumas variações dessas narrativas incluem o uso de enemas com substâncias tóxicas, como Dióxido de Cloro, reforçando práticas perigosas já amplamente condenadas por especialistas em saúde.

**Figura 10.** Exemplos de conspirações sobre curas do autismo.

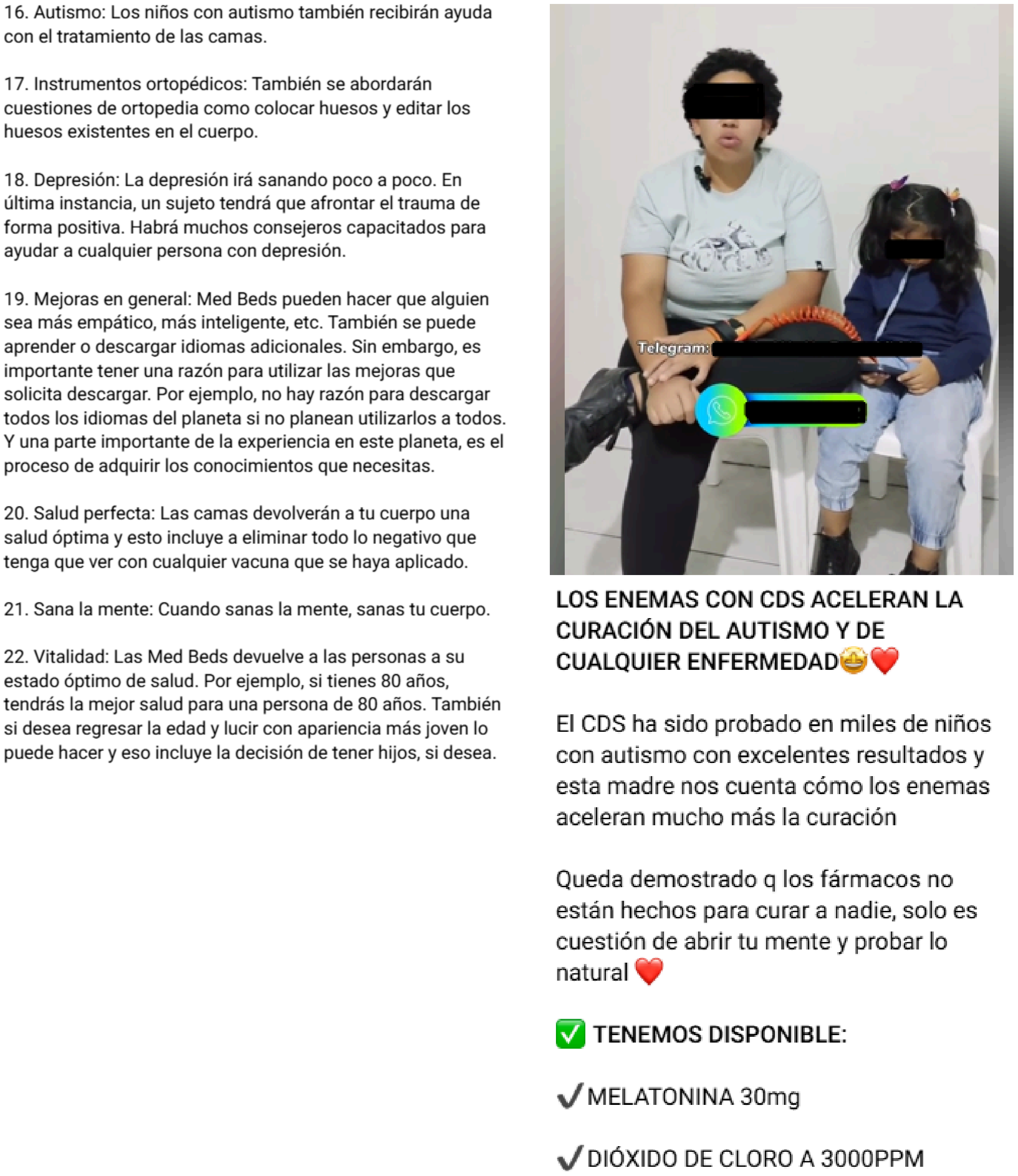

Fonte: Captura de tela (2025).

Em alguns grupos mais voltados ao "naturalismo", o autismo é tratado como um desequilíbrio energético que poderia ser corrigido com exposição a produtos químicos e luz infravermelha. Essas mensagens promovem a ideia de que o sol teria propriedades curativas específicas para o TEA, e que a combinação de certos suplementos vendidos pelos próprios anunciantes com "frequências eletromagnéticas" poderia restaurar o funcionamento do cérebro. Uma vertente ainda mais extrema inclui a promoção de dispositivos como "orgonites" e "biofiltros eletromagnéticos", alegando que ondas 5G e campos magnéticos artificiais são responsáveis pelo autismo. Além disso, algumas comunidades conspiratórias divulgam listas de antiparasitários como soluções para amenizar os sintomas, reforçando a desinformação já observada em protocolos de desparasitação. Essa abordagem pseudocientífica não apenas desvia a atenção de tratamentos baseados em evidências, mas também cria um mercado de produtos e terapias sem eficácia comprovada.

**Figura 11.** Exemplos de conspirações sobre curas do autismo.

O vermelho no arco iris curva a luz por ter mais eletrons mais massa do q o azul.

Libera na fricção a "massa"= eletrons = eletricidade porque somos seres elétricos, seres que podem se alimentar da luz do Sol. A Vita D a melanina e a melatonina não são as únicas expressões deste FATO.

Sal +água +luz e gordura animal = eletricidade =movimento.

Ainda estou testando ESSA COMBINAÇÃO Azul de Metileno e a Luz Infravermelha.

Mas já sei a resposta.

Entenda eu sou cientista.

Eu testo.

Para mim o clue é saber q ajuda no movimento acima de 4.000 metros de altura.

Carne e sol então

**Infravermelho é o Sol da manhã** e da tarde. Sem ele vc tem doenças. O Sol elimina o mofo q traz os íons negativos e impede a luz do sol vir a ser intramolecular.

Sem Sol não tem vida! Só mofo.

**Antiparasitários:**

1. Terebintina
2. CDS. Enemas de dióxido de cloro.
3. H2O2 10 gotas - 10 obtidas - 10 obtidas
4. Terapia com ozônio
5. Ártemis Anua
6. Infusão de Artemisa absentis (absinto)
7. Tintura de Nozes
8. Terra diatomácea de qualidade alimentar (não adequada para vermes ou oxiúros)
9. Comprimidos ou cápsulas de cravo
10. Limpeza do fígado
11. Juba de leão e Polyporus.. para proteger a microbiota do tratamento e melhorar a disbiose.
12. Clorofila líquida em parasitas hepáticos (dicrocoellium e fasciola)
13. Alho cru com o estômago vazio.
14. Extrato de cominho preto
15. Sementes de abóbora
16. Óleo de orégano
17. Casca de psyllium
18. Maria Treben. Saúde da farmácia, senhor.
19. Medicina Antroposófica. Carmelo Bizkarra
20. Óleos essenciais: canela, tea tree, louro, orégano grego.
21. As argilas Pancho, se ingeridas, também são antiparasitárias.
22. Disfania Ambrosioides (Epazote/Paico). Tome 2 folhas em infusão com o estômago vazio.

Fonte: Captura de tela (2025).

Em alguns grupos, o autismo é associado a um suposto "desalinhamento energético" que poderia ser corrigido com frequência de Tesla, eletrochoques e outras práticas sem qualquer fundamento científico. Algumas mensagens chegam ao absurdo de afirmar que atrás do cérebro existiria um "músculo cardíaco espiritual" que poderia ser "reativado" para curar o autismo. Essa narrativa mistura elementos esotéricos com promessas tecnológicas irreais, muitas vezes ligadas à venda de equipamentos ou sessões de "recalibração bioenergética". Essas teorias são especialmente preocupantes porque, ao afastar famílias de abordagens médicas baseadas em evidências, colocam crianças em risco ao promoverem práticas nocivas. A combinação de promessas milagrosas com mercantilização da esperança transforma essas comunidades em verdadeiros mercados de exploração emocional, onde a tragédia das famílias é explorada em benefício financeiro dos divulgadores dessas falsas curas.

[ versão português / english above ]

**Figura 12.** Exemplos de conspirações sobre curas do autismo.

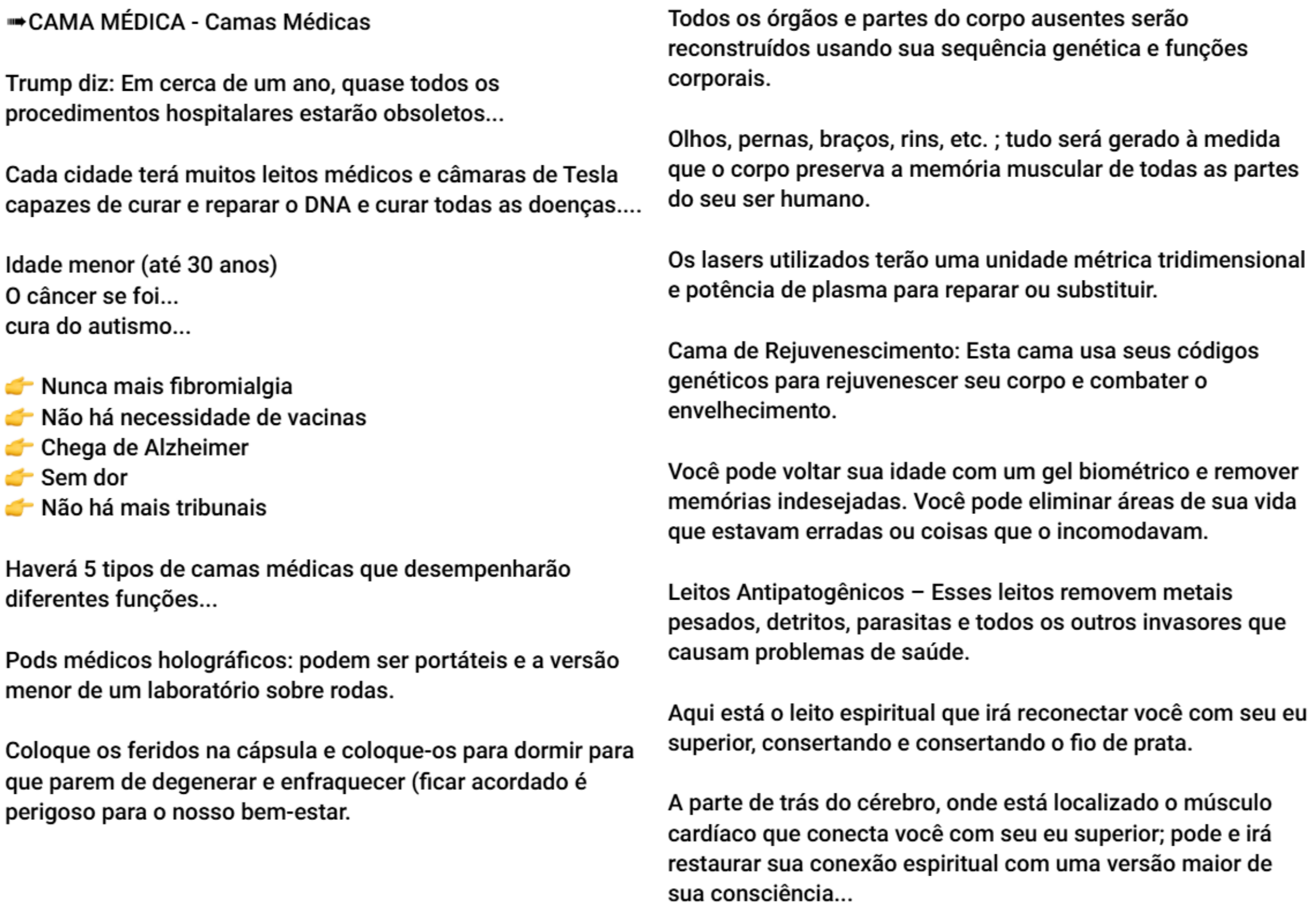

➡CAMA MÉDICA - Camas Médicas

Trump diz: Em cerca de um ano, quase todos os procedimentos hospitalares estarão obsoletos...

Cada cidade terá muitos leitos médicos e câmaras de Tesla capazes de curar e reparar o DNA e curar todas as doenças....

Idade menor (até 30 anos)
O câncer se foi...
cura do autismo...

👉 Nunca mais fibromialgia
👉 Não há necessidade de vacinas
👉 Chega de Alzheimer
👉 Sem dor
👉 Não há mais tribunais

Haverá 5 tipos de camas médicas que desempenharão diferentes funções...

Pods médicos holográficos: podem ser portáteis e a versão menor de um laboratório sobre rodas.

Coloque os feridos na cápsula e coloque-os para dormir para que parem de degenerar e enfraquecer (ficar acordado é perigoso para o nosso bem-estar.

Todos os órgãos e partes do corpo ausentes serão reconstruídos usando sua sequência genética e funções corporais.

Olhos, pernas, braços, rins, etc. ; tudo será gerado à medida que o corpo preserva a memória muscular de todas as partes do seu ser humano.

Os lasers utilizados terão uma unidade métrica tridimensional e potência de plasma para reparar ou substituir.

Cama de Rejuvenescimento: Esta cama usa seus códigos genéticos para rejuvenescer seu corpo e combater o envelhecimento.

Você pode voltar sua idade com um gel biométrico e remover memórias indesejadas. Você pode eliminar áreas de sua vida que estavam erradas ou coisas que o incomodavam.

Leitos Antipatogênicos – Esses leitos removem metais pesados, detritos, parasitas e todos os outros invasores que causam problemas de saúde.

Aqui está o leito espiritual que irá reconectar você com seu eu superior, consertando e consertando o fio de prata.

A parte de trás do cérebro, onde está localizado o músculo cardíaco que conecta você com seu eu superior; pode e irá restaurar sua conexão espiritual com uma versão maior de sua consciência...

Fonte: Captura de tela (2025).

**Quadro 02.** Lista de falsas curas e tratamentos do autismo mapeadas nas comunidades (A-Z).

| | Suposta cura e/ou tratamento | Exemplo |
|---|---|---|
| **1** | Ácido Alfa-Lipóico | Esse antioxidante é promovido como um "desintoxicante" que removeria metais pesados do corpo e "curaria" o autismo. No entanto, não há qualquer evidência científica de que o ácido alfa-lipóico tenha impacto nos sintomas do TEA. Seu uso excessivo pode causar efeitos colaterais, como queda drástica de açúcar no sangue e problemas neurológicos. |
| **2** | Ácido Chiquímico | Grupos pseudocientíficos afirmam que esse ácido, presente em algumas plantas, ajudaria na "reversão do autismo". No entanto, não há pesquisas científicas que sustentem essa alegação, e a crença de que compostos naturais podem "curar" o TEA ignora sua base neurodesenvolvimental e genética. |
| **3** | Ácido Cítrico | Há alegações de que o ácido cítrico presente em frutas cítricas ajudaria na desintoxicação e na "cura" do autismo. No entanto, o ácido cítrico é um composto comum na alimentação e não tem qualquer efeito sobre o TEA. Seu consumo excessivo pode causar irritação gástrica. |
| **4** | Ácido Etilenodiamino Tetra-Acético (EDTA) | O EDTA é usado na terapia de quelação, um tratamento legítimo para intoxicação por metais pesados, mas que é promovido de forma perigosa para crianças autistas. A quelação não trata nem cura o autismo e pode ser extremamente perigosa, causando danos renais, convulsões e até morte. |
| **5** | Ácido Fólico | Embora o ácido fólico seja essencial para o desenvolvimento fetal, não há nenhuma evidência de que sua suplementação "cure" o autismo. A deficiência dessa vitamina pode causar defeitos no tubo neural, mas não está associada ao TEA. |
| **6** | Ácido Gama Aminobutírico | Esse neurotransmissor é vendido como um suplemento para "acalmar" crianças autistas, mas não há comprovação científica de que o GABA tenha qualquer efeito terapêutico sobre o TEA. Além disso, a regulação de neurotransmissores no cérebro não é tão simples quanto tomar um suplemento. |
| **7** | Ácido Maleico | Supostamente promovido como "desintoxicante", o ácido maleico não tem qualquer relação com autismo e seu uso pode causar efeitos adversos gastrointestinais. |
| **8** | Ácido Ortosilícico (SiOH4) | Algumas alegações afirmam que esse composto ajudaria na "eliminação de metais pesados" e, consequentemente, no "tratamento" do autismo. No entanto, não há qualquer evidência científica que sustente essa ideia, e sua ingestão excessiva pode ser tóxica. |
| **9** | Aerochell | Esse termo parece estar relacionado a conspirações sobre "chemtrails" e substâncias químicas no ar. Algumas teorias afirmam que a exposição ao aerochell causaria autismo e que "purificações" com substâncias naturais reverteriam a condição. Isso é completamente falso e não há qualquer base científica nessa teoria da conspiração. |
| **10** | Água do Mar Hipertônica | Defensores dessa prática afirmam que a água do mar hipertônica "reprograma" o organismo e melhora o autismo. No entanto, não há estudos que demonstrem qualquer benefício real, e o consumo excessivo pode levar a desidratação, intoxicação por sódio e problemas renais. |
| **11** | Agulhas de Pinheiro | Algumas comunidades promovem extratos de agulhas de pinheiro como uma forma de "desintoxicar" o organismo e melhorar o autismo. No entanto, não há nenhuma comprovação científica dessa eficácia, e o consumo não supervisionado pode ser perigoso devido à presença de substâncias potencialmente tóxicas. |
| **12** | Alecrim | O alecrim é uma erva aromática utilizada na culinária e em alguns remédios naturais. Algumas alegações afirmam que ele melhora a "função cerebral" e reverte o autismo, mas não há estudos científicos que sustentem essa ideia. |
| **13** | Alho Cru | O alho é promovido como um "antibiótico natural" e "desintoxicante", sendo sugerido para eliminar "parasitas que causam autismo". Como já discutimos anteriormente, o autismo não é causado por parasitas, e comer alho cru não altera o neurodesenvolvimento. Além disso, o consumo excessivo pode causar irritação gastrointestinal severa. |
| **14** | Alho Macerado | Versão similar ao alho cru, promovida para "limpar" o organismo e eliminar supostas toxinas que causariam o autismo. Não há nenhuma evidência científica para essa alegação, e o excesso pode causar danos digestivos. |
| **15** | Alvejante | Essa é uma das "curas" mais perigosas e criminosas promovidas por grupos conspiracionistas. O dióxido de cloro, também conhecido como MMS (Miracle Mineral Solution) ou CDS (Chlorine Dioxide Solution), é um alvejante industrial altamente tóxico que esses grupos dizem "limpar o organismo" de substâncias que causariam autismo. Seu uso pode causar danos irreversíveis ao trato gastrointestinal, vômitos, diarreia severa, insuficiência hepática, problemas respiratórios e até morte. Agências de saúde, incluindo FDA (EUA), Anvisa (Brasil) e OMS, já emitiram alertas contra o uso do MMS/CDS, pois não há qualquer comprovação de que funcione para autismo, e sim provas de que é um veneno. |
| **16** | Aminoácidos | Embora aminoácidos sejam essenciais para o organismo, não há evidências de que a suplementação possa reverter ou tratar o autismo. Suplementos sem necessidade médica podem causar desequilíbrios metabólicos. |
| **17** | Argilas Pancho | Grupos conspiratórios afirmam que ingerir ou aplicar argilas "purificadoras" no corpo pode desintoxicar e curar o autismo. Isso não tem qualquer embasamento científico, e algumas argilas podem conter metais pesados que são tóxicos quando ingeridos. |
| **18** | Arginina | A arginina é um aminoácido essencial para diversas funções do corpo, mas não há nenhuma evidência científica de que sua suplementação tenha impacto no autismo. |
| **19** | Arruda | A arruda é promovida como um remédio natural para diversos problemas de saúde, incluindo o autismo. No entanto, não há qualquer evidência científica que comprove sua eficácia. Além disso, a arruda pode ser tóxica em grandes quantidades, causando danos hepáticos e irritação gastrointestinal severa. |
| **20** | Ártemis Anua | A planta Artemisia annua é usada no tratamento da malária, mas grupos pseudocientíficos a promovem como uma "cura natural" para o autismo. Não há nenhuma evidência científica que comprove essa alegação, e seu uso indevido pode ser tóxico. |
| **21** | Ashwagandha | A Withania somnifera, conhecida como ashwagandha, é uma erva usada na medicina ayurvédica para reduzir o estresse e melhorar a cognição. Alguns grupos afirmam que ela pode "regular neurotransmissores" e tratar o autismo. No entanto, não há evidências científicas de que a ashwagandha tenha qualquer efeito sobre o TEA. Além disso, o uso excessivo pode causar problemas gastrointestinais, sonolência intensa e interagir negativamente com medicamentos psiquiátricos. |
| **22** | Astaxantina | Esse antioxidante encontrado em algas e frutos do mar é promovido como um "protetor cerebral" que poderia ajudar no autismo. Embora antioxidantes sejam importantes para a saúde geral, não há evidências científicas de que a astaxantina tenha qualquer impacto no TEA. |
| **23** | Azul de Metileno | O azul de metileno é um corante químico com aplicações médicas, mas grupos pseudocientíficos afirmam que ele melhora a "oxigenação cerebral" e trata o autismo. Essa substância pode ser tóxica quando ingerida, causando danos neurológicos, pressão alta, arritmias cardíacas e problemas respiratórios. Seu uso não supervisionado pode ser extremamente perigoso. |

[ versão português / english above ]

| | | |
|---|---|---|
| **24** | Bed Meds | Esse termo pode se referir a diferentes medicamentos promovidos para uso antes de dormir, incluindo sedativos e ansiolíticos. Embora algumas pessoas autistas possam apresentar dificuldades para dormir, não há nenhuma "medicação para autismo", e qualquer prescrição deve ser feita por um profissional de saúde. O uso inadequado de sedativos pode levar a dependência, sonolência excessiva e risco de overdose. |
| **25** | Boro | Suplementos de boro são promovidos como uma forma de "melhorar a função cerebral" no autismo. No entanto, não há nenhuma evidência científica que sustente essa alegação, e o excesso de boro pode causar intoxicação, problemas digestivos e danos renais. |
| **26** | Brócolis | O brócolis é um alimento saudável e rico em antioxidantes, mas não é uma cura para o autismo. A alegação de que melhora o TEA vem de um estudo pequeno e preliminar sobre sulforafano, um composto presente no brócolis, mas não há comprovação robusta de que o consumo de brócolis tenha impacto no TEA. |
| **27** | Broto de Mostarda | Algumas dietas alternativas afirmam que brotos de mostarda "desintoxicam o corpo" e melhoram o autismo. No entanto, não há nenhuma evidência científica que comprove essa alegação. |
| **28** | Cálcio | O cálcio é essencial para a saúde óssea, mas não há qualquer relação entre deficiência de cálcio e autismo. O excesso de cálcio pode levar a problemas renais e calcificação de tecidos. |
| **29** | Caldo de Ossos | Promovido como uma "cura natural" para o autismo, o caldo de ossos é rico em colágeno e minerais, mas não há nenhuma evidência de que tenha impacto no TEA. Além disso, há preocupações sobre a presença de metais pesados, como chumbo, em caldos preparados a partir de ossos de animais. |
| **30** | Câmera Hiperbarica (HBOT) | A terapia com oxigênio hiperbárico envolve a inalação de oxigênio puro em alta pressão e é usada para tratar certas condições, como envenenamento por monóxido de carbono. No entanto, não há nenhuma evidência científica de que HBOT trate ou cure o autismo. O uso indevido pode ser perigoso, levando a danos pulmonares, convulsões e toxicidade por oxigênio. |
| **31** | Cardo Mariano | Essa erva é promovida como uma "desintoxicante do fígado", supostamente removendo substâncias que causariam autismo. No entanto, não há qualquer comprovação científica de que tenha impacto no TEA. |
| **32** | Carnitina | A carnitina é um composto envolvido no metabolismo energético e sua deficiência pode causar fraqueza muscular. No entanto, não há evidências de que sua suplementação tenha qualquer efeito sobre o autismo. |
| **33** | Carvão Ativado | O carvão ativado é promovido como um "desintoxicante" que removeria metais pesados e toxinas do organismo. No entanto, o carvão ativado pode interferir na absorção de nutrientes e medicamentos, além de causar obstrução intestinal e desidratação severa. Seu uso indevido pode ser perigoso. |
| **34** | Casca de Psyllium | O psyllium é uma fibra usada para regular o trânsito intestinal. Embora algumas pessoas autistas tenham problemas gastrointestinais, o psyllium não é uma cura para o autismo. O uso excessivo pode causar inchaço, gases e bloqueios intestinais. |
| **35** | Chá de Anis Estrelado | O anis estrelado contém substâncias que podem ser tóxicas em altas doses. Alguns grupos afirmam que ele "melhora o autismo", mas não há evidências científicas para essa alegação. O consumo excessivo pode causar convulsões, vômitos e efeitos neurotóxicos. |
| **36** | Chá de Funcho | O chá de funcho (erva-doce) é usado para aliviar problemas digestivos, mas não há evidências de que tenha qualquer efeito sobre o autismo. |
| **37** | Chelaton (Terapia de Quelação) | A terapia de quelação envolve o uso de substâncias como o EDTA para remover metais pesados do corpo. Grupos pseudocientíficos afirmam que isso "cura" o autismo, mas não há evidência científica que sustente essa alegação. Além disso, a quelação pode ser extremamente perigosa, causando danos renais, convulsões, alterações cardíacas e até morte. O FDA já alertou contra o uso da quelação para autismo. |
| **38** | Cisteína (N-Acetil-Cisteína) | A NAC é um antioxidante usado em algumas condições psiquiátricas, mas não há comprovação de que "cure" o autismo. |
| **39** | Citrato de Potássio | O citrato de potássio é usado para tratar problemas renais, mas não há nenhuma evidência de que tenha efeito sobre o TEA. O excesso pode causar arritmias cardíacas e problemas musculares. |
| **40** | Clorela (Chlorella Vulgaris) | A clorela é uma alga usada como suplemento e promovida como um "desintoxicante". No entanto, não há estudos que comprovem que ela trata o autismo. O uso excessivo pode causar problemas digestivos e reações alérgicas. |
| **41** | Clorofila Líquida | A clorofila líquida é promovida como um "desintoxicante" que removeria substâncias tóxicas associadas ao autismo. No entanto, não há nenhuma evidência científica que demonstre que a clorofila tenha qualquer impacto no TEA. Além disso, seu consumo excessivo pode causar fotossensibilidade e problemas digestivos. |
| **42** | Cobre | Algumas narrativas afirmam que a suplementação de cobre pode equilibrar neurotransmissores no autismo. No entanto, o excesso de cobre no organismo pode ser tóxico, causando danos hepáticos, neurológicos e metabólicos. |
| **43** | Coentro | O coentro é promovido como um "quelante natural" de metais pesados, alegadamente capaz de reverter o autismo. No entanto, não há nenhuma evidência científica de que ele tenha efeito sobre o TEA. Além disso, altas doses de coentro podem causar desconforto gastrointestinal. |
| **44** | Cominho Preto | O cominho preto (Nigella sativa) é utilizado na medicina tradicional para diversas finalidades, mas não há nenhuma evidência de que ele possa tratar ou curar o autismo. |
| **45** | Cravo | Assim como outras ervas promovidas como "purificadoras", o cravo não tem qualquer impacto comprovado sobre o TEA. Além disso, o óleo essencial de cravo pode ser tóxico em doses elevadas. |
| **46** | Creatina | A creatina é um suplemento usado para melhorar o desempenho muscular e energético, mas não há qualquer evidência de que trate ou melhore os sintomas do autismo. |
| **47** | Cristais / Quartzos | A crença de que cristais e quartzos podem equilibrar energias e curar o autismo é baseada em pseudociência. Não há qualquer evidência de que cristais tenham efeito terapêutico no TEA ou em qualquer condição médica. |
| **48** | Curcumina (Derivado da Cúrcuma) | A curcumina é um composto bioativo da cúrcuma com propriedades anti-inflamatórias. Embora estudos preliminares sugiram que possa ter benefícios gerais para a saúde, não há comprovação de que tenha efeito no tratamento do autismo. |
| **49** | Detox de Açúcares | A alegação de que eliminar açúcares da dieta "cura" o autismo não tem embasamento científico. Embora uma alimentação equilibrada possa melhorar a saúde geral, não há nenhuma evidência de que a retirada de açúcares trate o TEA. |
| **50** | Detox de Caseína | Essa prática envolve a eliminação de produtos lácteos da dieta sob a alegação de que a caseína interfere na função cerebral e causa autismo. No entanto, não há comprovação científica de que a caseína tenha qualquer relação com o TEA. |
| **51** | Detox de Glúten | Assim como no caso da caseína, a retirada do glúten da dieta é promovida como uma "cura para o autismo". Embora algumas pessoas autistas possam ter sensibilidade ao glúten, não há evidências de que a exclusão do glúten reverta ou trate o TEA. |
| **52** | Detox de Laticínios | A retirada de todos os laticínios da dieta como "tratamento" do autismo segue a mesma lógica do detox de caseína, sem qualquer evidência científica de eficácia. |
| **53** | Detox de Proteína Spike | Essa teoria antivacinas alega que pessoas que receberam vacinas contra COVID-19 precisam "eliminar" a proteína spike do organismo para evitar danos neurológicos, incluindo TEA. Não há qualquer base científica para essa teoria, e os produtos vendidos para essa "desintoxicação" podem ser prejudiciais à saúde. |
| **54** | Detox Vacinal | O "detox vacinal" é uma prática promovida por grupos antivacinas alegando que vacinas contêm toxinas que precisam ser eliminadas para "reverter o autismo". Essa prática não tem base científica e pode ser altamente perigosa, pois envolve o uso de substâncias tóxicas como dióxido de cloro (MMS/CDS), carvão ativado e quelantes agressivos. As vacinas NÃO causam autismo, e qualquer tratamento que afirme "desintoxicar" vacinas é uma farsa perigosa. |
| **55** | Dieta | Embora uma alimentação saudável seja essencial para o bem-estar geral, não existe "dieta para curar o autismo". Dietas restritivas sem acompanhamento médico podem causar deficiências nutricionais graves. |
| **56** | Dimetilglicina | Esse suplemento é promovido como um "melhorador cognitivo" no autismo. No entanto, não há evidências científicas que comprovem sua eficácia. |
| **57** | Dióxido de Cloro | O dióxido de cloro, também conhecido como MMS (Miracle Mineral Solution) ou CDS, é um alvejante industrial altamente tóxico que conspiracionistas promovem como "cura para o autismo". Seu uso pode causar danos irreversíveis ao trato gastrointestinal, insuficiência hepática, problemas respiratórios, convulsões e até morte. Organizações como OMS, FDA e Anvisa já alertaram contra essa prática perigosa. |
| **58** | Dióxido Vibracional Mono Atômico | Esse nome não tem qualquer base na ciência e é promovido por vendedores de falsas curas. Não existe comprovação de que essa substância tenha qualquer efeito sobre o autismo. |
| **59** | Disfania Ambrosioides (Epazote/Paico) | Essa erva é tradicionalmente usada como vermífugo e alguns grupos afirmam que ela pode "curar o autismo removendo parasitas". Como já discutimos anteriormente, o TEA não é causado por parasitas, e o uso indevido dessa erva pode ser tóxico para o fígado. |
| **60** | EGCG (Catequina do Chá-Verde) | O EGCG, um antioxidante do chá-verde, é promovido como uma substância que "melhora o autismo". Embora seja um composto saudável, não há nenhuma evidência científica de que tenha impacto sobre o TEA. |

[ versão português / english above ]

| | | |
|---|---|---|
| **61** | Enemas | Os enemas são promovidos por grupos conspiracionistas como uma forma de "remover toxinas" ou "eliminar parasitas que causam autismo". Algumas práticas incluem o uso de enemas de café, dióxido de cloro (MMS/CDS), bicarbonato de sódio e ervas agressivas. O uso frequente e sem supervisão médica pode causar irritação intestinal severa, desidratação, inflamações, sangramentos e perfuração do reto, colocando a vida da criança em risco. Não há qualquer base científica que justifique o uso de enemas para tratar o TEA. |
| **62** | Enzimas Digestivas | Algumas dietas promovem o uso de enzimas digestivas para "melhorar a absorção de nutrientes" em autistas. Embora enzimas possam ser benéficas para pessoas com problemas gastrointestinais específicos, não há evidências de que elas tenham qualquer impacto sobre o autismo. |
| **63** | Extrato de Cominho Preto | O cominho preto é promovido como um "anti-inflamatório natural" que supostamente poderia reduzir os sintomas do TEA. No entanto, não há nenhuma comprovação científica que sustente essa ideia. |
| **64** | Folha de Dente-de-Leão | O dente-de-leão é usado em práticas de medicina alternativa para "purificar o fígado", mas não há evidência de que tenha qualquer efeito sobre o autismo. |
| **65** | Gengibre | O gengibre tem propriedades antioxidantes e anti-inflamatórias, mas não há estudos que comprovem que ele tenha impacto no TEA. |
| **66** | Gliadina | A gliadina é uma proteína presente no glúten, e alguns grupos afirmam que ela está ligada ao autismo. Embora pessoas com doença celíaca possam se beneficiar de uma dieta sem glúten, não há comprovação de que a retirada da gliadina melhore ou cure o TEA. |
| **67** | Glicina | A glicina é um aminoácido envolvido na neurotransmissão, e algumas teorias afirmam que sua suplementação pode "corrigir deficiências cerebrais" em autistas. No entanto, não há evidências científicas que comprovem essa relação. |
| **68** | Glifosato | Grupos conspiracionistas afirmam que o glifosato, um herbicida usado na agricultura, estaria ligado ao autismo. Embora o impacto de agrotóxicos na saúde seja um tema relevante, não há estudos científicos que provem que o glifosato cause TEA. |
| **69** | Glutationa Injetável | A glutationa é um antioxidante natural, e alguns grupos promovem sua aplicação injetável como uma "terapia de desintoxicação" para o autismo. Não há nenhuma evidência científica que comprove essa prática, e a administração de glutationa injetável sem necessidade pode causar efeitos colaterais, incluindo reações alérgicas e desequilíbrios metabólicos. |
| **70** | Groselha-Negra | A groselha-negra é promovida como um antioxidante, mas não há evidências científicas que indiquem que seu consumo tenha qualquer impacto no TEA. |
| **71** | Hesperidina | A hesperidina, um flavonoide encontrado em frutas cítricas, é promovida como um "neuroprotetor". No entanto, não há estudos que sustentem sua eficácia no autismo. |
| **72** | Hidroxicloroquina | Após a Pandemia da COVID-19, a hidroxicloroquina foi promovida como um suposto tratamento para várias condições, incluindo o autismo. Não há nenhuma evidência científica que comprove essa alegação, e o uso indiscriminado desse medicamento pode causar arritmias cardíacas, problemas hepáticos e danos oculares. |
| **73** | Homeopatia | Tratamentos homeopáticos são amplamente promovidos para diversas condições, incluindo o autismo. No entanto, não há nenhuma base científica que comprove sua eficácia. A homeopatia se baseia na diluição extrema de substâncias, o que significa que os "remédios" homeopáticos não contêm ingredientes ativos em quantidades detectáveis. |
| **74** | Ibuprofeno Inalado | Algumas teorias sugerem que o ibuprofeno inalado pode "reduzir a inflamação cerebral" e tratar o autismo. Não há nenhuma evidência científica que sustente essa ideia, e a inalação de ibuprofeno pode ser extremamente tóxica para os pulmões. |
| **75** | Infusão de Artemisa absentis (Absinto) | O absinto é uma planta usada na medicina herbal, mas seu uso em altas doses pode ser tóxico. Algumas alegações sugerem que ele "limpa o organismo" e melhora o autismo, mas não há qualquer comprovação científica para essa alegação. Além disso, o absinto contém tuiona, uma substância que pode causar convulsões e danos neurológicos. |
| **76** | Ivermectina | A ivermectina, promovida erroneamente durante a Pandemia da COVID-19, também é sugerida como um tratamento para o autismo. Alguns grupos afirmam que ela eliminaria "parasitas que causam TEA", mas essa teoria é completamente falsa, pois o autismo não é causado por infecções parasitárias. O uso inadequado de ivermectina pode levar a danos hepáticos, convulsões e até coma. |
| **77** | Juba de Leão | Esse cogumelo medicinal é promovido como um "regenerador neuronal" que poderia melhorar o autismo. Embora haja estudos preliminares sobre seu impacto na cognição, não há nenhuma evidência robusta de que ele tenha efeito no TEA. |
| **78** | Lactoferrina | A lactoferrina é uma proteína presente no leite materno e em alguns suplementos. Algumas alegações afirmam que sua suplementação poderia "regular o sistema imunológico" e tratar o autismo, mas não há comprovação científica para essa afirmação. |
| **79** | Leite de Camelo | O leite de camelo é promovido como um tratamento para autismo por alguns grupos alternativos, devido à sua composição rica em proteínas e anticorpos. No entanto, não há nenhuma evidência científica de que o consumo desse leite tenha impacto no TEA. Além disso, seu consumo não pasteurizado pode representar riscos à saúde, como infecções bacterianas. |
| **80** | Limão | Algumas narrativas sugerem que o consumo de limão "desintoxica o organismo" e melhora o autismo. Embora seja um alimento saudável, não há evidência de que o limão tenha qualquer impacto no TEA. |
| **81** | Limpeza do Fígado | A "limpeza do fígado" é uma prática alternativa que afirma eliminar toxinas que causariam autismo. Esse método geralmente envolve o consumo de grandes quantidades de óleos, sais de magnésio e sucos cítricos, o que pode causar diarreia severa, desidratação e distúrbios metabólicos. Não há qualquer evidência científica de que o autismo seja causado por toxinas hepáticas. |
| **82** | Losna | A Artemisia absinthium, conhecida como losna, é promovida por alguns grupos como um vermífugo "capaz de curar o autismo". No entanto, não há nenhuma base científica que comprove essa alegação. Além disso, a losna contém tuiona, um composto que pode causar convulsões, danos hepáticos e efeitos neurotóxicos. |
| **83** | Luz Infravermelha | Há alegações de que a terapia com luz infravermelha melhora "funções cerebrais" em autistas. Embora a luz infravermelha tenha aplicações médicas legítimas, não há evidências científicas de que ela tenha impacto no TEA. |
| **84** | Maçã | Maçãs são alimentos saudáveis, mas não curam autismo. Algumas alegações afirmam que seu teor de pectina ajuda a "desintoxicar metais pesados", mas não há qualquer evidência científica que sustente essa teoria. |
| **85** | Magnésio | O magnésio é um mineral essencial, e algumas alegações sugerem que a deficiência desse nutriente causaria autismo. Não há comprovação científica de que a suplementação de magnésio tenha qualquer efeito no TEA. O excesso de magnésio pode causar diarreia, queda da pressão arterial e arritmias cardíacas. |
| **86** | Manteiga GHEE Orgânica | O GHEE é uma gordura clarificada usada na culinária e na medicina ayurvédica. Algumas narrativas afirmam que ele "melhora o cérebro e trata o autismo", mas não há estudos que sustentem essa ideia. |
| **87** | Mebendazol | O mebendazol é um vermífugo, e algumas comunidades pseudocientíficas afirmam que "eliminar parasitas cura o autismo". Essa teoria é completamente falsa, pois o autismo não é causado por infecções parasitárias. O uso inadequado de mebendazol pode causar problemas hepáticos e distúrbios gastrointestinais severos. |
| **88** | Medicina Antroposófica | A medicina antroposófica é uma abordagem alternativa baseada em conceitos espirituais e holísticos. Embora algumas terapias possam ser úteis como suporte, não há evidências científicas de que a medicina antroposófica trate ou cure o autismo. |
| **89** | Melatonina Transdérmica | A melatonina é usada para regular o sono e pode ser benéfica para pessoas autistas com dificuldades para dormir. No entanto, não é uma cura para o TEA. Além disso, o uso transdérmico (absorvido pela pele) não tem comprovação científica de eficácia. |
| **90** | Microbioma / Transplante Fecal | Algumas pesquisas sugerem que o microbioma intestinal pode influenciar aspectos do neurodesenvolvimento, levando grupos pseudocientíficos a promover o transplante fecal como uma cura para o autismo. Embora o transplante fecal seja um tratamento legítimo para infecções intestinais graves, não há evidências científicas que comprovem sua eficácia para o TEA. Além disso, esse procedimento pode ser extremamente perigoso, expondo os pacientes a infecções e complicações graves. |
| **91** | NAC (N-Acetilcisteína) | A N-acetilcisteína é um antioxidante estudado para várias condições psiquiátricas, e alguns estudos exploram seu potencial para autismo. No entanto, não há evidência de que seja uma cura para o TEA, e seu uso deve ser feito apenas sob supervisão médica. |
| **92** | NAD+ IV | O NAD+ (nicotinamida adenina dinucleotídeo) é promovido por clínicas de terapias alternativas como uma "cura celular" para o autismo. No entanto, não há nenhuma evidência científica que sustente essa alegação. Além disso, infusões intravenosas de NAD+ podem causar efeitos colaterais severos, como pressão arterial baixa e reações adversas. |
| **93** | Nim-da-Índia (Neem) | O óleo de nim é promovido como um "limpador do organismo" e "cura para o autismo", mas não há qualquer evidência científica que comprove essa afirmação. O consumo excessivo de neem pode ser tóxico para o fígado e causar distúrbios gastrointestinais graves. |
| **94** | Óleo de Avestruz | O óleo de avestruz é promovido por alguns grupos como um suplemento anti-inflamatório para o TEA. No entanto, não há estudos que comprovem sua eficácia no tratamento do autismo. |
| **95** | Óleo de Canela | Óleo de canela é promovido como um "regulador metabólico" que poderia tratar o autismo. No entanto, não há nenhuma evidência científica que sustente essa alegação. Além disso, o consumo excessivo pode causar irritação gástrica e reações alérgicas. |
| **96** | Óleo de Coco | Embora seja uma gordura saudável, o óleo de coco não tem nenhuma propriedade que trate ou cure o autismo. O uso excessivo pode levar a ganho de peso e aumento do colesterol. |
| **97** | Óleo de Louro | O óleo de louro é promovido por alguns grupos como um purificador do organismo, mas não há evidências científicas que sustentem essa afirmação. |
| **98** | Óleo de Orégano | O óleo de orégano tem propriedades antimicrobianas e antifúngicas, mas não há nenhuma comprovação de que ele tenha impacto no autismo. O uso excessivo pode causar problemas gastrointestinais e irritação nas mucosas. |
| **99** | Orgonites (Cristais) | Orgonites são objetos de cristal e resina promovidos como "reguladores de energia" para "curar o autismo". Isso não tem qualquer base científica e faz parte de crenças esotéricas sem eficácia comprovada. |
| **100** | ORMUS (Cristais de Sal Marinho e Sal do Himalaia Triturados) | ORMUS é um conceito pseudocientífico que sugere que sais marinhos triturados contêm "elementos monoatômicos" com propriedades quânticas que curariam o autismo. Isso é uma fraude sem qualquer embasamento científico. Não há nenhuma evidência de que ORMUS tenha qualquer efeito no TEA. |
| **101** | Orotato de Lítio | O lítio é usado em tratamentos psiquiátricos para transtorno bipolar, e algumas alegações sugerem que sua versão em orotato poderia "corrigir desequilíbrios cerebrais" em autistas. Não há qualquer evidência científica que sustente essa afirmação, e o uso inadequado de lítio pode causar toxicidade grave, danos renais e alterações cardíacas. |
| **102** | Ouro Coloidal | O ouro coloidal é promovido como um "melhorador cognitivo" para o autismo, mas não há evidências científicas que comprovem essa alegação. O consumo excessivo pode causar acúmulo de metais no organismo e toxicidade. |
| **103** | Ouro Monoatômico | Esse conceito, popular em grupos esotéricos, afirma que "partículas especiais de ouro" podem alterar a consciência e melhorar a cognição de autistas. Isso não tem nenhuma base científica e é uma fraude. |
| **104** | Óxido Nítrico | O óxido nítrico é um gás que ajuda na regulação vascular e neurológica, e algumas alegações sugerem que sua suplementação poderia melhorar o autismo. No entanto, não há estudos científicos que sustentem essa teoria, e o uso excessivo pode causar alterações na pressão arterial e riscos cardiovasculares. |
| **105** | Oxigênio Hiperbárico | A terapia com oxigênio hiperbárico envolve a inalação de oxigênio puro sob alta pressão e é promovida como uma cura para o autismo. Não há qualquer evidência científica que comprove essa eficácia, e o uso indevido pode levar a convulsões, toxicidade por oxigênio e danos pulmonares. |
| **106** | Ozônio | A terapia com ozônio é promovida por grupos pseudocientíficos como uma forma de "purificar o organismo", mas não há qualquer comprovação de que isso funcione para o autismo. O ozônio pode ser tóxico quando inalado ou injetado, causando danos pulmonares, inflamações graves e risco de embolia. |
| **107** | Peróxido de Hidrogênio (H2O2) | O uso de peróxido de hidrogênio para "curar autismo" é uma das terapias mais perigosas e irresponsáveis promovidas por grupos alternativos. Algumas pessoas incentivam o consumo oral ou a injeção dessa substância, alegando que mataria parasitas ou eliminaria "toxinas". Isso é uma prática extremamente perigosa que pode causar queimaduras internas, danos severos ao trato digestivo e até levar à morte. |
| **108** | Picnogenol | O picnogenol é um antioxidante natural extraído da casca do pinheiro marítimo. Embora tenha propriedades antioxidantes, não há evidências científicas que comprovem seu efeito no autismo. |
| **109** | Polyporus | Esse cogumelo medicinal é promovido por algumas comunidades como um modulador do sistema imunológico para autistas, mas não há comprovação científica de que tenha qualquer efeito no TEA. |
| **110** | Prata Coloidal | A prata coloidal é promovida como um "antisséptico natural" capaz de eliminar agentes que causariam autismo. No entanto, não há qualquer evidência científica que comprove essa alegação, e o consumo prolongado pode causar intoxicação grave e acúmulo de prata nos tecidos (argiria), deixando a pele azulada permanentemente. |
| **111** | Probióticos | Embora o microbioma intestinal possa ter influência em aspectos da saúde geral, não há evidências de que probióticos curem o autismo. Alguns estudos sugerem que podem ajudar em sintomas gastrointestinais associados ao TEA, mas não alteram a condição neurodesenvolvimental. |
| **112** | Protocolos de Desintoxicação | Os "protocolos de desintoxicação" incluem práticas como quelação, enemas agressivos, dietas restritivas e consumo de substâncias perigosas como dióxido de cloro (MMS/CDS). Esses métodos não têm base científica e podem ser extremamente nocivos, causando intoxicação, desidratação severa e até falência hepática. |
| **113** | Protocolos de Desparasitação | Os "protocolos de desparasitação" partem da falsa premissa de que o autismo é causado por infecções parasitárias. Esses tratamentos incluem o uso de vermífugos perigosos, enemas de dióxido de cloro e dietas extremas. O uso excessivo de antiparasitários pode causar danos hepáticos, neurotoxicidade e desequilíbrios graves no organismo. |
| **114** | Prunella Vulgaris (Erva Auto-Curável) | Essa planta medicinal é promovida como um "modulador do sistema imunológico" para autistas. No entanto, não há estudos científicos que sustentem essa alegação. |
| **115** | Pulseira de Cobre de Neodímio | Algumas pessoas acreditam que pulseiras magnéticas de cobre melhoram a função cerebral e curam o autismo. Isso não tem nenhuma base científica e se trata apenas de um placebo sem efeito real. |
| **116** | Quercetina | A quercetina é um flavonoide antioxidante encontrado em frutas e vegetais. Algumas alegações sugerem que sua suplementação ajudaria no autismo, mas não há estudos científicos robustos que comprovem essa relação. |
| **117** | Resina de Pinus | A resina de pinheiro é promovida como um "desintoxicante natural", mas não há nenhuma evidência científica de que ela tenha qualquer impacto no autismo. Além disso, seu consumo inadequado pode ser tóxico e causar reações alérgicas graves. |
| **118** | Rotenona | A rotenona é um pesticida natural que, paradoxalmente, já foi associada ao aumento do risco de doenças neurodegenerativas. Algumas teorias sugerem que sua exposição pode estar relacionada a sintomas de autismo, mas não há evidências científicas de que seja um tratamento para o TEA. A rotenona pode ser altamente tóxica quando ingerida, causando danos neurológicos irreversíveis. |
| **119** | Selênio | O selênio é um mineral essencial para o metabolismo, mas não há evidências de que sua suplementação trate ou cure o autismo. O excesso de selênio pode ser tóxico, causando queda de cabelo, problemas gastrointestinais e danos neurológicos. |
| **120** | Sementes de Abóbora | Algumas alegações sugerem que as sementes de abóbora eliminam "parasitas causadores de autismo". Essa afirmação não tem base científica, e o consumo de sementes de abóbora não tem qualquer efeito sobre o TEA. |
| **121** | Serotonina | Algumas teorias sugerem que o autismo estaria relacionado a um "desequilíbrio de serotonina" e que a suplementação ou manipulação desse neurotransmissor poderia curá-lo. Embora medicamentos que atuam na serotonina possam ser prescritos para tratar sintomas específicos, não há nenhuma evidência de que serotonina cure o autismo. A administração inadequada de substâncias que alteram a serotonina pode causar síndrome serotoninérgica, levando a confusão mental, convulsões e arritmias cardíacas. |
| **122** | Silício | O silício é um mineral essencial para a formação óssea, mas não há nenhuma evidência de que ele tenha qualquer efeito sobre o autismo. |
| **123** | Sulfato de Colesterol | Algumas alegações sugerem que o sulfato de colesterol regula processos metabólicos e que sua deficiência estaria ligada ao TEA. No entanto, não há comprovação científica de que a suplementação desse composto tenha qualquer impacto no autismo. |
| **124** | Sulforafano | O sulforafano, presente no brócolis, tem sido estudado por seus potenciais efeitos neuroprotetores, mas não há evidência suficiente para afirmar que ele cura ou trata o autismo. |
| **125** | Suor (Sauna) | Algumas práticas alternativas afirmam que "suar toxinas" em saunas poderia curar o autismo. Isso não tem qualquer embasamento científico. Expor crianças a saunas pode ser perigoso, levando a desidratação severa, hipertermia e danos renais. |
| **126** | Superóxido Dismutase | A SOD é uma enzima antioxidante promovida como um suplemento para combater o "estresse oxidativo" no autismo. No entanto, não há evidência científica de que sua suplementação tenha efeito terapêutico no TEA. |
| **127** | Suramina | A suramina é um antiparasitário que tem sido investigado em estudos preliminares para possíveis efeitos no autismo. No entanto, não há comprovação científica robusta de que seja segura ou eficaz para o TEA. O uso indevido pode causar efeitos colaterais graves, incluindo toxicidade renal, supressão do sistema imunológico e risco de morte. |
| **128** | Taurina | A taurina é um aminoácido presente em suplementos energéticos e algumas dietas. Algumas teorias sugerem que ajudaria na cognição, mas não há nenhuma evidência científica de que a suplementação de taurina trate ou cure o autismo. |
| **129** | Ter Fé em Deus | A fé pode ser um elemento positivo para muitas pessoas, mas o autismo não é uma condição que pode ser "curada" pela fé, oração ou práticas religiosas. Essa narrativa é perigosa porque desvia o foco de abordagens baseadas em evidências e pode levar à negação de tratamentos e apoios adequados. |

[ versão português / english above ]

| | | |
|---|---|---|
| **130** | Terebintina | A terebintina (óleo derivado da resina de pinheiros) é promovida como um "purificador do organismo" para autistas, mas não há qualquer embasamento científico para essa alegação. A ingestão ou inalação de terebintina pode causar intoxicação grave, falência hepática, danos neurológicos e morte. |
| **131** | Terra de Diatomáceas | A terra de diatomáceas é uma substância usada como inseticida natural, mas alguns grupos afirmam que sua ingestão "elimina parasitas" e "cura o autismo". Não há comprovação científica disso, e o consumo pode causar irritação severa no trato digestivo e pulmonar. |
| **132** | Tintura das Agulhas de Pinheiro | Esse extrato vegetal é promovido como um "desintoxicante", mas não há evidências de que tenha qualquer impacto no autismo. |
| **133** | Tintura de Nozes | A tintura de nozes é sugerida como um antiparasitário para autismo. Como já discutido anteriormente, o autismo não é causado por parasitas, e o consumo dessa tintura sem controle pode ser tóxico. |
| **134** | Tintura do Açafrão da Terra | O açafrão-da-terra (cúrcuma) tem propriedades antioxidantes, mas não há nenhuma evidência de que sua tintura tenha qualquer efeito no autismo. |
| **135** | Tintura do Melão de São Caetano | O melão-de-são-caetano é promovido como um "limpador do organismo", mas não há comprovação de que tenha qualquer efeito sobre o TEA. Seu consumo excessivo pode ser tóxico para o fígado e causar distúrbios gastrointestinais. |
| **136** | Tirosina | A tirosina é um aminoácido essencial para a produção de neurotransmissores, mas não há evidência de que sua suplementação altere a manifestação do TEA. |
| **137** | Tratamento com Frequência Tesla | Esse tratamento se baseia na ideia pseudocientífica de que frequências eletromagnéticas podem "ajustar o cérebro" de autistas. Não há qualquer evidência científica que sustente essa alegação. |
| **138** | Trementina | A trementina, um solvente químico, é promovida como um "eliminador de parasitas" e "desintoxicante". Sua ingestão pode causar intoxicação severa, falência renal e morte. |
| **139** | Vitamina A | Embora a vitamina A seja essencial para diversas funções corporais, não há evidências científicas de que sua suplementação cure ou trate o autismo. O excesso pode ser tóxico, causando problemas hepáticos e danos neurológicos. |
| **140** | Vitamina B1 | Algumas alegações sugerem que a deficiência de vitamina B1 estaria ligada ao autismo. Não há comprovação científica dessa relação, e a suplementação excessiva pode causar efeitos adversos neurológicos. |
| **141** | Vitamina B12 | A B12 é essencial para a função cerebral, mas não há evidência de que sua suplementação cure o autismo. |
| **142** | Vitamina B2 | A riboflavina é importante para o metabolismo, mas não há nenhuma evidência de que sua suplementação tenha impacto no TEA. |
| **143** | Vitamina B6 | Embora a vitamina B6 esteja envolvida na neurotransmissão, não há comprovação de que sua suplementação altere o autismo. |
| **144** | Vitamina B8 | A biotina é essencial para o metabolismo energético, mas não há estudos que comprovem qualquer relação entre deficiência de biotina e TEA. |
| **145** | Vitamina B9 | O ácido fólico é essencial para o desenvolvimento fetal, mas não há evidência de que sua suplementação após o nascimento altere o autismo. |
| **146** | Vitamina C | Embora a vitamina C seja importante para o sistema imunológico, não há evidências científicas de que trate ou cure o autismo. |
| **147** | Vitamina D | Embora alguns estudos investiguem a relação entre vitamina D e desenvolvimento neurológico, não há provas de que sua suplementação cure o autismo. |
| **148** | Vitamina E | A vitamina E é um antioxidante importante, mas não há estudos que comprovem seu impacto no TEA. |
| **149** | Zeólita | A zeólita é promovida como um "eliminador de metais pesados" para autistas, mas não há evidências de que funcione. |
| **150** | Zinco | O zinco é essencial para o metabolismo, mas não há evidência de que sua suplementação trate o autismo. |

Fonte: Elaboração própria (2025).

### 4.2. Rede

Os gráficos (Figuras 13 e 14) e tabelas (05 e 06) a seguir apresentam uma perspectiva da coexistência de usuários (autores) de publicações sobre autismo e conteúdos sobre autismo nas comunidades de teorias da conspiração entre diferentes países. O objetivo é identificar padrões de compartilhamento e conexões entre nações a partir da repetição de IDs de usuários e da disseminação de conteúdos semelhantes. Para isso, foram geradas redes de coautoria e de similaridade de conteúdos, bem como matrizes que quantificam essas interações.

A Figura 13 representa a rede de conexões entre países baseada na coexistência de IDs de usuários que postaram sobre autismo dentro dessas comunidades. Cada nó do gráfico representa um país, e o tamanho do nó é proporcional ao número de usuários que publicaram sobre o tema e que também apareceram em outros países. As arestas, ou conexões entre os nós, indicam a frequência com que os mesmos usuários foram encontrados publicando sobre autismo em múltiplos países. Quanto mais espessa a aresta, maior a quantidade de usuários compartilhados entre as nações.

O que chama atenção nessa rede é que nenhum país fica isolado — há conexões em toda a estrutura, o que demonstra a transnacionalidade dessas narrativas conspiratórias. Comunidades categorizadas como Transnacional, além de países México, Colômbia e Argentina possuem os maiores nós e as conexões mais espessas, o que significa que concentram um número expressivo de usuários que publicam sobre autismo em comunidades de diferentes países. O nó da categoria Transnacional, por exemplo, é o maior da rede, indicando que esse grupo engloba usuários ativos simultaneamente em diversas nações, funcionando como um ponto de convergência para a disseminação dessas narrativas.

**Figura 13.** Gráfico de redes de IDs de usuários coexistentes entre países (autores).

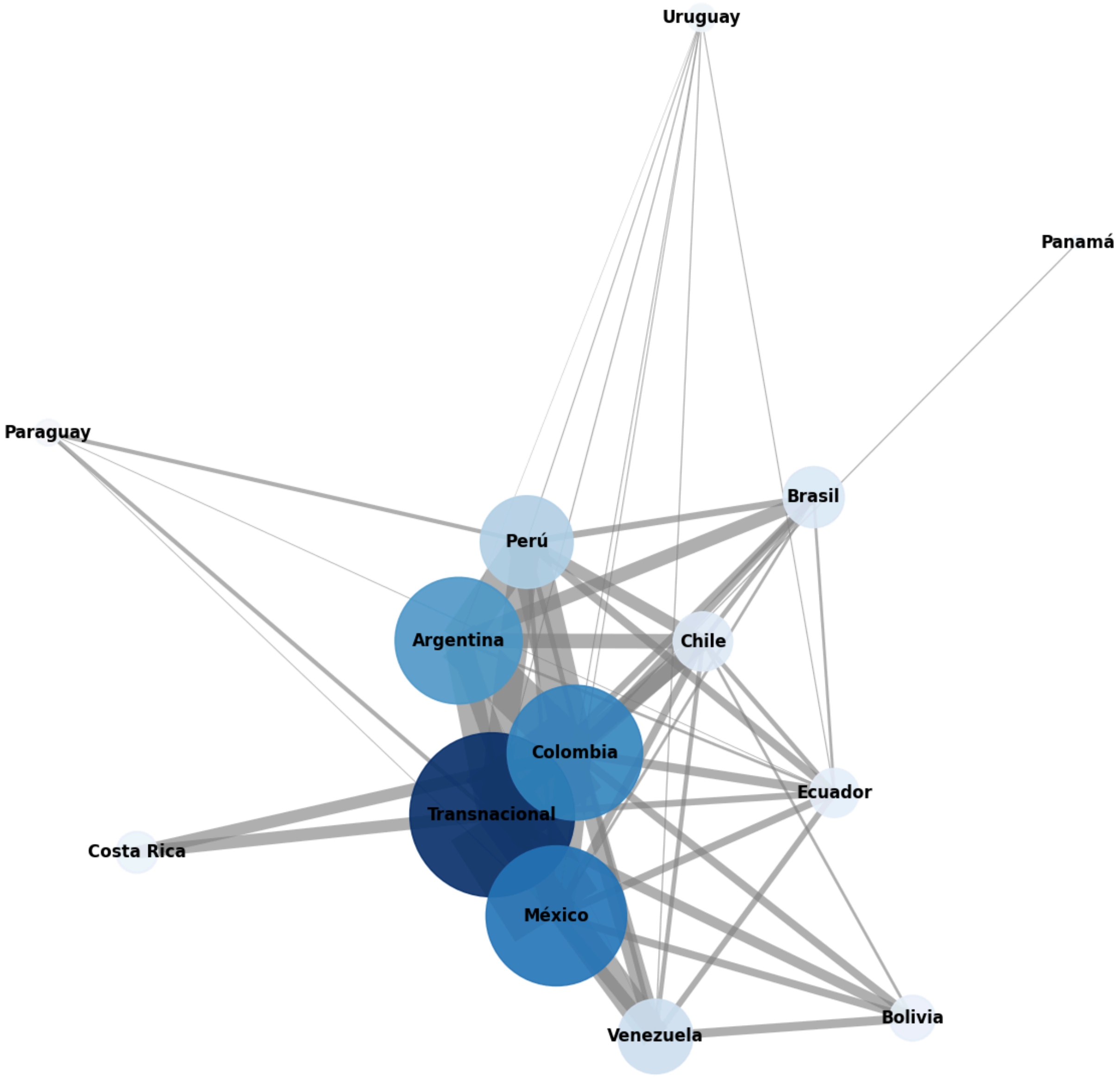

Fonte: Elaboração própria (2025).

A Tabela 05 detalha numericamente essas conexões, apresentando uma matriz de coexistência de usuários entre os países. A diagonal principal da tabela mostra quantas vezes os mesmos usuários postaram sobre autismo dentro do próprio país, enquanto os valores fora da diagonal indicam a quantidade de vezes que esses usuários também publicaram sobre o tema em outro país. Essa métrica permite compreender melhor quais nações possuem maior volume de usuários engajados na disseminação dessas narrativas e quais são os principais fluxos de interconexão entre os países.

Os números revelam padrões intrigantes. Os grupos do Brasil apresentam 10.280 ocorrências de IDs postando sobre autismo dentro dos próprios grupos, o que o coloca como o maior polo da rede. Comunidades Transnacionais (4.970), México (1.129) e Colômbia (1.490) também aparecem com volumes altos dentro de suas próprias comunidades. Esses dados mostram que há grupos que funcionam como agregadores de usuários, mantendo uma base fixa, mas ao mesmo tempo exportando essas narrativas para outros países.

A coexistência de usuários entre diferentes países também segue essa tendência. Por exemplo, México e comunidades Transnacionais compartilham 697 IDs de usuários, enquanto Colômbia e comunidades Transnacionais têm 623 IDs em comum. Além disso, México e Colômbia compartilham 244 IDs, sugerindo que esses países possuem uma interseção significativa nas narrativas difundidas. Esses números confirmam que algumas comunidades servem como amplificadores, com usuários que transitam entre diferentes espaços digitais e replicam as mesmas publicações em múltiplos locais.

**Tabela 05.** Coexistência dos mesmos usuários dentre países (autores).

| | Perú | Transnacional | México | Colombia | Panamá | Argentina | Bolivia | Venezuela | Chile | Ecuador | Brasil | Costa Rica | Uruguay | Paraguay |
|---|---|---|---|---|---|---|---|---|---|---|---|---|---|---|
| **Perú** | 672 | 202 | 32 | 277 | 0 | 245 | 0 | 36 | 89 | 60 | 49 | 0 | 8 | 30 |
| **Transnacional** | 202 | 4.970 | 697 | 623 | 0 | 204 | 74 | 209 | 129 | 48 | 50 | 89 | 9 | 30 |
| **México** | 32 | 697 | 1.129 | 244 | 0 | 152 | 56 | 84 | 55 | 56 | 20 | 0 | 8 | 6 |
| **Colombia** | 277 | 623 | 244 | 1.490 | 9 | 317 | 58 | 126 | 109 | 67 | 50 | 89 | 8 | 0 |
| **Panamá** | 0 | 0 | 0 | 9 | 9 | 0 | 0 | 0 | 0 | 0 | 0 | 0 | 0 | 0 |
| **Argentina** | 245 | 204 | 152 | 317 | 0 | 1.488 | 0 | 3 | 105 | 20 | 100 | 0 | 3 | 0 |
| **Bolivia** | 0 | 74 | 56 | 58 | 0 | 0 | 184 | 64 | 20 | 0 | 0 | 0 | 0 | 0 |
| **Venezuela** | 36 | 209 | 84 | 126 | 0 | 3 | 64 | 261 | 32 | 43 | 0 | 0 | 9 | 0 |
| **Chile** | 89 | 129 | 55 | 109 | 0 | 105 | 20 | 32 | 580 | 33 | 34 | 0 | 0 | 0 |
| **Ecuador** | 60 | 48 | 56 | 67 | 0 | 20 | 0 | 43 | 33 | 119 | 20 | 0 | 8 | 6 |
| **Brasil** | 49 | 50 | 20 | 50 | 0 | 100 | 0 | 0 | 34 | 20 | 10.280 | 0 | 0 | 0 |
| **Costa Rica** | 0 | 89 | 0 | 89 | 0 | 0 | 0 | 0 | 0 | 0 | 0 | 159 | 0 | 0 |
| **Uruguay** | 8 | 9 | 8 | 8 | 0 | 3 | 0 | 9 | 0 | 8 | 0 | 0 | 56 | 0 |
| **Paraguay** | 30 | 30 | 6 | 0 | 0 | 0 | 0 | 0 | 0 | 6 | 0 | 0 | 0 | 73 |

Fonte: Elaboração própria (2025).

A Figura 14 traz uma perspectiva complementar, analisando a circulação de conteúdos sobre autismo dentro dessas redes. Neste caso, cada nó representa um país, e seu tamanho reflete a quantidade de conteúdos repetidos internacionalmente. As arestas indicam a frequência com que os mesmos conteúdos foram compartilhados entre os países, com espessuras variáveis conforme a intensidade do intercâmbio de postagens. Diferente da análise de usuários, que mede a autoridade e influência de quem publica, este gráfico destaca os padrões de disseminação das narrativas, ou seja, como determinados textos, imagens e vídeos circulam transnacionalmente, com o mesmo conteúdo em comunidades diferentes.

Novamente, observa-se que nenhum país está isolado, o que sugere que as mesmas peças de desinformação aparecem simultaneamente em diferentes comunidades ao redor do mundo. Comunidades de caráter Transnacional, além de países como Colômbia e México emergem como centrais na disseminação de conteúdos, com conexões especialmente fortes com Chile, Peru e Brasil. O nó Brasil, por exemplo, tem uma concentração expressiva de conteúdos compartilhados dentro do país (10.591), mas também participa ativamente de trocas de postagens com outras nações.

**Figura 14.** Gráfico de redes de IDs de conteúdos coexistentes entre países

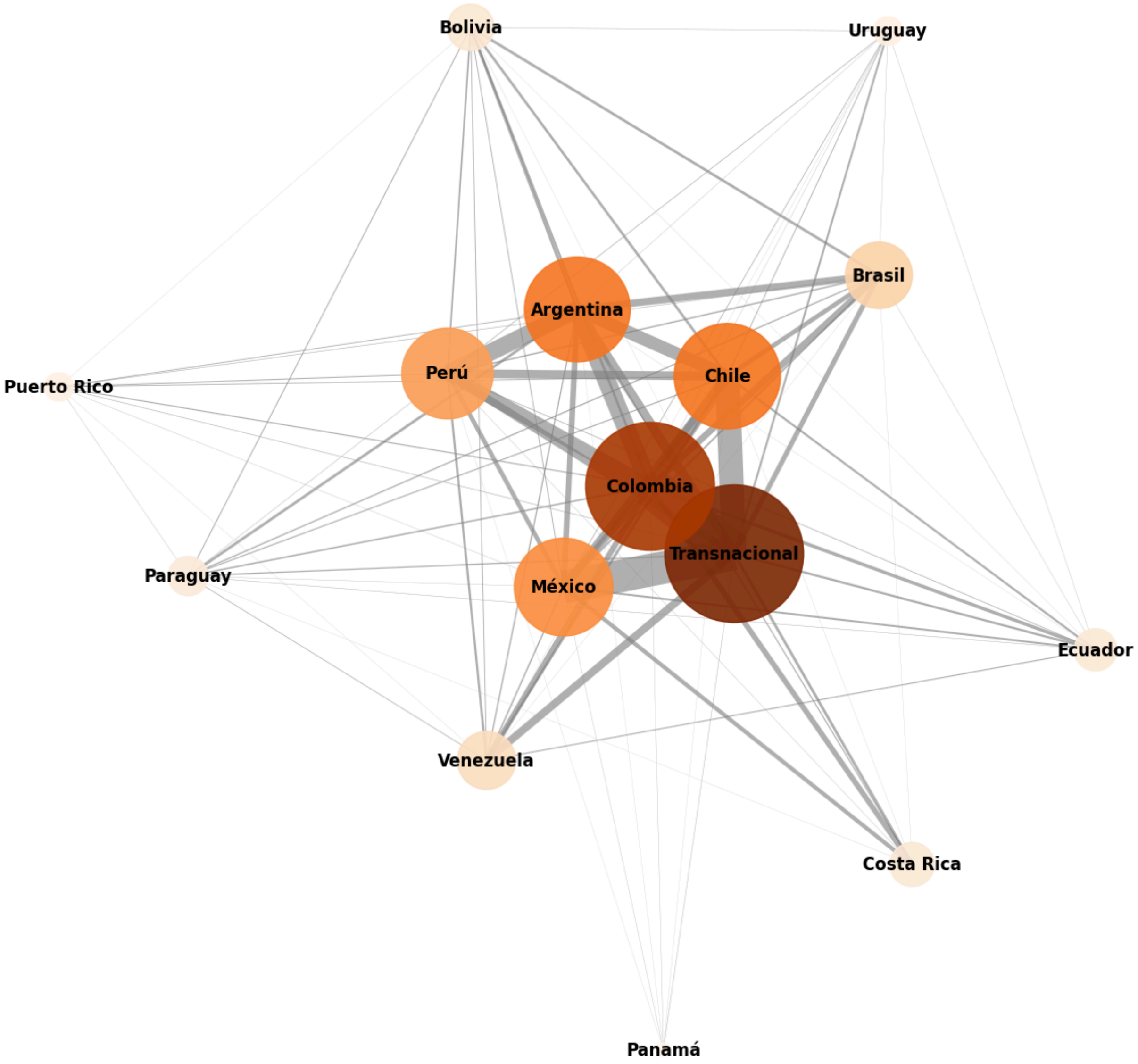

Fonte: Elaboração própria (2025).

A Tabela 06 complementa essa visualização ao quantificar a coexistência de conteúdos entre os países. Assim como na matriz de usuários, a diagonal principal apresenta a quantidade de postagens repetidas dentro do próprio país, enquanto os valores fora da diagonal indicam quantas vezes um mesmo conteúdo apareceu em diferentes nações.

Os dados mostram que o Brasil e os grupos Transnacionais são os principais nós da disseminação de conteúdos. O Brasil registra 10.591 repetições internas de conteúdos, enquanto o Transnacional aparece com 5.039. No caso das conexões internacionais, Colômbia e México compartilham 567 conteúdos idênticos, enquanto Argentina e Transnacional possuem 64 conteúdos replicados entre si. O volume dessas conexões indica que há uma dinâmica estruturada de circulação de postagens, em que determinados conteúdos são constantemente reaproveitados e adaptados para diferentes públicos.

Além disso, países que poderiam ser considerados secundários nesse ecossistema, como Costa Rica, Paraguai e Puerto Rico, também aparecem conectados à rede, demonstrando que as narrativas não estão limitadas a grandes centros populacionais ou a países de um mesmo eixo geopolítico. Isso sugere que há uma rede ativa de tradução e adaptação de conteúdos, permitindo que as desinformações ultrapassem barreiras linguísticas e alcancem audiências diversas.

**Tabela 06.** Coexistência dos mesmos conteúdos dentre países.

| | Argentina | Transnacional | Colombia | México | Chile | Perú | Brasil | Venezuela | Bolivia | Costa Rica | Ecuador | Paraguay | Uruguay | Puerto Rico | Panamá |
|---|---|---|---|---|---|---|---|---|---|---|---|---|---|---|---|
| **Argentina** | 1.517 | 64 | 130 | 40 | 91 | 122 | 52 | 9 | 20 | 8 | 1 | 19 | 2 | 6 | 1 |
| **Transnacional** | 64 | 5.039 | 291 | 234 | 173 | 55 | 35 | 55 | 1 | 19 | 16 | 9 | 14 | 4 | 3 |
| **Colombia** | 130 | 291 | 1.567 | 50 | 85 | 96 | 47 | 40 | 16 | 38 | 24 | 11 | 2 | 8 | 2 |
| **México** | 40 | 234 | 50 | 1.178 | 54 | 31 | 12 | 10 | 6 | 29 | 14 | 2 | 5 | 2 | 2 |
| **Chile** | 91 | 173 | 85 | 54 | 582 | 62 | 32 | 19 | 18 | 1 | 15 | 8 | 6 | 6 | 1 |
| **Perú** | 122 | 55 | 96 | 31 | 62 | 675 | 10 | 16 | 12 | 2 | 5 | 1 | 3 | 7 | 1 |
| **Brasil** | 52 | 35 | 47 | 12 | 32 | 10 | 10.591 | 1 | 19 | 1 | 2 | 9 | 2 | 4 | 0 |
| **Venezuela** | 9 | 55 | 40 | 10 | 19 | 16 | 1 | 262 | 6 | 0 | 8 | 5 | 1 | 1 | 0 |
| **Bolivia** | 20 | 1 | 16 | 6 | 18 | 12 | 19 | 6 | 184 | 0 | 1 | 7 | 3 | 1 | 0 |
| **Costa Rica** | 8 | 19 | 38 | 29 | 1 | 2 | 1 | 0 | 0 | 160 | 0 | 1 | 0 | 0 | 0 |
| **Ecuador** | 1 | 16 | 24 | 14 | 15 | 5 | 2 | 8 | 1 | 0 | 126 | 3 | 2 | 0 | 0 |
| **Paraguay** | 19 | 9 | 11 | 2 | 8 | 1 | 9 | 5 | 7 | 1 | 3 | 73 | 2 | 2 | 0 |
| **Uruguay** | 2 | 14 | 2 | 5 | 6 | 3 | 2 | 1 | 3 | 0 | 2 | 2 | 56 | 0 | 0 |
| **Puerto Rico** | 6 | 4 | 8 | 2 | 6 | 7 | 4 | 1 | 1 | 0 | 0 | 2 | 0 | 23 | 0 |
| **Panamá** | 1 | 3 | 2 | 2 | 1 | 1 | 0 | 0 | 0 | 0 | 0 | 0 | 0 | 0 | 9 |

Fonte: Elaboração própria (2025).

Ao analisar os dados combinados, fica evidente que a desinformação sobre autismo não está restrita a um único país ou idioma, mas sim faz parte de um ecossistema transnacional interconectado, no qual usuários engajados e conteúdos replicados contribuem para a consolidação de falsas narrativas. O fato de nenhum país estar isolado nesses gráficos e matrizes indica que essas redes são altamente organizadas e eficientes na disseminação de teorias da conspiração, aproveitando a dinâmica das plataformas digitais para atingir públicos distintos. Essa estrutura transnacional reforça a importância de estratégias de contenção coordenadas entre países, visando mitigar os impactos dessas desinformações e proteger populações vulneráveis de crenças e práticas prejudiciais associadas ao autismo.

### 4.3. Séries temporais

Conforme vemos na Figura 15, o volume de publicações sobre autismo dentro das comunidades conspiratórias apresentou um crescimento exponencial nos últimos anos. Em janeiro de 2019, havia apenas 4 postagens mensais sobre o tema, número que saltou para 35 postagens em janeiro de 2020, um aumento de 775% em um ano. Com o início da Pandemia da COVID-19, o crescimento foi ainda mais expressivo: em janeiro de 2021, o volume de publicações alcançou 260 postagens, um aumento de 635% em relação ao ano anterior. Esse pico reflete a intensificação das narrativas sobre autismo durante a Pandemia, muitas vezes associadas a discursos antivacinas e tratamentos alternativos fraudulentos.

O aumento continuou nos anos seguintes, atingindo 438 publicações em janeiro de 2022 e 365 postagens em janeiro de 2023, demonstrando que a desinformação sobre autismo permaneceu relevante mesmo após a fase crítica da Pandemia. Em 2024, o número de postagens chegou a 587, e em 2025, atingiu o recorde de 611 postagens mensais, consolidando um crescimento de mais de 15.000% (x151) desde 2019. Esses números evidenciam como o autismo se tornou um tema central dentro dessas comunidades, sendo explorado para fomentar desinformação e teorias conspiratórias.

**Figura 15.** Gráfico de série temporal de conteúdos sobre autismo (total).

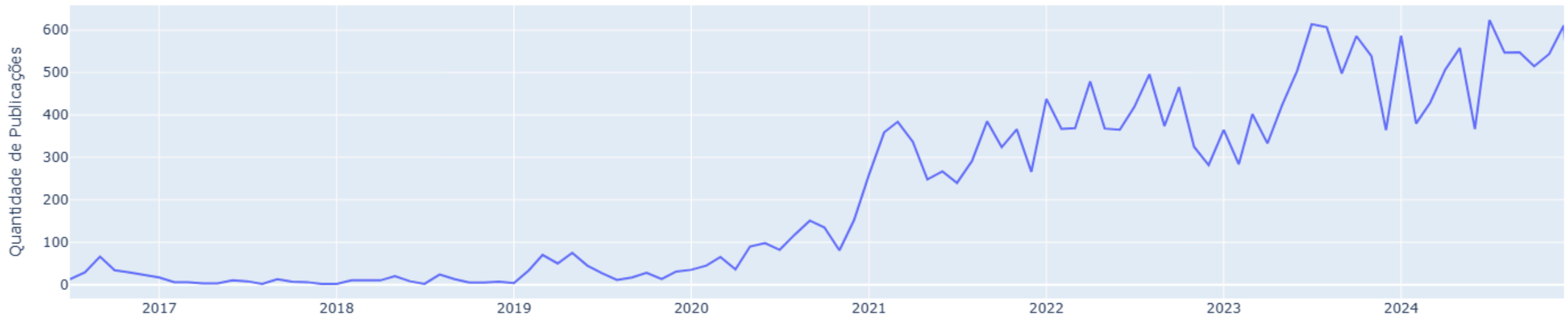

Fonte: Elaboração própria (2025).

Com a Figura 16, podemos ver a segmentação por países. A análise revela que o Brasil lidera em volume de postagens sobre autismo dentro dessas comunidades, com picos superiores a 300 publicações mensais a partir de 2023. O Brasil se destaca como o maior disseminador de conteúdos relacionados ao tema, tanto por sua base populacional ativa nas redes quanto pelo histórico de influência de movimentos antivacinas no país. Outro ponto de destaque são os grupos Transnacionais, que aparecem como uma das principais categorias no tema, indicando que há uma grande intersecção de conteúdos que circulam entre diferentes países sem estarem atrelados diretamente a um contexto nacional específico.

Países como México, Colômbia e Argentina também apresentam crescimento significativo nas postagens ao longo dos anos, ainda que em menor escala. A partir de 2021, observa-se um aumento constante no volume de publicações nesses países, com picos em momentos estratégicos, como períodos de campanhas de vacinação ou lançamentos de novas pesquisas científicas sobre o autismo. A presença de países menores na rede, como Costa

Rica, Paraguai e Venezuela, sugere que a desinformação sobre autismo conseguiu penetrar em diversas esferas, extrapolando grandes centros populacionais e alcançando audiências antes menos impactadas por essas narrativas.

**Figura 16.** Gráfico de série temporal de conteúdos sobre autismo (por países).

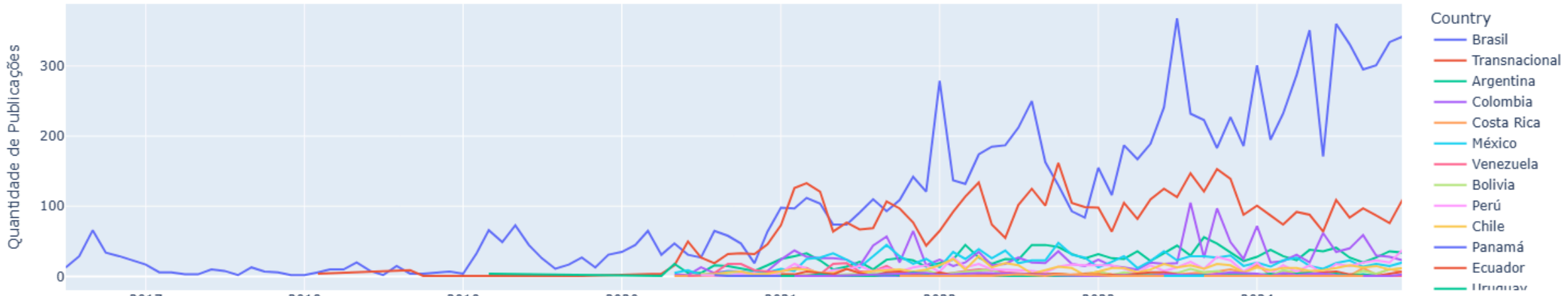

Fonte: Elaboração própria (2025).

Já por meio da Figura 17, a seguir, vemos a segmentação das postagens por categoria. Essas nos demonstram quais narrativas específicas impulsionaram a desinformação sobre autismo nos últimos anos. Entre as categorias analisadas, as que mais cresceram foram *Off Label and Quackery* (medicamentos não autorizados e curas alternativas), *Antivax* e *QAnon*, todas diretamente associadas a discursos conspiratórios que exploram o medo e a insegurança dos pais de crianças autistas. Em 2019, a categoria *Off Label and Quackery* era praticamente inexistente, mas passou a registrar mais de 200 publicações mensais a partir de 2021, consolidando um crescimento abrupto desse tipo de conteúdo.

A categoria *Antivax* também apresentou uma escalada notável, com picos acima de 150 postagens por mês em momentos de alta circulação de campanhas antivacinas. A associação entre autismo e vacinas, já desmentida amplamente pela comunidade científica, continua sendo um dos pilares centrais das narrativas disseminadas nessas comunidades de teorias da conspiração. Além disso, a presença de *QAnon*, uma teoria conspiratória originalmente focada em política e controle global, indica que essas redes têm integrado o tema do autismo em narrativas mais amplas sobre manipulação da saúde e da ciência.

Outras categorias, como terraplanismo, negacionismo às mudanças climáticas e conspirações envolvendo alienígenas, aparecem em menor escala, mas demonstram que há uma intersecção entre diversas teorias da conspiração, indicando que a desinformação sobre autismo não está isolada, mas faz parte de um ecossistema maior de desinformação que se adapta conforme as circunstâncias. O crescimento dessas categorias ao longo dos anos reforça que a narrativa sobre o autismo tem sido explorada de maneira oportunista, aproveitando eventos globais e contextos de crise para amplificar conteúdos enganosos.

**Figura 17.** Gráfico de série temporal de conteúdos sobre autismo (por categorias).

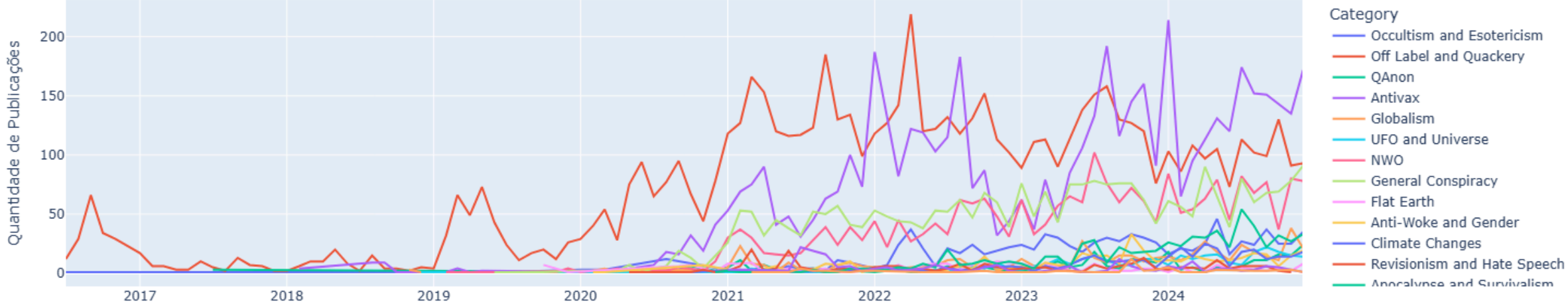

Fonte: Elaboração própria (2025).

Os gráficos evidenciam o crescimento exponencial da desinformação sobre autismo em comunidades conspiratórias no Continente, especialmente após a Pandemia da COVID-19. O Brasil se destaca como o maior polo de disseminação desses conteúdos com fluxo mensal superior aos demais, seguido por redes transnacionais e países latino-americanos como México, Colômbia e Argentina. As narrativas mais difundidas estão ligadas a tratamentos não autorizados, teorias antivacinas e conspirações globais, demonstrando como o autismo tem sido instrumentalizado por diferentes vertentes da desinformação.

### 4.4. Análise de conteúdo

Já aprofundadas as análises sobre a progressão temporal e geográfica das postagens, esta análise de conteúdo busca complementar o entendimento ao explorar as palavras mais frequentes no corpus coletado. Como vemos na nuvem a seguir (Figura 18), a predominância dos termos “autismo” (26.845 menções), “vacinas” (7.677), “vacuna” (9.530), “autista” (4.153) e “crianças” (4.532) evidencia o foco das discussões em narrativas ligadas ao espectro autista e sua relação com imunização. O uso recorrente das palavras em diferentes idiomas sugere que essas narrativas circulam amplamente em contextos multilíngues, reforçando sua natureza transnacional. A forte presença de links, como indicado pela elevada frequência de “https” (23.763 menções) e “www” (6.878), aponta para uma estratégia de disseminação baseada no compartilhamento de fontes externas, muitas das quais possivelmente vinculadas a pseudociência, fóruns conspiratórios e sites alternativos de “saúde natural”.

Outro aspecto marcante é a correlação entre autismo e substâncias químicas, com destaque para “alumínio” (3.300), “cloro” (2.068), “mercúrio” (1.266) e “timerosal” (1.133), substâncias frequentemente mencionadas em teorias da conspiração sobre autismo. A presença desses termos sugere que um volume significativo das postagens associa o transtorno a supostos agentes tóxicos presentes em vacinas ou no ambiente, uma alegação amplamente refutada por estudos científicos. A palavra “dióxido” (1.626), frequentemente associada ao dióxido de cloro, reforça a presença de discursos que promovem tratamentos perigosos e fraudulentos, como o MMS (Miracle Mineral Solution), uma substância vendida ilegalmente sob a falsa promessa de “curar” o autismo.

A análise também revela a inserção do tema autismo dentro de narrativas mais amplas sobre controle populacional e desconfiança em relação à medicina. Termos como “CDC” (2.551), “gov” (1.780), “NIH” (1.427) e “comusav” (2.190) indicam um discurso que frequentemente questiona instituições científicas e governamentais, fomentando desinformação sobre políticas de saúde pública. Além disso, palavras como “Bill Gates” (embora não apareça nos termos mais frequentes, é visível na nuvem de palavras) sugerem a interseção com teorias conspiratórias que atribuem o aumento dos casos de autismo a supostas agendas de vacinação obrigatória promovidas por elites globais. Essa interconexão com narrativas de conspiração mais abrangentes reforça o caráter instrumentalizado do autismo dentro desse ecossistema desinformativo.

Por fim, a análise lexical sugere um discurso alarmista, reforçado por termos como “problemas” (3.999), “risco” (embora não esteja na lista principal, aparece destacado na nuvem), “causar” (2.191) e “doenças” (2.743). A escolha de palavras que evocam medo e insegurança indica uma estratégia de engajamento baseada na exploração de emoções negativas, o que pode facilitar a adesão de novos participantes às comunidades conspiratórias. O uso de expressões como “cura” (1.447), “tratamento” e “verdade” (1.439) sugere que esses grupos também se posicionam como alternativas à medicina tradicional, promovendo soluções pseudocientíficas para o autismo. Em conjunto, esses padrões reforçam a importância de monitoramento e resposta coordenada para mitigar o impacto dessas narrativas, que podem influenciar negativamente a percepção pública sobre o autismo e comprometer decisões informadas sobre tratamentos e políticas de inclusão.

**Figura 18.** Nuvem de palavras mais frequentes nas publicações sobre autismo.

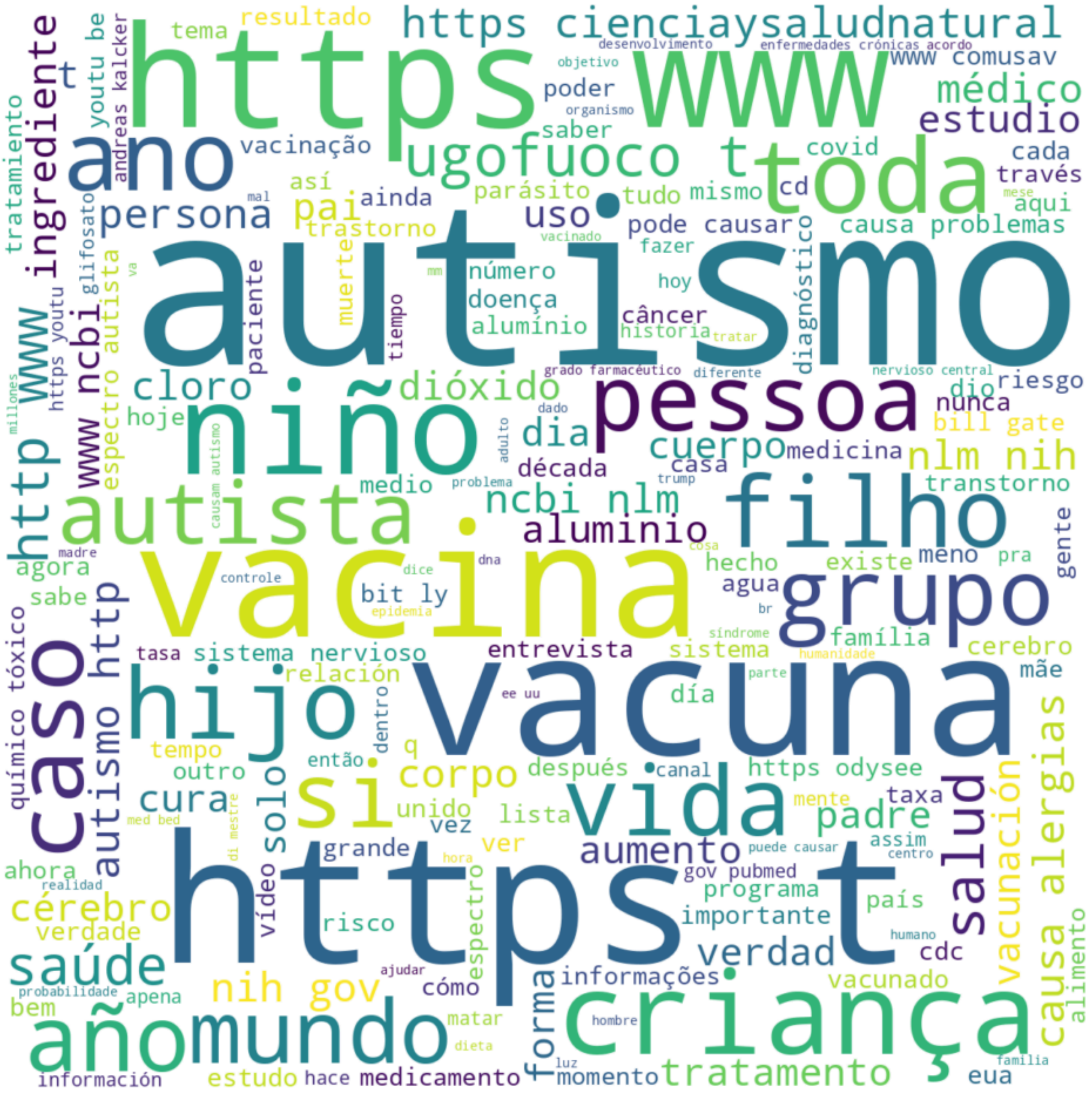

Fonte: Elaboração própria (2025).

## 5. Reflexões e trabalhos futuros

Para responder à questão de pesquisa — ***"Como as comunidades de teorias da conspiração na América Latina e no Caribe estruturam, articulam e sustentam a disseminação da desinformação sobre o autismo?"*** — este estudo adotou uma abordagem metodológica consistente com uma série mais ampla de investigações sobre teorias da conspiração no Telegram. Após uma investigação extensiva, foi identificado um total de **1.649 comunidades de teorias da conspiração** em toda a América Latina e no Caribe, abrangendo uma ampla gama de temas de desinformação. No total, essas comunidades publicaram **58.637.137 conteúdos** entre dezembro de 2015 (primeiros registros) e janeiro de 2025 (período do estudo), com um total de **5.345.332 usuários**.

Para analisar sistematicamente a dinâmica dessas comunidades, foram aplicadas quatro principais abordagens metodológicas: **(i) Análise descritiva das alegações relacionadas ao autismo** – Foi realizada uma classificação estruturada de 150 supostas causas e 150 supostas curas para o autismo, categorizando-as por origem temática e tipo de tratamento proposto para identificar padrões narrativos; **(ii) Análise de redes** – Um algoritmo proprietário foi desenvolvido para mapear interconexões entre comunidades, rastreando convites para o Telegram (t.me/ links) compartilhados entre grupos e canais, revelando como as comunidades reforçam narrativas internas e promovem outras teorias da conspiração; **(iii) Análise de séries temporais** – Utilizando a biblioteca Pandas (McKinney, 2010) para a estruturação dos dados e a biblioteca Plotly (Plotly Technologies Inc., 2015) para a visualização, padrões temporais foram analisados para examinar tendências na produção de conteúdo e flutuações no engajamento ao longo do tempo; **(iv) Análise de conteúdo** – Técnicas de análise textual foram aplicadas para examinar frequências de palavras e variações temáticas, fornecendo insights sobre a persistência e a evolução das narrativas sobre autismo.

As seções a seguir apresentam os principais achados, com avaliação das características estruturais e dos padrões comportamentais das comunidades de teorias da conspiração na América Latina e no Caribe, seguidos de recomendações para pesquisas futuras.

### 5.1. Principais reflexões

O mercado de influencers e *coachs* de autismo que promovem uma visão patologizante da condição se relaciona com o fortalecimento de comunidades conspiratórias que operam no lucrativo mercado da "cura do autismo". Esses influenciadores frequentemente retratam o autismo como uma condição trágica e debilitante, o que alimenta estigmas e preconceitos, além de criar um terreno fértil para a disseminação de soluções milagrosas e pseudocientíficas. Ao reforçar a ideia de que o autismo é uma doença a ser combatida ou curada, esses influencers abrem espaço para práticas perigosas, como o uso de dióxido de cloro (MMS), transplante fecal e até abordagens espirituais que prometem "libertar" o indivíduo do autismo por meio de exorcismo. Essas soluções não possuem respaldo científico e podem colocar em risco a saúde e o bem-estar das pessoas autistas.

O discurso patologizante também gera desinformação, contribuindo para a formação de comunidades conspiratórias que desconfiam da medicina baseada em evidências e rejeitam abordagens que promovem a aceitação da neurodiversidade. Esses grupos frequentemente compartilham narrativas alarmistas sobre vacinas, toxinas ambientais e supostos complôs da indústria farmacêutica, alimentando teorias da conspiração que ganham força nas redes sociais. Influenciadores e usuários mal intencionados que lucram com o desespero e a desinformação dos familiares de autistas frequentemente se posicionam como salvadores ou especialistas autoproclamados, oferecendo cursos, mentorias e terapias alternativas com promessas de cura. Ao criarem um senso de urgência e medo, eles atraem um público vulnerável e disposto a pagar por qualquer promessa de normalidade.

Ao patologizar o autismo e apresentar soluções fáceis que visam sua "cura" ou "erradicação", esses discursos perpetuam a ideia de que características neurodivergentes são defeitos a serem corrigidos, ecoando a lógica eugenista de exclusão e conformidade a um ideal normativo. Isso desumaniza as pessoas autistas ao tratá-las como erros biológicos, legitimando intervenções invasivas e potencialmente perigosas. Esse cenário não só reforça preconceitos contra autistas, como também desvia o foco de políticas públicas e abordagens baseadas em direitos humanos, que priorizam a inclusão social, o respeito à neurodiversidade e o acesso a suportes adequados. Portanto, a patologização do autismo não apenas alimenta mercados lucrativos e conspirações pseudocientíficas, como também sustenta um projeto social eugenista que visa normalizar corpos e mentes, apagando as diferenças neurocognitivas em prol de uma uniformidade social capacitista.

Um importante ponto que vale reforçar, é como o mercado de curas milagrosas e remédios alternativos não possui fronteiras e atua de forma articulada, promovendo redes de entregas de produtos químicos pronta-entrega, com fornecedores distribuídos por todo o continente. Algumas capturas de tela das comunidades ajudam a ilustrar (Figura 19), pois não apenas fornecem contato direto, mas chegam a colocar uma criança para ensinar como preparar "colírios de imunoterapia" em uma das propostas de "cura" de autismo. No mesmo vídeo com uma criança preparando o químico, existe um link para compra do Kit de preparo.

**Figura 19.** Exemplos de comercialização de curas milagrosas para autismo.

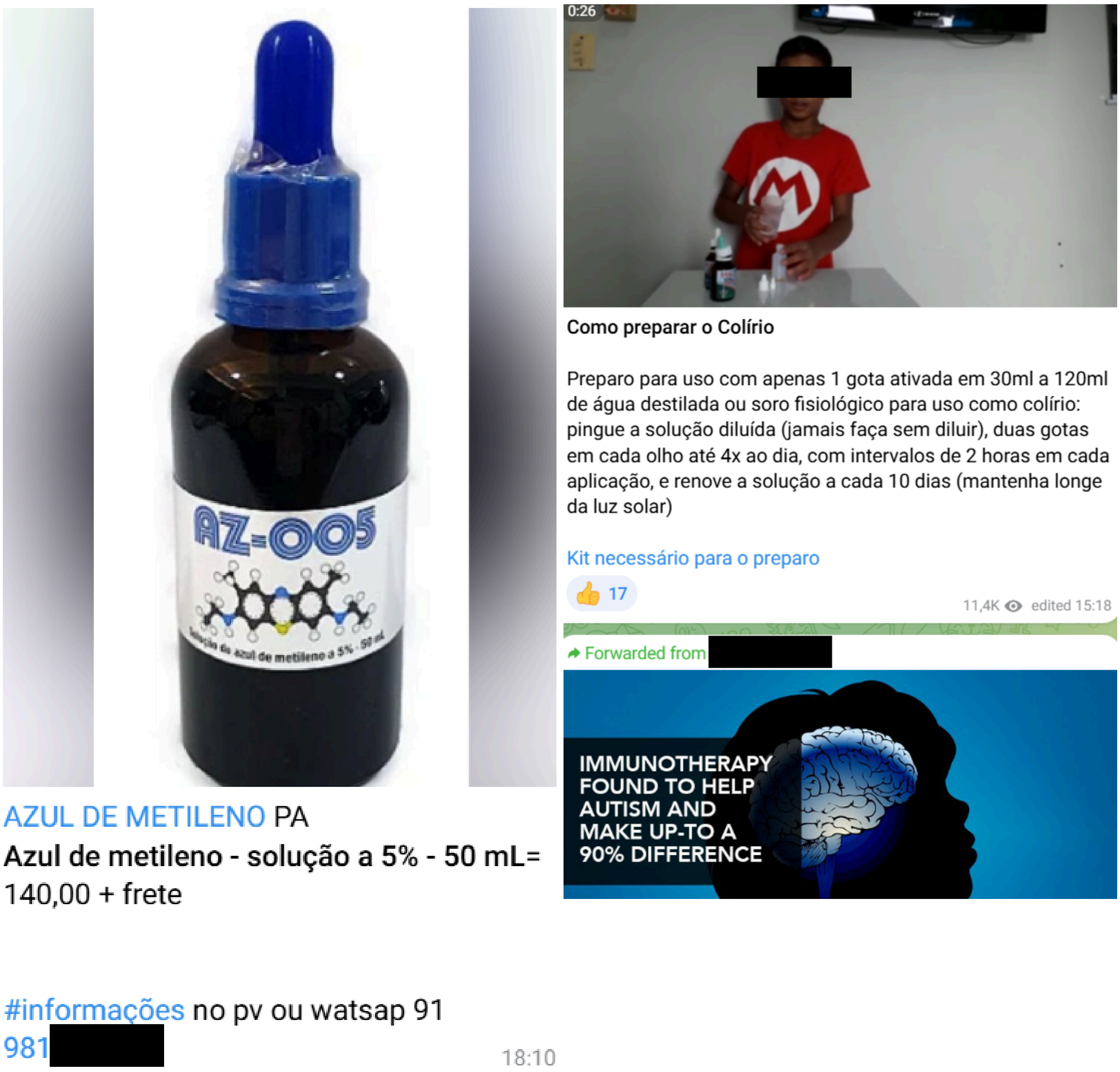

Fonte: Elaboração própria (2025).

Nesse contexto, apontam-se as principais descobertas:

**Cerca de 100 milhões de visualizações e 4 milhões de usuários:** Teorias da conspiração envolvendo autismo alcançaram pelo menos 4.186.031 usuários na América Latina e no Caribe entre 2015 e 2025, totalizando 99.318.993 visualizações, 107.880 reações em 47.261 publicações mapeadas e categorizadas no continente.

**Brasil em primeiro lugar no Continente:** As comunidades brasileiras de teorias da conspiração somam 46% dos conteúdos sobre autismo no continente, totalizando 22.007 publicações, alcançando até 1.726.364 usuários e 13.944.477 visualizações. Seguido do Brasil, Argentina, México, Venezuela e Colômbia também ocupam posição de destaque dentre os países que mais produzem conteúdos conspiracionistas sobre autismo.

**150 falsas causas do autismo, de parasitas a Doritos:** Entre as versões mapeadas, encontram-se desde explicações para "causa" do autismo como deficiência de serotonina e exposição a alumínio, até alegações como consumo de Doritos, inversão do campo magnético da Terra e influência de chemtrails. Outras teorias recorrem ao pânico moral e ao negacionismo científico, atribuindo o autismo ao 5G, Wi-Fi, microondas e até às vacinas.

**150 falsas curas do autismo, de dióxido de cloro a eletrochoque de Tesla:** A falsa promessa de cura para o autismo se tornou um negócio lucrativo, impulsionado por desinformação e oportunismo. Entre as 150 "curas" identificadas, destacam-se práticas perigosas, como o consumo de dióxido de cloro (CDS), conhecido como "MMS", uma substância tóxica promovida como solução milagrosa. Além disso, métodos absurdos como ozonioterapia, terapia de eletrochoque de Tesla e até a ingestão de prata coloidal e azul de metileno são vendidos como tratamentos supostamente eficazes. Muitos desses produtos e práticas são comercializados abertamente por grupos que exploram o desespero de familiares, lucrando com a monetização da mentira e colocando vidas em risco.

**Crescimento de 15.000% no conspiracionismo sobre autismo pós-Pandemia:** A Pandemia da COVID-19 foi a porta de entrada para a explosão da desinformação sobre autismo no Continente. Entre 2019 e 2024 (cinco anos), o volume de conteúdos enganosos cresceu mais de 15.000% (x151), com um aumento expressivo de 635% (x7,35) apenas durante o período pandêmico (2020-2021). Esse crescimento acelerado demonstra como a crise sanitária abriu espaço para narrativas conspiratórias que continuaram se expandindo nos anos seguintes.

### 5.2. Trabalhos futuros

Apesar do escopo abrangente desta pesquisa, algumas limitações merecem destaque e podem ser abordadas em estudos futuros:

**Ampliação para outras plataformas digitais:** O estudo focou exclusivamente no Telegram, mas investigações futuras podem analisar a disseminação de desinformação sobre autismo em outras redes sociais, como X (antigo Twitter), Facebook, Instagram, YouTube e TikTok, para verificar diferenças nas estratégias e alcance dessas narrativas.

**Impacto da desinformação na tomada de decisão das famílias:** Embora este estudo tenha mapeado a disseminação da desinformação, ainda há uma lacuna sobre como essas crenças influenciam decisões médicas e terapêuticas de pais e responsáveis por crianças autistas. Pesquisas futuras podem explorar os impactos diretos dessas narrativas no acesso a tratamentos baseados em evidências.

**Estratégias de monetização e lucro com desinformação:** A comercialização de falsas curas para o autismo é um aspecto crítico da desinformação. Estudos futuros podem investigar a economia da desinformação, identificando quem se beneficia financeiramente com a venda de produtos fraudulentos e como essas práticas são estruturadas.

**Estratégias de combate à desinformação:** Compreender como agências reguladoras, profissionais de saúde e a comunidade científica reagem a essas narrativas pode auxiliar no desenvolvimento de campanhas mais eficazes de combate à desinformação. Estudos podem explorar abordagens de comunicação e a efetividade de intervenções em espaços online.

**Variações regionais na adoção de narrativas conspiratórias:** A pesquisa revelou que diferentes países da América Latina e do Caribe apresentam dinâmicas próprias na disseminação da desinformação sobre autismo, ainda que conectados entre si. Estudos futuros podem aprofundar as diferenças regionais, explorando fatores sociopolíticos e culturais que favorecem ou inibem essas crenças em distintos contextos nacionais.

Por fim, este trabalho busca contribuir com reflexões objetivas à Estratégia Saúde da Família (ESF), ao Programa Nacional de Imunizações (PNI) e à Política Nacional de Humanização (PNH) do Ministério da Saúde (MS). Dessa forma, apresentam-se reflexões direcionadas ao Programa Saúde com Ciência, ao Comitê de Enfrentamento da Desinformação sobre o Programa Nacional de Imunizações e as Políticas de Saúde Pública, do Governo Federal, ao Instituto Brasileiro de Informação em Ciência e Tecnologia (IBICT) e à Frente Parlamentar em Defesa da Vacina da Câmara dos Deputados.

## 6. Referências

## 7. Biografia dos autores

**Ergon Cugler de Moraes Silva** é autista. Possui Mestrado em Administração Pública e Governo pela Fundação Getulio Vargas (FGV), MBA em Data Science & Analytics pela Universidade de São Paulo (USP) e é Bacharel em Gestão de Políticas Públicas pela USP. Colabora com o Observatório Interdisciplinar de Políticas Públicas (OIPP USP), o Grupo de Estudos em Tecnologia e Inovação na Gestão Pública (GETIP USP), o Monitor do Debate Político no Meio Digital (Monitor USP) e o Grupo de Trabalho em Estratégia, Dados e Soberania do Grupo de Estudos e Pesquisa em Segurança Internacional (GEPSI UnB), vinculado ao Instituto de Relações Internacionais da Universidade de Brasília (UnB). Foi pesquisador do Instituto Brasileiro de Informação em Ciência e Tecnologia (IBICT), onde trabalhou com estratégias contra a desinformação para o Governo Federal. Atualmente, está vinculado à Fundação Getulio Vargas (FGV) e ao Conselho Nacional de Desenvolvimento Científico e Tecnológico (CNPq) por meio do Laboratório de Estudos sobre Desordem Informacional e Políticas Públicas (DesinfoPop), com o Centro de Estudos em Administração Pública e Governo (CEAPG/FGV/EAESP). É também diretor de Políticas Públicas, Pesquisa e Incidência da Associação Nacional para Inclusão das Pessoas Autistas (Autistas Brasil). Além disso, está vinculado no programa de pós-graduação em Data Science for Social and Business Analytics da Universitat de Barcelona.Website: https://ergoncugler.com/. Contato: contato@ergoncugler.com.

**Arthur Ataide Ferreira Garcia** é autista. Começou sua atuação como ativista na defesa dos direitos das pessoas autistas aos seus 11 anos de idade, participando e mobilizando ações, palestras e eventos em prol da inclusão dentro de todos os segmentos da sociedade. É palestrante, e vice-presidente da Associação Nacional para Inclusão das Pessoas Autistas (Autistas Brasil), entidade em que foi um dos fundadores. Graduando em Medicina pela Universidade Metropolitana de Santos (UNIMES) e único jovem brasileiro a palestrar na Cúpula de Neurodiversidade de Stanford. Também redigiu e é idealizador da lei estadual 17.759/2023, do “Protocolo Individualizado de Avaliação”. Colabora com diversas iniciativas do Ministério da Educação e do Ministério da Saúde. Website: www.autistas.org.br. Contato: arthur.afg.2003@gmail.com.

**Guilherme de Almeida** é autista. Doutorando (bolsista CAPES de Excelência Acadêmica) e Mestre em Educação pela Faculdade de Educação da Universidade Estadual de Campinas (UNICAMP). Bacharel em Direito pela Pontifícia Universidade Católica do Paraná (PUC/PR). Atualmente, é Presidente da Associação Nacional para Inclusão das Pessoas Autistas (Autistas Brasil). Único pesquisador brasileiro membro do Stanford Neurodiversity Project, onde atua nos Comitês de Inclusão no Ensino Superior e Inclusão no Mercado de Trabalho. Também é membro do Grupo de Estudos e Pesquisas PAIDEIA da Faculdade de Educação da UNICAMP e do Comitê dos Direitos de Pessoas com Deficiência no âmbito Judicial do Conselho Nacional de Justiça (CNJ). Website: www.autistas.org.br. Contato: g229669@dac.unicamp.br.

**Julie Ricard** é pesquisadora, doutoranda em Administração Pública e Governo pela Fundação Getulio Vargas (FGV) e mestre em Relações Internacionais pelo Sciences Po (França) e em Estudos de Gênero pela Université Paris 7. Trabalha com os temas de desinformação, políticas públicas e desigualdades de gênero, com ênfase em saúde, democracia e tecnologias. Atualmente, é bolsista do Conselho Nacional de Desenvolvimento Científico e Tecnológico (CNPq), atuando como pesquisadora do Centro de Estudos em Administração Pública e Governo (CEAPG/FGV/EAESP) por meio do Laboratório de Estudos sobre Desordem Informacional e Políticas Públicas (DesinfoPop). Atuou como consultora da UNESCO, apoiando os esforços de enfrentamento à desinformação sobre saúde pública da Secretaria de Comunicação Social da Presidência da República e como Diretora do programa Tecnologia e Democracia na Data-Pop Alliance. É fundadora da plataforma Eureka, que promove espaços de formação política e cultural por meio de clubes de leitura e cinema em parceria com organizações da sociedade civil e universidades. Website: https://www.eureka.club. Contato: juliec.ricard@gmail.com.